\documentclass[aps,prd,a4paper,onecolumn,showpacs,eqsecnum,showkeys,preprintnumbers,amsmath,amssymb,nofootinbib,usenames,dvipsnames]{revtex4-2}
\usepackage{graphicx}
\usepackage{bm}
\usepackage{dcolumn}
\usepackage{color}
\usepackage{multirow}

\usepackage[titletoc]{appendix}

\usepackage{mathrsfs} 
\usepackage{txfonts} 

\newcommand{\bea}{\begin{eqnarray}}
\newcommand{\eea}{\end{eqnarray}}

\def\beq{\begin{equation}}
\def\eeq{\end{equation}}

\newcommand{\rd}{\mathrm{d}}  

\makeatletter
\def\p@subsection{}
\def\p@subsubsection{}
\makeatother

\def\al{\alpha}
\def\be{\beta}
\def\ga{\gamma}

\def\de{\delta}
\def\De{\Delta}

\def\th{\theta}

\def\la{\lambda}

\def\rh{\rho}
\def\si{\sigma}

\def\ta{\tau}
\def\up{\upsilon}

\begin{document}

\title{Black holes and other exact solutions in six-derivative gravity
\vspace{0.2cm}}

\author{Breno L. Giacchini}
\email{breno.giacchini@matfyz.cuni.cz}
\affiliation{
{\small Institute of Theoretical Physics, Faculty of Mathematics and Physics, Charles University, V Hole{\v s}ovi{\v c}k{\'a}ch 2, 180 00 Prague 8, Czech Republic}
}

\author{Ivan Kol\'a\v{r}}
\email{ivan.kolar@matfyz.cuni.cz}
\affiliation{
{\small Institute of Theoretical Physics, Faculty of Mathematics and Physics, Charles University, V Hole{\v s}ovi{\v c}k{\'a}ch 2, 180 00 Prague 8, Czech Republic}
}

\date{\small\today}

\begin{abstract} 
\noindent

We study exact static spherically symmetric vacuum solutions in generic six-derivative gravity (\textit{i.e.}, without assuming specific relations between the coupling constants). Using modified Schwarzschild coordinates, we systematically classify solutions through Frobenius expansions, determining their number of free parameters and confirming previously known cases, such as the regular solutions at the origin. Importantly, we identify novel solutions absent in four-derivative gravity, including those with (double-degenerate) extreme horizons (and their near-horizon limits) that exist without matter sources, which may indicate the existence of regular black holes. We also find asymptotically (anti-)de Sitter spacetimes, giving rise to an effective cosmological constant. The solutions can be classified into six main classes, and, when possible, we provide the description in standard Schwarzschild coordinates, in which they split into thirteen main solution classes.

\end{abstract}

\maketitle
\noindent


\section{Introduction}

It was recently proved that, for a generic six-derivative gravity, all static spherically symmetric vacuum solutions admitting an expansion in Frobenius series 
(in the standard spherically symmetric coordinates and with exponents that do not depend on the parameters of the model) 
are necessarily regular at $r=0$~\cite{Praha1}. This result is in sharp contrast to gravity models with fewer derivatives: the singular Schwarzschild metric is a solution of both general relativity and four-derivative gravity; moreover, the latter model admits other families of singular solutions~\cite{Stelle78}. 
It is natural, therefore, to wonder what the solutions in six-derivative gravity expanded around other points look like, which families of solutions this model admits, and how many parameters characterize them. The purpose of the present work is to answer these questions and provide more details of the exact solutions in six-derivative gravity.

Although the literature on exact solutions in four-derivative gravity is vast (see, \textit{e.g.},~\cite{Stelle78,Holdom:2002,Stelle15PRL,Stelle15PRD,Podolsky:2018pfe,Svarc:2018coe,Podolsky:2019gro,Pravda:2020zno,Pravda:2024uyv,Holdom:2022zzo,Holdom:2016nek,Cai:2015fia,Feng:2015sbw,Kokkotas:2017zwt,Kehagias:2015ata,Bonanno:2019rsq,Salvio:2019llz,Daas:2022iid,Silveravalle:2022wij,Bonanno:2022ibv,Lu:2012xu,Held:2022abx,Pravdova:2023nbo,Huang:2022urr} and references therein), having received considerable increment in the last decade, studies of classical solutions beyond the linear approximation in models with sixth- and higher-order field equations are relatively rare and in an early stage of consideration~\cite{Holdom:2002,Pawlowski:2023dda,Daas:2024pxs,Rodrigues-da-Silva:2020cpd,Praha1}. 
Among the six-derivative gravities, there has been more research on solutions to particular models with cubic curvature terms (\textit{e.g.},~\cite{Daas:2023axu,Daas:2025lzr,Hennigar:2016gkm,Bueno:2016lrh,Hennigar:2017ego}), which still yield fourth-order field equations. For example, in the cases of higher-dimensional quasi-topological and Einsteinian cubic gravities, the models are built with the requirement of having a reduction of differential order in the field equations for geometries with certain symmetries, which makes them unique if compared to other gravitational actions with the same number of derivatives. The first work to explore the space of solutions in generic gravity models with six, eight and ten derivatives was~\cite{Holdom:2002}, which reported only finding regular solutions. The explanation for this result in the context of a general six-derivative gravity was provided in~\cite{Praha1}, together with the first, preliminary description of some classes of exact solutions. 

The appealing idea that higher derivatives might be able to smooth out the singularities present in general relativity is also inspired by quantum considerations. For instance, an ultraviolet completion of general relativity is expected to resolve the classical singularities, and higher-derivative terms are relevant in many routes towards quantum gravity.\footnote{See, \textit{e.g.},~\cite{Giacchini:2021pmr} for a brief review of the role of higher-derivative terms in quantum gravity.} It is known from the semi-classical approach that terms with up to four metric derivatives are required to have a consistent renormalizable quantum field theory on a curved background~\cite{UtDW}. If the same terms are included in the gravitational action, the corresponding quantum gravity theory is renormalizable~\cite{Stelle77} (in opposition to Einstein gravity, which is non-renormalizable~\cite{hove,GorSag12}). Terms with more than four metric derivatives can improve the convergence of loop integrals even more, yielding super-renormalizable gravity models~\cite{AsoreyLopezShapiro}. Such terms also occur in the low-energy regime of string theory~\cite{Candelas:1985en,Accioly:2016qeb} and as counterterms in the perturbative quantization of general relativity.

How higher-derivative terms are treated varies according to the context in which they appear. In the effective approach, for example, it is customary to take them as interaction vertices only and to discard some terms by applying field redefinitions. In this vein, the solutions obtained ought to be regarded as small perturbations against the solutions of general relativity (see, \textit{e.g.},~\cite{Anselmi:2013wha,deRham:2020ejn,Daas:2023axu,Pawlowski:2023dda,Glavan:2024cfs}). Here, we do not follow this approach; instead, we treat the higher-derivative terms on the same footing as the Einstein--Hilbert one. This procedure is also used when searching for solutions in four-derivative gravity~\cite{Stelle78,Holdom:2002,Stelle15PRL,Stelle15PRD,Svarc:2018coe,Podolsky:2019gro}, and it could be behind a mechanism of singularity resolution in models with more derivatives~\cite{Holdom:2002,Praha1}. One might wonder about the stability of such solutions, for higher derivatives usually are associated with Ostrogradsky instability (in non-degenerate systems). Several proposals to understand stable and meta-stable configurations in classical higher-derivative systems, interactions with ghosts, and recovery of unitarity (from the quantum point of view, including quantum gravity) have been put forward in recent years; see~\cite{Bender:2007wu,Salvio:2015gsi,Salvio:2019ewf,Salvio:2019wcp,Salvio:2024joi,Damour:2021fva,Deffayet:2021nnt,Asorey:2024oxw,Anselmi:2017ygm,Donoghue:2019fcb,ModestoShapiro16,Kuntz:2024rzu,Held:2023aap,Held:2025ckb,Deffayet:2025lnj,CastardellidosReis:2019fgu} and references therein. However, the study of the stability of the solutions we obtain here is beyond the scope of the present work. Our goal is to derive exact solutions to the most general six-derivative gravity model, regardless of whether such an action is the fundamental or emergent one at a certain energy scale.

Compared to our previous work~\cite{Praha1}, which focused on expansions around the origin and proved the regularity of solutions at ${\bar{r} = 0}$ in standard Schwarzschild coordinates (for coupling-independent exponents), the present paper provides a comprehensive classification of exact solutions in six-derivative gravity. Using modified Schwarzschild coordinates, we analyse Frobenius expansions around arbitrary points ${r = r_0}$ and at infinity, which enables a systematic and rigorous counting of free parameters (thanks to the field equations being autonomous in these coordinates). This leads to the discovery of new solution classes, including wormholes, extreme horizons and asymptotically (anti-)de Sitter [(A)dS] spacetimes, and an improved understanding of the ones preliminary reported in~\cite{Praha1}.

This work is organised as follows: 
in Sec.~\ref{Sec.GenSixderG}, we introduce the six-derivative gravity model that is the subject of our study. The types of solutions that we consider are defined in Sec.~\ref{Sec.ClassPosSol}, together with a preliminary discussion of how they relate when expressed in standard and modified Schwarzschild coordinates. In this section, we also present the classes of solutions admitted by the model, with the technical details provided in Appendix~\ref{App}. Each class of solutions corresponding to expansions in powers of $\Delta=r-r_0$ is discussed in Sec.~\ref{Sec.SolExpr0}, whereas the ones in powers of $r^{-1}$ are discussed in Sec.~\ref{Sec.SolExpInvr}. 
In Sec.~\ref{Sec.SumCon}, we summarise the main results and draw our conclusions.
In addition to the aforementioned Appendix~\ref{App}, in Appendix~\ref{App-Lin} we briefly present some results concerning solutions of linearised six-derivative gravity, while Appendix~\ref{app2} contains the explicit (and long) expressions of some quantities defined throughout the work.

\section{General six-derivative gravity: action and field equations}
\label{Sec.GenSixderG}

The gravity model we consider is the most general extension of the Einstein--Hilbert action by terms with a total of four and six derivatives of the metric. Although there are four independent Riemann-polynomial scalars with four metric derivatives, and seventeen with six metric derivatives~\cite{Fulling:1992vm}, in a four-dimensional spacetime there is no loss of generality in working with the action
\beq
\label{mostgeneralaction}
\begin{split}
S  = & \int \rd^4 x \sqrt{-g} \Big[  \al  R  
+ \be_1 R^2 +  \be_2 R_{\mu\nu}^2 
+ \ga_1 R \Box R 
+ \ga_2 R_{\mu\nu} \Box R^{\mu\nu} 
+ \ga_3 R^3 
+ \ga_4 R R_{\mu\nu} R^{\mu\nu} 
\\
&
+ \, \ga_5 R_{\mu\nu} R^\mu{}_\rh R^{\rh \nu}
+ \ga_6 R_{\mu\nu} R_{\rh\si} R^{\mu\rh\nu\si} 
+ \ga_7 R R_{\mu\nu\rh\si} R^{\mu\nu\rh\si}
+ \, \ga_8 R_{\mu\nu\rh\si} R^{\mu\nu\ta\up} R^{\rh\si}{}_{\ta\up} 
\Big] 
,
\end{split}
\eeq 
where the constants $\al$, $\be_{1,2}$ and $\ga_{1,\ldots,8}$ are, respectively, the coefficients of the terms with a total of two, four and six metric derivatives.
All the other terms can be expressed as combinations of the terms in~\eqref{mostgeneralaction} and boundary or topological terms (that do not contribute to the equations of motion)~\cite{Decanini:2007gj}. Three of these combinations are specific to a four-dimensional spacetime: the Gauss--Bonnet topological identity,
\beq
\int \rd^4 x \sqrt{-g} R_{\mu\nu\al\be}^2 \, = \, \int \rd^4 x \sqrt{-g} \big[   4 R_{\mu\nu}^2  - R^2  \big] + \chi ,
\eeq
where $\chi$ is a topological invariant,
and Xu's identities~\cite{Xu:1987pz,Harvey:1995},
\bea
R_{\mu\nu} R^\mu{}_{\rho\si\ta} R^{\nu\rho\si\ta}  & = & \frac{1}{4} R^3 - 2 R R_{\mu\nu} R^{\mu\nu} + 2 R_{\mu\nu} R^\mu{}_\rh R^{\rh \nu} + 2 R_{\mu\nu} R_{\rh\si} R^{\mu\rh\nu\si}  + \frac{1}{4} R R_{\mu\nu\rh\si} R^{\mu\nu\rh\si} ,
\\
R_{\mu\al\nu\be} R^\mu{}_\rh{}^\nu{}_\si R^{\rh\al\si\be} & = & - \frac{5}{8} R^3 + \frac{9}{2} R R_{\mu\nu} R^{\mu\nu} - 4 R_{\mu\nu} R^\mu{}_\rh R^{\rh \nu} - 3 R_{\mu\nu} R_{\rh\si} R^{\mu\rh\nu\si}  - \frac{3}{8} R R_{\mu\nu\rh\si} R^{\mu\nu\rh\si} + \frac{1}{2} R_{\mu\nu\rh\si} R^{\mu\nu\ta\up} R^{\rh\si}{}_{\ta\up} .
\eea
Throughout this work, we use the term ``general six-derivative gravity'' to refer to the model~\eqref{mostgeneralaction} with the assumption that all the coefficients $\al$, $\be_{1,2}$ and $\ga_{1,\ldots,8}$ are non-zero and completely unrelated, \textit{i.e.}, there is no special relation between them. 
In addition, we assume 
\beq
\label{ga1ga2}
\ga_2(3  \ga_1 + \ga_2 ) \neq 0
\eeq
to guarantee that there is no reduction of the total differential order of the field equations that we shall solve.\footnote{It might be also instructive to recall that, from the quantum gravity point of view,~\eqref{ga1ga2} is necessary for the model to be renormalizable~\cite{AsoreyLopezShapiro}.}
This also means that the model propagates two pairs of spin-0 and spin-2 massive excitations when linearised around the flat background (see Appendix~\ref{App-Lin} for further details).

Applying the variational principle to the action~\eqref{mostgeneralaction}, we obtain the vacuum field equations
\beq
\label{EoM}
\mathcal{E}_{\mu\nu} \equiv \frac{1}{\sqrt{-g}}  \frac{\de S}{\de g^{\mu\nu}} = 0,
\eeq
that satisfy the generalised Bianchi identity
\beq
\label{Bianchi0}
\nabla^\nu \mathcal{E}_{\mu\nu} = 0.
\eeq

To study static and spherically symmetric solutions of~\eqref{EoM}, it is convenient to work in modified Schwarzschild coordinates such that
\beq
\label{metric}
\rd s^2 = - H(r) \rd t^2 + \frac{\rd r^2}{H(r)} + F^2(r) \left( \rd \th^2 + \sin^2 \th \rd \phi^2 \right) 
.
\eeq
A metric in this form admits the residual gauge transformations
\beq
\label{ResGauge}
t \longrightarrow t^\prime = \la t , \qquad r \longrightarrow r^\prime = \la^{-1} r + \nu ,
\eeq
with two parameters $\la$ and $\nu$. 
The main advantage of these coordinates is that the field equations~\eqref{EoM} become an autonomous system, \textit{i.e.}, the only dependence on $r$ is through $F(r)$ and $H(r)$. This property is extremely useful for proving that the functions $F(r)$ and $H(r)$ can be obtained as series with coefficients following recursive relations and to determine the number of parameters characterizing a solution; in addition, it also expedites the calculations.
The same result can be achieved using a metric written in conformal-to-Kundt form, as it was successfully applied to four-derivative gravity~\cite{Podolsky:2018pfe,Svarc:2018coe,Podolsky:2019gro}. Nevertheless, the structure of the field equations in the six-derivative gravity does not allow all the beneficial simplifications that such coordinates have in that model, and a metric in the form~\eqref{metric} is enough for our purposes.

The solutions in the modified Schwarzschild coordinates~\eqref{metric} might be mapped to solutions in standard Schwarzschild coordinates,
\beq
\label{metric-Standard}
\rd s^2 = - B(\bar{r}) \rd t^2 + A(\bar{r}) \rd \bar{r}^2 + \bar{r}^2 \left( \rd \th^2 + \sin^2 \th \rd \phi^2 \right) 
,
\eeq
via the correspondence
\beq
\label{Fr}
\bar{r} = F(r),
\eeq
which gives
\beq
\label{DicAB}
A(\bar{r}) = \frac{1}{H(r) F^{\prime 2}(r)} , \qquad B(\bar{r}) = H(r) ,
\eeq
in which $r=r(\bar{r})$ should be obtained by inverting~\eqref{Fr}. 
Notice that the metric~\eqref{metric-Standard} admits a 1-parameter residual gauge freedom, namely, a re-scaling of the time coordinate
\beq
\label{ResGauge-Standard}
t \longrightarrow t^\prime = \la t.
\eeq
Although most of the considerations in the present work are carried out in the coordinates~\eqref{metric}, for each solution found we shall discuss if and how it is expressed in the form~\eqref{metric-Standard}, which not only provides a more direct physical interpretation but also facilitates comparison with known solutions in four- and six-derivative gravity~\cite{Stelle78,Stelle15PRD,Podolsky:2019gro,Praha1}.

For a metric in the form~\eqref{metric}, the field equations~\eqref{EoM} are  diagonal,
\beq
\mathcal{E}_{\mu\nu} = \text{diag} \left( \mathcal{E}_{tt}(r) , \, \mathcal{E}_{rr}(r), \, \mathcal{E}_{\th\th}(r) , \,  \mathcal{E}_{\th\th}(r) \sin^2\th \right) ,
\eeq
and the generalised Bianchi identity~\eqref{Bianchi0} provides a constraint between the components,
\beq
\label{Bianchi1}
\left( \frac{2  H F'}{F}+\frac{3}{2}  H' \right) \mathcal{E}_{rr} + H \mathcal{E}_{rr}' +\frac{H'}{2 H^2} \mathcal{E}_{tt}  -\frac{2 F'}{F^3} \mathcal{E}_{\theta\theta} = 0.
\eeq
Hence, there are only two independent quantities among $\lbrace \mathcal{E}_{tt} , \, \mathcal{E}_{rr}, \, \mathcal{E}_{\th\th} \rbrace$. Despite this fact, the structure of~\eqref{Bianchi1}, with non-trivial prefactors for each term $\mathcal{E}_{\mu\nu}$, prevents us from simply choosing two components to solve order by order using the Frobenius technique.\footnote{The situation here is different from what happens in the standard spherically symmetric coordinates~\eqref{metric-Standard}, for which~\eqref{Bianchi1} can be solved for $\mathcal{E}_{\th\th}$ without any further assumption on the form of the functions in the action (see, \textit{e.g.},~\cite{Stelle15PRD,Holdom:2002}). In that case, in the Frobenius technique, it is sufficient to solve $\mathcal{E}_{tt}=0$ and $\mathcal{E}_{\bar{r}\bar{r}}=0$ order by order, as in~\cite{Praha1}.} In certain cases, it might be necessary to check whether the third component of the field equations --- or, equivalently, the identity~\eqref{Bianchi1} --- holds.
This situation will become clear when we consider each family of solutions individually.
For the following discussion, it also useful to rewrite the identity~\eqref{Bianchi1} in terms of the quantities $\mathcal{E}^{\mu}{}_{\nu}$,
\beq
\label{bia}
{\mathcal{E}^{r}{}_{r}}^\prime + \frac{2 (\mathcal{E}^{r}{}_{r} - \mathcal{E}^{\th}{}_{\th} ) F' }{F } + \frac{(\mathcal{E}^{r}{}_{r} - \mathcal{E}^{t}{}_{t} ) H'}{2 H} = 0.
\eeq

We calculated the field equations~\eqref{EoM} for a metric of the form~\eqref{metric} using the package \textsc{xAct}~\cite{xAct,Brizuela:2008ra,xCoba} for \textit{Mathematica}~\cite{Mathematica}. We do not write down the complete expressions due to their length, but we shall refer to relevant aspects of them when necessary. For the moment, we mention that $\mathcal{E}_{tt}(r)$ and $\mathcal{E}_{\th\th}(r)$ are of sixth order in derivative in both $F(r)$ and $H(r)$, while $\mathcal{E}_{rr}(r)$ is of fifth order in both $F(r)$ and $H(r)$. The terms of highest order in derivative are proportional to the parameters $\ga_1$ and $\ga_2$ of the action~\eqref{mostgeneralaction}, for the terms $R\Box R$ and $R_{\mu\nu}\Box R^{\mu\nu}$ contain the largest number of derivatives acting on a single metric component. As already mentioned, the field equations depend on the coordinate $r$ only through $F(r)$ and $H(r)$; this is the crucial advantage of the coordinates~\eqref{metric} over the standard spherically symmetric ones~\eqref{metric-Standard}.

\section{Classification of possible solutions}
\label{Sec.ClassPosSol}

The solutions we consider in this work are of two types: Frobenius series around a point $r=r_0$ and asymptotic solutions for $r \to \infty$ in the form of Frobenius series for the variable $1/r$. In the former case, we assume that $F(r)$ and $H(r)$ can be expressed as
\beq
\label{FrobeniusHF}
F(r) = \De^\si \sum_{n=0}^\infty f_{n} \De^n , \qquad  H(r) = \De^\tau \sum_{n=0}^\infty h_{n} \De^n   , \quad \text{with} \quad  \De \equiv r - r_0 \quad \text{and} \quad f_0,h_0 \neq 0.
\eeq
Here, $\si$ and $\tau$ are real parameters that have yet to be determined and are assumed to be independent of the particular values of the couplings $\alpha,\beta_{1,2}$ and $\gamma_{1,\ldots,8}$. The condition $f_0,h_0 \neq 0$ guarantees that the leading order of the expansions is governed by the values of $\si$ and $\tau$. Because of the freedom to shift the coordinate $r$, given by the residual gauge transformation~\eqref{ResGauge}, the point $r_0$ in this local analysis does not have a particular physical meaning. Therefore, we can group all the solutions of the form~\eqref{FrobeniusHF} in classes labelled by the pair $\lbrace \si,\tau \rbrace$.\footnote{This choice of notation using curly brackets is reminiscent of the practice in quadratic gravity, for which Frobenius solutions in standard Schwarzschild coordinates are denoted by round brackets~\cite{Stelle78,Stelle15PRD}, while square brackets are used for those in Kundt coordinates~\cite{Podolsky:2018pfe,Podolsky:2019gro}.}

On the other hand, for the asymptotic solutions, we adopt the form
\beq
\label{FrobeniusHF-Infinity}
F(r) = r^\si \sum_{n=0}^\infty f_{n} r^{-n} , \qquad  H(r) = r^\tau \sum_{n=0}^\infty h_{n} r^{-n}   , \quad \text{with}  \quad f_0,h_0 \neq 0,
\eeq
where it is assumed that $r \to \infty$. We denote solution classes of this type by the pair $\lbrace \si,\tau \rbrace^\infty$, with the superscript indicating that it should be regarded as an expansion around infinity, with decreasing powers of $r$. The leading behaviour of the solution is again dictated by the parameters $\si$ and $\tau$.

\subsection{Relation to solutions in standard Schwarzschild coordinates}
\label{Sec.Corresp}

The relation between a solution written in modified coordinates~\eqref{metric} and in the standard Schwarzschild coordinates~\eqref{metric-Standard} via the correspondence~\eqref{Fr} depends on whether it is an expansion around a finite $r=r_0$ or an asymptotic expansion, and on the indicial structure of the solution. It might also depend on the first coefficients of the function $F(r)$. 
Indeed, from Eq.~\eqref{Fr} and assuming that $F(r)$ is not a constant function, it follows that:
\begin{itemize}
\item[i.] In the case of the expansions~\eqref{FrobeniusHF}, if ${r\to r_0 \geqslant 0}$, then
\begin{equation}
\label{map1}
    \begin{aligned}
        \sigma>0 &\implies \bar{r}\to0 ,
        \\
        \sigma=0 &\implies \bar{r}\to\bar{r}_0>0 ,
        \\
        \sigma<0 &\implies \bar{r}\to\infty .
    \end{aligned}
\end{equation}
\item[ii.] In the case of the expansions~\eqref{FrobeniusHF-Infinity}, if ${r\to\infty}$, then
\begin{equation}
\label{map2}
    \begin{aligned}
        \sigma>0 &\implies \bar{r}\to\infty ,
        \\
        \sigma=0 &\implies \bar{r}\to\bar{r}_0>0 ,
        \\
        \sigma<0 &\implies \bar{r}\to0 .
    \end{aligned}
\end{equation}
\end{itemize}
An exception to the reasoning above occurs if $\sigma = 0$ and $F(r)=f_0$ is constant, since the relation~\eqref{Fr} cannot be inverted to define $r=r(\bar{r})$. As a consequence, in this case, the metric~\eqref{metric} is a direct product of two 2-dimensional metrics and cannot be expressed in the form~\eqref{metric-Standard}.

Having identified how the indicial structures $\lbrace \si,\tau \rbrace$ and $\lbrace \si,\tau \rbrace^\infty$ define 
whether the expansion in the coordinate $\bar{r}$ is around the origin, a certain $\bar{r}_0 \neq 0$, or infinity,
let us consider these three possibilities separately.
In the case of expansions around $\bar{r}=0$, we assume a representation in Frobenius series
\beq
\label{FrobeniusAB0}
A(\bar{r}) = \bar{r}^s \sum_{n=0}^\infty a_{n} \, \bar{r}^n , \qquad  B(\bar{r}) = b_{0} \bar{r}^t \left( 1 + \sum_{n=1}^\infty b_{n} \, \bar{r}^n  \right)  ,  \qquad a_0,b_0 \neq 0,
\eeq
and denote the solution class of this type by $(s,t)_0$. A similar expansion can be used around a point $\bar{r}_0 \neq 0$,
\beq
\label{FrobeniusABDe}
A(\bar{r}) = \bar{\De}^s \sum_{n=0}^\infty a_{n} \, \bar{\De}^n , \qquad  B(\bar{r}) = b_{0} \bar{\De}^t \left( 1 + \sum_{n=1}^\infty b_{n} \, \bar{\De}^n  \right)  , \qquad \bar{\De} \equiv \bar{r} - \bar{r}_0 , \qquad a_0,b_0 \neq 0,
\eeq
although in this case it might be more convenient to define
\beq
\label{FrobeniusC}
A(\bar{r}) = \frac{1}{C(\bar{r})} , \qquad
C(\bar{r}) = \bar{\De}^{-s} \sum_{n=0}^\infty c_{n} \, \bar{\De}^n , \qquad \bar{\De} \equiv \bar{r} - \bar{r}_0 , \qquad c_0 \neq 0.
\eeq
In consonance with a tradition in the field (see, \textit{e.g.}~\cite{Stelle78,Stelle15PRD,Podolsky:2019gro}), we shall denote this class of solutions by $(-s,t)_{\bar{r}_0}$.
Finally, for asymptotic expansions in decreasing powers of $r$, we write
\beq
\label{FrobeniusAB-inf}
A(\bar{r}) = \frac{1}{C(\bar{r})} , \qquad
C(\bar{r}) = \bar{r}^{-s} \sum_{n=0}^\infty c_{n} \, \bar{r}^{-n} , \qquad  B(\bar{r}) = b_{0} \bar{r}^t \left( 1 + \sum_{n=1}^\infty b_{n} \, \bar{r}^{-n}  \right) \qquad c_0,b_0 \neq 0 ,
\eeq
and denote by $(-s,t)^\infty$.

Using the relations~\eqref{Fr} and~\eqref{DicAB}, one can obtain the correspondence between the solution written in the two coordinates at leading order:\footnote{The mapping between the two series in different coordinates was verified at the higher orders for the relevant solutions.}
\begin{itemize}
\item If $\si \neq 0$, we obtain
\beq
\label{st1}
s = \frac{2 - 2\si - \tau}{\si} , \qquad t = \frac{\tau}{\si} .
\eeq
According to~\eqref{map1} and~\eqref{map2}, this scenario could correspond to expansions around $\bar{r} = 0$ or asymptotic expansions as $\bar{r}\to\infty$.

\item If $\sigma=0$ and $r\to r_0\geqslant 0$, then
\beq
\label{st2}
s=-\tau , \qquad t = \tau , \qquad \text{ if } \qquad f_1\neq 0.
\eeq
However, in the general case when there exists an integer  $N \in \lbrace 2,3,\ldots \rbrace$ such that $f_{1,\ldots,N-1} = 0$ and $f_N \neq 0$, \textit{i.e.},
\beq
F(r) = f_0 + \sum_{n=0}^\infty f_{N+n} \De^{N+n} ,
\eeq
then the solutions in the class $\lbrace 0,\tau \rbrace$ might be mapped to solutions that are of non-Frobenius type in the standard coordinates~\eqref{metric-Standard}. This happens because the identification~\eqref{Fr} yields
\beq
\bar{\De} \sim f_N \De^N .
\eeq
In other words, the formal series of $C(\bar{r})$ and $B(\bar{r})$ increase by non-integer steps $\bar{\De}^{1/N}$,
\beq
\label{Non-FrobeniusAB}
C(\bar{r}) = \bar{\De}^{-s} \sum_{n=0}^\infty c_{n} \bar{\De}^{\frac{n}{N}} , \qquad  B(\bar{r}) = b_{0} \bar{\De}^t \left( 1 + \sum_{n=1}^\infty b_{n} \bar{\De}^{\frac{n}{N}}  \right) .
\eeq
In such cases, the correspondence between the indicial structure of the solutions is via
\beq
\label{st3}
s = - \frac{\tau}{N} - \frac{2(N-1)}{N}   , \qquad t = \frac{\tau}{N} .
\eeq
In particular, note that for $N=1$ the above formulas recover the correspondence~\eqref{st2} when $f_1\neq 0$, and we have a Frobenius series. We shall identify the classes of solutions that have the structure~\eqref{Non-FrobeniusAB} by the label $(-s,t)_{\bar{r}_0,1/N}$.

\item If $\sigma=0$ and $r\to \infty$, then the relation between $s$, $t$, $\sigma$ and $\tau$ also depends on the order of first non-zero coefficient $f_N$. We shall not discuss this scenario due to the fact that we did not encounter a solution in this category. The only solution of type $\lbrace 0,\tau\rbrace^\infty$ that we find has $f_n = 0$ for all $n = 1,2,\ldots$ and thence cannot be described by the standard spherically symmetric coordinates.
\end{itemize}

\subsection{Solutions of the indicial equations}

The first step to identify the possible classes $\lbrace \si,\tau \rbrace$ and $\lbrace \si,\tau \rbrace^\infty$ of solutions admitted by the general six-derivative gravity is to verify for which values of the parameters $\sigma$ and $\tau$ the field equations can be solved at the leading order in the expansion. 
To this end, it is convenient to work with the field equations in the form
\beq
\label{FieldEqUD}
\mathcal{E}^{\mu}{}_{\nu} = 0,
\eeq
so that all the components have the same leading behaviour.

By substituting
\beq
F(r) \sim \De^\sigma , \qquad H(r) \sim \De^\tau
\eeq
into~\eqref{FieldEqUD}, we notice that there are nine types of structures that might contribute to the leading-order expansion of the field equations, depending on how $\sigma$ and $\tau$ relate to each other. 
Contributions coming from the Einstein--Hilbert term (proportional to $\al$) are of the form
\begin{equation} \label{9TermsA}
T_1  \propto  \De^{-2 \sigma }, \qquad T_2  \propto  \De^{\tau -2},
\end{equation} 
the ones originated from the four-derivative terms (proportional to $\beta_{1,2}$) are
\begin{equation} \label{9TermsB}
T_3 \propto  \De^{-4 \sigma }, \qquad T_4  \propto  \De^{-2 \sigma +\tau -2}, \qquad T_5  \propto  \De^{2 \tau -4},
\end{equation} 
while those from six-derivative terms (proportional to $\gamma_{1,\ldots,8}$) are
\begin{equation} \label{9TermsC}
T_6  \propto  \De^{-6 \sigma }, \qquad T_7  \propto  \De^{-4 \sigma +\tau -2}, \qquad T_8  \propto  \De^{-2 \sigma +2 \tau -4} , \qquad T_9  \propto  \De^{3 \tau -6}  .
\end{equation} 
Therefore, for the expansions~\eqref{FrobeniusHF} around $r=r_0$, at the leading (lowest) order, we have
\beq
\label{pfeld}
\mathcal{E}^{\mu}{}_{\nu}(r) \sim \De^{p(\si,\ta)},
\eeq
where
\beq
p(\si,\ta) = \min\lbrace -2 \sigma , -4 \sigma , -6 \sigma , \tau -2 , -2 \sigma +\tau -2 , -4 \sigma +\tau -2 , 2 \tau -4 , -2 \sigma +2 \tau -4 , 3 \tau -6 \rbrace.
\eeq

Regarding the asymptotic expansions, since the field equations do not explicitly depend on the coordinate $r$, the types of structures that occur when substituting
\beq
F(r) \sim r^\sigma , \qquad H(r) \sim r^\tau 
\eeq
into the field equations~\eqref{FieldEqUD} are also given by Eqs.~\eqref{9TermsA}--\eqref{9TermsC} via the replacement $\De \mapsto r$. Now, the leading term of the expansion of $\mathcal{E}^{\mu}{}_{\nu}$ is the highest-order one,
\beq
\label{pfeld2}
\mathcal{E}^{\mu}{}_{\nu}(r) \sim r^{p(\si,\ta)},
\eeq
with
\beq
p(\si,\ta) = \max\lbrace -2 \sigma , -4 \sigma , -6 \sigma , \tau -2 , -2 \sigma +\tau -2 , -4 \sigma +\tau -2 , 2 \tau -4 , -2 \sigma +2 \tau -4 , 3 \tau -6 \rbrace.
\eeq

\begin{table}[t]
    \centering
    \begin{tabular}{|c|c|c|c|c|}
            \hline
            \multirow{2}{*}{Case} & Leading power & Terms that contribute to   & \multicolumn{2}{|c|}{Relevant scenario for:}   \\ 
                   & $p(\si,\ta)$ & the indicial equations & Expansions around $r=r_0$  & Asymptotic expansions \\ 
            \hline
            I & $-2 \sigma$ & $T_1$ & $\sigma < 0$ and $\tau > 2-2 \sigma$  & $\sigma >0$ and $\tau <2-2 \sigma$ \\ 
            \hline
            II  & $-2 \sigma$ & $T_1$, $T_2$ & $\sigma < 0$ and $\tau =2-2 \sigma$  & $\sigma >0$ and $\tau = 2-2 \sigma$ \\ 
            \hline
            III  & $\tau-2$ & $T_2$ & $\sigma < 0$ and $2<\tau <2-2 \sigma$  & $\sigma > 0$ and $2-2 \sigma <\tau <2$ \\ 
            \hline
            IV  & 0 & $T_2$, $T_5$, $T_9$ & $\sigma < 0$ and $\tau =2$  & $\sigma > 0$ and $\tau =2$ \\ 
            \hline
            \multirow{2}{*}{V}  & \multirow{2}{*}{$3 \tau -6$} & \multirow{2}{*}{$T_9$} & $\sigma < 0$ and $\tau < 2$,  & $\sigma > 0$ and $\tau > 2$, \\ 
               & 	 & 	 & $\sigma \geqslant 0$ and $\tau <2-2 \sigma$  & $\sigma \leqslant 0$ and $\tau >2-2 \sigma$ \\ 
            \hline
            VI  & 0 & $T_1$, $T_3$, $T_6$ & $\sigma = 0$ and $\tau > 2-2 \sigma =2$  & $\sigma = 0$ and $\tau <2-2 \sigma =2$ \\ 
            \hline
            VII  & 0 & $T_1,\ldots, T_9$ & $\sigma = 0$ and $\tau =2-2 \sigma =2$  & $\sigma = 0$ and $\tau =2-2 \sigma =2$ \\ 
            \hline
            VIII  & $-6\sigma$ & $T_6$ & $\sigma >0$ and $\tau >2-2 \sigma$  & $\sigma < 0$ and $\tau <2-2 \sigma$ \\ 
            \hline
            IX  & $-6\sigma$ & $T_6,\ldots, T_9$ & $\sigma >0$ and $\tau =2-2 \sigma$  & $\sigma < 0$ and $\tau =2-2 \sigma$ \\ 
            \hline
        \end{tabular}    
        \caption{\small Summary of the nine possible cases for the system of indicial equations: the leading power of the expansion of the field equations~\eqref{pfeld} and~\eqref{pfeld2}, the types of terms~\eqref{9TermsA}, \eqref{9TermsB} and~\eqref{9TermsC} that are relevant at this order, and which ranges of values of $\sigma$ and $\tau$ are described by each case.
         }
\label{Tab0}
\end{table}

Since solving the field equations at leading order allows us to identify the possible 
indicial structures $\lbrace \sigma,\tau \rbrace$ and $\lbrace \sigma,\tau \rbrace^\infty$
for which a solution might exist, we shall refer to the field equations at leading order as \emph{indicial equations}.
Depending on the relation between $\sigma$ and $\tau$, different terms among~\eqref{9TermsA}--\eqref{9TermsC} can contribute to the leading-order term of the field equations. In Table~\ref{Tab0} we summarise the possible scenarios for expansions around $r=r_0$ and for asymptotic expansions, while in Appendix~\ref{App} we present the explicit expressions and detailed analysis of the indicial equations in each one of those scenarios.

It is important to stress that our analysis of the indicial equations assumes that its solutions $\sigma$ and $\tau$ are independent of the particular values of the couplings of the model.\footnote{This assumption is analogous to the one usually applied when searching for solutions in a generic quadratic gravity model, where solutions with indices depending on $\beta_{1}$ and $\beta_{2}$ are known to exist~\cite{Stelle15PRD}, but are yet to be studied.} This means that, in principle, the classes of solutions $\lbrace \sigma,\tau \rbrace$ and $\lbrace \sigma,\tau \rbrace^\infty$ identified here are common to all the six-derivative models given by the action~\eqref{mostgeneralaction}. We used the term ``in principle'' because some classes of solutions do depend on weak constraints between the couplings of the model, in the form of inequalities. To explain this issue better, let us distinguish
between the following four possible situations for the indicial equations and their solutions:
\begin{itemize}
\item[i.] In Cases I, III, V and VIII (see Table~\ref{Tab0}), the leading-order term of the field equations essentially depends only on $\sigma$, $\tau$ and the parameters of the model; all the dependence on the coefficients $f_0$ and $h_0$ are in the form of an irrelevant multiplicative factor. 
The requirement that the indicial equations are solved for $\sigma$ and $\tau$ irrespective of the values of the parameters $\alpha$, $\beta_{1,2}$ and $\gamma_{1,\ldots,8}$ translates into a system of equations whose only solutions correspond to the indicial structures $\lbrace 0,0 \rbrace$ and $\lbrace 0,1 \rbrace$, as we show in Appendix~\ref{App}.

\item[ii.] Cases II and IX are characterized by a relation between $\sigma$ and $\tau$ (namely, $\tau = 2 - 2\sigma$) and the indicial equations depend on the coefficients $f_0$ and $h_0$ in a non-trivial way through the factor $z \equiv f_0^2 h_0$. Therefore, assuming the independence of the index $\sigma$ on the parameters $\alpha$, $\beta_{1,2}$ and $\gamma_{1,\ldots,8}$, we obtain a system that might be solved for $\sigma$ and $z$. The solution for $z$ fixes the relation between the coefficients $f_0$ and $h_0$ in a way that does not depend on the parameters of the model. For instance, in Appendix~\ref{App} we show that the only solutions in these cases correspond to the indicial structures $\lbrace 1,0 \rbrace$ and $\lbrace 1,0 \rbrace^\infty$, in both cases with $f_0^2 h_0 = 1$.

\item[iii.] In Cases IV ($\tau = 2$) and VI ($\sigma = 0$), one parameter is free while the other is fixed to a constant value. In addition to this free parameter, the leading term of the field equations depends on the coefficient $h_0$ (Case IV) or $f_0$ (Case VI) in a non-trivial way. Therefore, solving the indicial equations requires fixing $h_0$ or $f_0$ as a function of the parameters of the model. This might only be possible if the parameters $\alpha$, $\beta_{1,2}$ and $\gamma_{1,\ldots,8}$ satisfy certain constraints. Indeed, as shown in Appendix~\ref{App}, the only solution in theses cases has indicial structure $\lbrace 1,2 \rbrace^\infty$ and it requires that
\beq
\label{cons12}
\frac{1}{\alpha} (144 \gamma_{3} +36  \gamma_{4} +9  \gamma_{5} +9  \gamma_{6} +24  \gamma_{7} +4  \gamma_{8}) > 0.
\eeq
In this regard, this solution of the indicial equations is slightly different from the ones mentioned before, since it presumes a relation between the parameters of the model (in the form of an inequality, though). 

\item[iv.] Finally, in the special Case VII, both parameters $\sigma$ and $\tau$ are fixed, and the ``indicial equations'' are, actually, equations for the first coefficients of the expansions $f_0$ and $h_0$. Similar to the cases described in the previous item, these solutions also assume certain inequality relations between the parameters of the model (see Appendix~\ref{App}), which
hold for a significant portion of the parameter space. The corresponding indicial structures are $\lbrace 0,2 \rbrace$ and $\lbrace 0,2 \rbrace^\infty$, and the explicit solutions for $f_0$ and $h_0$ can be found in Eq.~\eqref{SolCaseVII}.
\end{itemize}

To summarise the analysis of the indicial equations for the general six-derivative gravity, the possible classes of solutions of type~\eqref{FrobeniusHF} expanded around $r=r_0$ are
\beq
\label{ListSolr0}
\boxed{
\lbrace 1,0 \rbrace , \quad \lbrace 0,0 \rbrace , \quad \lbrace 0,1 \rbrace , \quad \lbrace 0,2 \rbrace ,
}
\eeq
whereas for asymptotic expansions of the type~\eqref{FrobeniusHF-Infinity} we have
\beq
\label{ListSolrInf}
\boxed{
\lbrace 1,0 \rbrace^\infty , \quad \lbrace 1,2 \rbrace^\infty , \quad  \lbrace 0,2 \rbrace^\infty .
}
\eeq
In the next two sections, we shall discuss each of these classes of solutions and how they are expressed in standard spherically symmetric coordinates.

We close this section by recalling that any specific six-derivative gravity might admit solutions belonging to classes not listed in~\eqref{ListSolr0} and~\eqref{ListSolrInf}. This happens because the indicial equations also admit solutions that explicitly depend on the couplings of the model, in the form $\sigma=\sigma(\al,\beta_{1},\beta_{2},\gamma_{1},\ldots,\gamma_{8})$ and $\tau=\tau(\al,\beta_{1},\beta_{2},\gamma_{1},\ldots,\gamma_{8})$. Thus, if some of the couplings $\al,\beta_{1,2},\gamma_{1,\ldots,8}$ are switched off or taken in particular combinations, the space of solutions of the indicial equations might change. An analysis of the complete space of solutions is extraordinarily complicated, given that it would depend on the eleven coefficients $\al$, $\be_{1,2}$ and $\ga_{1,\ldots,8}$. Although here we shall not discuss solutions to particular models, some examples can be found in~\cite{Praha1,Daas:2023axu}.

\section{Solutions expanded around $r=r_0$}
\label{Sec.SolExpr0}

\subsection{Solutions in the class $\lbrace 1,0 \rbrace$}
\label{Sec-10}

The indicial equations for solutions of type $\lbrace 1,0 \rbrace$ fix the relation between the coefficients $f_0$ and $h_0$ such that $f_0^2 h_0 = 1$. Using this condition, the expansion of the field equations in powers of $\De$ starts at order $\De^{-4}$,
\beq
\label{ExpEoM0010}
\mathcal{E}^{\mu}{}_{\nu} = \sum_{i=-4}^\infty \left( \mathcal{E}^{\mu}{}_{\nu} \right)_{i} \De^i  = 0 .
\eeq
Substituting this expansion of $\mathcal{E}^{\mu}{}_{\nu}$ into the generalised Bianchi identity~\eqref{bia}, at the lowest order we obtain
\beq
\label{bia10}
- \frac{ 2 }{\De^5}  \left[ \big(\mathcal{E}^{r}{}_{r} \big)_{-4} + \big({\mathcal{E}^{\theta}}_{\theta} \big)_{-4}
\right] 
+ O (\De^{-4}) = 0 \quad \Longrightarrow \quad \big({\mathcal{E}^{\theta}}_{\theta} \big)_{-4} = - \big(\mathcal{E}^{r}{}_{r} \big)_{-4}.
\eeq
More generally, if the field equations are solved up to an order $\De^N$, \textit{i.e.}, $\big(\mathcal{E}^{\mu}{}_{\nu} \big)_{n} = 0$ for all $n \in \lbrace -4, -3, \ldots , N \rbrace$, then~\eqref{bia} yields
\beq
\label{bia10n}
\left[  (N+3) \big(\mathcal{E}^{r}{}_{r} \big)_{N+1} - 2 \big({\mathcal{E}^{\theta}}_{\theta} \big)_{N+1} \right]   \De^N + O(\De^{N+1}) = 0
\quad \Longrightarrow \quad  \big({\mathcal{E}^{\theta}}_{\theta} \big)_{N+1} = \frac{N+3}{2} \big(\mathcal{E}^{r}{}_{r} \big)_{N+1} .
\eeq
Therefore, by induction it follows that for solutions of type $\lbrace 1,0 \rbrace$ it is sufficient to solve the field equations $\mathcal{E}^{t}{}_{t} = 0$ and $\mathcal{E}^{r}{}_{r} = 0$ order by order.\footnote{Alternatively, one can solve $\mathcal{E}^{t}{}_{t} = 0$ and $\mathcal{E}^{\th}{}_{\th} = 0$ order by order, and verify if $\mathcal{E}^{r}{}_{r} = 0$ is solved at order $\De^{-2}$. This last requirement is necessary because for $N=-3$ the last equation in~\eqref{bia10n} would be satisfied even if $\big(\mathcal{E}^{r}{}_{r} \big)_{-2} \neq 0$.}

Expanding the field equations, we find, at the lowest order,
\beq
\begin{aligned}
\label{Sys10-4}
\big(\mathcal{E}^{t}{}_{t} \big)_{-4}  & =  -
\left( \mathfrak{a}_1   f_1^2 +  \mathfrak{a}_2   f_0^3  f_1  h_1 + \mathfrak{a}_3   f_0^6  h_1^2 \right) \frac{1 }{ f_0^8} 
= 0,
\\
\big(\mathcal{E}^{r}{}_{r} \big)_{-4}  & =  
\left( \mathfrak{a}_1^\prime   f_1^2 + \mathfrak{a}_2^\prime   f_0^3  f_1  h_1 + \mathfrak{a}_3^\prime   f_0^6  h_1^2 \right) \frac{1 }{ f_0^8} 
= 0,
\end{aligned}
\eeq
where we define the quantities $\mathfrak{a}_{1,2,3}$ and $\mathfrak{a}_{1,2,3}^\prime$ whose explicit expressions can be found in Eq.~\eqref{osas123elinha}.
The only solution for~\eqref{Sys10-4} (without assuming any particular relation between the coefficients $\gamma_{1,\ldots,8}$)  is
\beq
\label{10f1h1}
f_1 = h_1 = 0.
\eeq

At order $\De^{-3}$, 
the condition~\eqref{10f1h1} automatically solves
\beq
\big(\mathcal{E}^{t}{}_{t} \big)_{-3}  = \frac{1}{ f_0^9}
\left(\mathfrak{b}_1  f_1^3 + \mathfrak{b}_2 f_0^3 f_1^2  h_1 + \mathfrak{b}_3 f_0^6 f_1  h_1^2 + \mathfrak{b}_4 f_0^9 h_1^3 \right) = 0 ,
\eeq
because the terms in $\big(\mathcal{E}^{t}{}_{t} \big)_{-3}$ are proportional to $f_1$ or $h_1$ (here, $\mathfrak{b}_{1,\ldots,4}$  are combinations of $\ga_{1,\ldots,8}$). For the other component we have
\beq
0 = \big(\mathcal{E}^{r}{}_{r} \big)_{-3}  =  
- \frac{16 }{f_0^7} \left[ 4 (8 \ga_1 +3 \ga_2) f_3 + (10 \ga_1 + 3 \ga_2)  f_0^3 \, h_3 \right] ,
\eeq
which defines a relation between $f_3$ and $h_3$, namely,
\beq
\label{Constfh10}
4 (8 \ga_1 + 3 \ga_2 ) f_3 = - (10 \ga_1 + 3 \ga_2) f_0^3 \, h_3 .
\eeq
Thus, one parameter among $f_3$ and $h_3$ is free.

At the next order $\De^{-2}$, we find that
\beq
\label{Sys10-2}
\big(\mathcal{E}^{t}{}_{t} \big)_{-2}  = 0,
\qquad
\big(\mathcal{E}^{r}{}_{r} \big)_{-2}  = 0 
\eeq
are automatically satisfied once~\eqref{10f1h1} and~\eqref{Constfh10} are imposed.

Starting at order $\De^{-1}$, each order of the field equations fixes two parameters as functions of the previous ones and, in principle, a recursive relation can be obtained. Indeed, at given order $\Delta^N$ ($N = -1,0,1,\ldots$), the structure of the field equations is such that
\beq
\label{Sys10-N}
\begin{aligned}
\big( \mathcal{E}^{t}{}_{t} \big)_{N}  & =  
\tfrac{(N+2) (N+3) (N+4) (N+5) (N+7)}{ f_0 ^7} 
\left\lbrace  \left[ 2 (N+8) \ga_1  +  (N+7) \ga_2   \right]  f_0 ^3 \,  h_{N+6} + 2 (N+7) (4  \ga_1 + \ga_2 )  f_{N+6}  + \Phi_{tt,N+5} \right\rbrace     = 0,
\\
\big( \mathcal{E}^{r}{}_{r} \big)_{N}  & =  
\tfrac{(N+2) (N+4) (N+5) (N+7)}{ f_0 ^7} 
\left\lbrace    \left[ 4 (N+8)  \ga_1 + (N+9) \ga_2  \right]  f_0 ^3 \, h_{N+6} + 2 (N+7) (8  \ga_1 +3  \ga_2 )  f_{N+6}  + \Phi_{rr,N+5}  \right\rbrace    = 0,
\end{aligned}
\eeq
where $\Phi_{tt,N+5}$ and $\Phi_{rr,N+5}$ denote terms depending on the coefficients $f_n$ and $h_n$ with $n\in\lbrace 0,1,2,\ldots,N+5\rbrace$. Hence, the system~\eqref{Sys10-N} can be solved for $f_{N+6}$ and $h_{N+6}$:
\beq
\label{rec10}
\begin{aligned}
f_{N+6} & =  \frac{\left[ 4 ( N + 8)  \ga_1   + ( N + 9) \ga_2 \right]  \Phi_{tt,N+5} - \left[ 2 ( N + 8)  \ga_1  + ( N + 7) \ga_2 \right]  \Phi_{rr,N+5}   }{4  ( N + 7)  ( N + 6) \ga_2 (3  \ga_1 + \ga_2 )},
\\
h_{N+6} & =  - \frac{  \left( 8  \ga_1  + 3  \ga_2  \right) \Phi_{tt,N+5}  - \left( 4  \ga_1  + \ga_2  \right) \Phi_{rr,N+5}}{2  f_0^3 ( N + 6) \ga_2  (3  \ga_1 + \ga_2 )}.
\end{aligned}
\eeq

Finally, we can identify the free parameters of the solution: $f_0$, $f_2$, $h_2$, one among $f_3$ and $h_3$ [the other is fixed  by~\eqref{Constfh10}], $f_4$, $h_4$ and $r_0$. So, we have a total of seven parameters --- five of which are physical, after taking into account the residual gauge symmetry~\eqref{ResGauge}. 
The solution reads
\bea
\begin{aligned}
F(r) & =  f_0 \De + f_2 \De^3 + f_3 \De^4 + f_4 \De^5 + O(\De^6),
\\
H(r) & =  \frac{1}{f_0^2} + h_2 \De^2 - \dfrac{4 (8 \ga_1 + 3 \ga_2 ) f_3}{(10 \ga_1 + 3 \ga_2) f_0^3} \De^3 + h_4 \De^4 + O(\De^5),
\end{aligned}
\eea
if $10 \ga_1 + 3 \ga_2 \neq 0$, otherwise,
\bea
\begin{aligned}
F(r) & =  f_0 \De + f_2 \De^3  + f_4 \De^5 + O(\De^6),
\\
H(r) & =  \frac{1}{f_0^2} + h_2 \De^2 + h_3 \De^3 + h_4 \De^4 + O(\De^5).
\end{aligned}
\eea
We do not write the solutions for $f_5$ and $h_5$ in explicit form because they are very long, but we remark that they are proportional to $\left[ \ga_2 (3  \ga_1 + \ga_2 )\right]^{-1}$, as suggested by~\eqref{rec10}. This indicates that the limit $\ga_2 (3  \ga_1 + \ga_2 ) \to 0$ is not smooth. In particular, the solutions obtained here do not reproduce the solutions of models that do not contain sixth-order metric derivatives in both its spin-0 and spin-2 sectors.

The conditions $f_0^2 h_0 = 1$ and $f_1=h_1=0$ guarantee that all the solutions in the class $\lbrace 1,0 \rbrace$ have a regular Kretschmann scalar at $r=r_0$,
\beq
\label{Kret10}
R_{\mu\nu\al\be} R^{\mu\nu\al\be} \underset{r \to r_0}{=}  24 \left(\frac{6 f_2 h_2}{f_0^3}+\frac{18 f_2^2}{f_0^6}+h_2^2\right)  + O(\De) .
\eeq
Moreover, since these solutions are static and spherically symmetric, it follows that any curvature invariant built by contracting an arbitrary number of Riemann and metric tensors is regular at $r=r_0$~\cite{Bronnikov:2012wsj}.

On the other hand, the regularity of scalar invariants containing derivatives of curvatures generally depends on relations involving higher-order coefficients~\cite{Giacchini:2021pmr}. As an example, if $f_3,h_3\neq 0$, the scalars
\beq
\label{CD01}
\begin{aligned}
\Box R & \underset{r \to r_0}{=} - \frac{8 ( 16 f_3 + 5 f_0^3 h_3 )}{f_0^5 \De} - \frac{12 (15 f_0^4 h_4+50 f_0f_4 +24 f_0^3 f_2 h_2-15 f_2^2)}{f_0^6} + \ldots,
\\
R_{\mu\nu\al\be} \Box R^{\mu\nu\al\be} & \underset{r \to r_0}{=} 16 \frac{12 f_2 (8 f_3 + f_0^3 h_3) + f_0^3 h_2 (16 f_3 + 5 f_0^3 h_3)}{f_0^8 \De} + \ldots
\end{aligned}
\eeq
might diverge as $r\to r_0$; similar behaviour is also observed for the scalars $R_{\mu\nu} \Box R^{\mu\nu}$ and $R \Box R$.

\subsubsection*{Description in standard spherically symmetric coordinates and interpretation}

Using the relation~\eqref{st1}, it follows that the solutions in the class $\lbrace 1,0 \rbrace$ correspond to the class $(0,0)_0$ in standard spherically symmetric coordinates~\eqref{metric-Standard}. Since they are the only solutions in powers of $\Delta$ with $\sigma > 0$ [see Eq.~\eqref{ListSolr0}], from the discussion in Sec.~\ref{Sec.Corresp} it follows that, among the solutions presented in this work, they are the only ones describing the origin $\bar{r}=0$ in standard coordinates. The uniqueness of this family of solutions was also verified by analysing the indicial equations directly in the coordinates~\eqref{metric-Standard}~\cite{Praha1}.

In particular, the conditions $f_0^2 h_0 = 1$ and $f_1=h_1=0$ guarantee that
\beq
a_0 = 1, \qquad a_1=b_1=0
\eeq
for the expansion~\eqref{FrobeniusAB0}, which are the necessary conditions for the regularity of the solution at $\bar{r}=0$. Moreover, the relation~\eqref{Constfh10} translates into a similar relation between $a_3$ and $b_3$,
\beq
(8 \ga_1 + 3 \ga_2 ) a_3 = 3 (4 \ga_1 + \ga_2) b_3 .
\eeq
Therefore, the solution
\beq
\label{Sol00AB}
\begin{aligned}
A(\bar{r}) & =   1 
+ a_2 \bar{r}^2 
+ a_3  \bar{r}^3
+ a_4 \bar{r}^4 
+ {a}_5 \bar{r}^5
+  O(\bar{r}^6),
\\
\frac{B(\bar{r})}{b_0} & = 1 
+ b_2 \bar{r}^2
+ b_3 \bar{r}^3
+ b_4 \bar{r}^4
+ {b}_5 \bar{r}^5 + O(\bar{r}^6)
\end{aligned}
\eeq
is characterised by a total of six parameters, $a_2$, $a_3$ (or $b_3$), $a_4$, $b_0$, $b_2$ and $b_4$, among which five are physical [$b_0$ corresponds to the 1-parameter residual gauge freedom to re-scale the time coordinate; see Eq.~\eqref{ResGauge-Standard}]. The subsequent coefficients $a_{5,\ldots}$ and $b_{5,\ldots}$ are determined by the previous ones. These results are in agreement with the solutions obtained in~\cite{Praha1} by working directly in standard coordinates.
However, while in that work the number of free parameters was identified by solving the field equations up to a certain order and noticing that no new free parameter appeared, using the modified coordinates~\eqref{metric} we were able to actually prove that the recursive relations exist, yielding a precise count of free parameters.

Since all the solutions in this class $(0,0)_0$ are regular at $\bar{r}=0$ and contain an (anti-)de~Sitter-like or a Minkowski core,\footnote{To be precise, we call an (anti-)de~Sitter-like core the situation in which $a_2,b_2\neq 0$ but not necessarily $a_2=-b_2$, while for the Minkowski core we require $a_2,b_2= 0$. There is also the possibility of having only one non-zero parameter among $a_2$ and $b_2$, but this does not affect the discussion.}
they define a regime of sufficiently small $\bar{r}$ that should also be described by the linearised field equations. 
Therefore, it is important to compare~\eqref{Sol00AB} with the linearised solutions, which are briefly summarised in Appendix~\ref{App-Lin}. 
Note that the domain of validity of the weak-field approximation is given by $a_2\bar{r}^2 + O(\bar{r}^3) \ll 1$ and $b_2 \bar{r}^2+ O(\bar{r}^3)\ll 1$; the more coefficients drop out, the closer the spacetime is to Minkowski.

The global vacuum linearised solution~\eqref{A0B0linSC} also has (anti-)de~Sitter-like or Minkowski cores, but it is characterised by five parameters --- one less than the exact solution~\eqref{Sol00AB}. The difference is caused by the parameter of the $O(\bar{r}^3)$ term: while in the exact solution it is possible to have $a_3,b_3\neq 0$, the linearised one has $\bar{a}_3=\bar{b}_3= 0$ [see Eq.~\eqref{A0B0lin}]. This term might indicate the breakdown of the linear approximation. For instance, the solutions with $a_3=b_3=0$ have not just regular curvature invariants but also curvature-derivative invariants, thus being better approximated by the linearised ones (up to order $\bar{r}^4$) with a matching number of free parameters. On the other hand, there exist exact solutions with $a_3,b_3\neq 0$ and, in this case, the validity range of the linear approximation is smaller, with differences appearing already at $O(\bar{r}^3)$.\footnote{The situation here is different from the four-derivative gravity, since there it happens that the linearised vacuum solution and the exact $(0,0)_0$ solutions have precisely the same number of parameters~\cite{Stelle15PRD}. Moreover, in that case there are other classes of (singular) solutions $(s,t)_0$ describing expansions around $\bar{r}=0$.}
Last but not least, another possible comparison of the solutions with $a_3,b_3\neq 0$ is with linearised solutions sourced by Dirac deltas and its derivatives, as they can generate terms cubic in $\bar{r}$. For example, the linearised solution~\eqref{PhiPsi-delta} sourced by a delta has a de~Sitter core and non-trivial coefficients of $\bar{r}^3$. (Nevertheless, there might be a bit of a heuristic in comparing solutions of the linearised field equations with distributional sources that are defined globally and local exact solutions defined by the Frobenius expansion around certain points, especially since the weak-field regime is certainly broken further from $\bar{r}=0$.)

\subsection{Solutions in the class $\lbrace 0,0 \rbrace$}
\label{Sec-00}

For solutions in the class $\lbrace 0,0 \rbrace$, expanding the field equations in powers of $\De$,
\beq
\label{ExpEoM0000}
\mathcal{E}^{\mu}{}_{\nu} = \sum_{i=0}^\infty \left( \mathcal{E}^{\mu}{}_{\nu} \right)_{i} \De^i  = 0 ,
\eeq
and substituting into the generalised Bianchi identity~\eqref{bia}, we find
\beq
\label{bia001}
\big(\mathcal{E}^{r}{}_{r} \big)_{1} + 
 \frac{2 f_1}{f_0} \left[ \big(\mathcal{E}^{r}{}_{r} \big)_{0} - \big(\mathcal{E}^{\th}{}_{\th} \big)_{0} \right] 
+ 
 \frac{h_1}{2 f_0} \left[ \big(\mathcal{E}^{r}{}_{r} \big)_{0} - \big(\mathcal{E}^{t}{}_{t} \big)_{0} \right] 
+ O(\De) = 0 .
\eeq
Therefore, at lowest order in $\De$ one can solve the three components of the field equations,
\beq
\label{00astres}
\big( \mathcal{E}^{t}{}_{t} \big)_{0} = 0 , \qquad \big( \mathcal{E}^{r}{}_{r} \big)_{0} = 0 , \qquad  \big( \mathcal{E}^{\th}{}_{\th} \big)_{0} = 0 ,
\eeq
and Eq.~\eqref{bia001} will guarantee that the $rr$ component will be automatically solved at order $\De^1$,
\beq
\big(\mathcal{E}^{r}{}_{r} \big)_{1} = 0.
\eeq
For the next orders, using~\eqref{bia} it is not difficult to verify that if the field equations are solved up to order $\De^N$, \textit{i.e.}, $\big(\mathcal{E}^{\mu}{}_{\nu} \big)_{n}=0$ for all $n\in \lbrace 0,1,\ldots, N\rbrace$, then
\beq
\big(\mathcal{E}^{r}{}_{r} \big)_{N+1} = 0.
\eeq
Hence, solutions in the class $\lbrace 0,0 \rbrace$ can be found by solving~\eqref{00astres} (at lowest order) and the two components
\beq
\label{eom00}
\mathcal{E}^{t}{}_{t} = 0 \, , \qquad \mathcal{E}^{\th}{}_{\th} = 0
\eeq
order by order.\footnote{Alternatively, one can solve $\mathcal{E}^{r}{}_{r} = 0$ and $\mathcal{E}^{\th}{}_{\th} = 0$ (or $\mathcal{E}^{t}{}_{t} = 0$) to all orders. However, while both $\mathcal{E}^{t}{}_{t}$ and $\mathcal{E}^{\th}{}_{\th}$ are of the sixth order in $H(r)$ and $F(r)$,  $\mathcal{E}^{r}{}_{r}$ is of the fifth order. Due to this mismatch of differential order between the equations and the indicial structure of the solutions $\lbrace 0,0 \rbrace$, the analogue of the system~\eqref{aquela00} for any of these two possibilities would involve $\left( \mathcal{E}^{r}{}_{r} \right)_{N+1}$ instead of $\left( \mathcal{E}^{r}{}_{r} \right)_{N}$.
}

The terms in~\eqref{eom00} have the structure
\beq
\begin{aligned}
\mathcal{E}^{t}{}_{t} & =  \frac{H^2}{F} \left[ 2 (4  \ga_1 + \ga_2 )  F^{(6)}  H+(2  \ga_1 + \ga_2 ) F H^{(6)}\right] + \cdots ,
\\
\mathcal{E}^{\th}{}_{\th} & =  \frac{(8  \ga_1 +3  \ga_2 )  F^{(6)}  H^3}{F}+\frac{1}{2} (4  \ga_1 + \ga_2 ) H^{(6)} H^2 + \cdots ,
\end{aligned}
\eeq
where the ellipsis denotes terms depending on the lower-order derivatives $F^{(5)},F^{(4)},\ldots,F$, and $H^{(5)},H^{(4)},\ldots,H$. 
Expanding the field equations in powers of $\De$, at order $\De^N$ we find
\beq
\label{aquela00}
\begin{aligned}
\left( \mathcal{E}^{t}{}_{t} \right)_{N}  & =  \frac{(N+6)! \, h_0 ^2 }{N! \, f_0} \left[ 2  h_0  (4  \ga_1 + \ga_2 ) f_{N+6} + f_0  (2  \ga_1 + \ga_2 ) h_{N+6} + \Phi_{tt,N+5} \right]   = 0,
\\
\left( \mathcal{E}^{\th}{}_{\th} \right)_{N}  & =  \frac{(N+6)! \, h_0 ^2 }{2 N! \, f_0} \left[ 2 h_0 (8 \ga_1 + 3 \ga_2) f_{N+6} + f_0 (4 \ga_1 + \ga_2 ) h_{N+6} + \Phi_{\th\th,N+5} \right]   = 0,
\end{aligned}
\eeq
where $\Phi_{tt,N+5}$ and $\Phi_{\th\th,N+5}$ are functions of the coefficients $h_i$ and $f_i$ with $i = 0,\ldots, N+5$ [although we use the same notation, the functions that appear here are different from the ones that occur in Eq.~\eqref{Sys10-N}].
This linear system for the coefficients $h_{N+6}$ and $f_{N+6}$ can be easily solved,
\beq
\label{Recur00}
\begin{aligned}
f_{N+6} & =  -\frac{ ( 4  \ga_1 + \ga_2 )  \Phi_{tt,N+5}  +  ( 2  \ga_1   + \ga_2 )  \Phi_{\th\th,N+5}  }{4  \ga_2 (3  \ga_1 + \ga_2 )  h_0 },
\\
h_{N+6} & =  \frac{  (8  \ga_1  + 3  \ga_2 )  \Phi_{tt,N+5} +  ( 4  \ga_1   + \ga_2 )  \Phi_{\th\th,N+5} }{2  \ga_2  (3  \ga_1 + \ga_2 ) f_0 },
\end{aligned}
\eeq
provided that $\ga_2  (3  \ga_1 + \ga_2 ) \neq 0$, \textit{i.e.}, if the field equations are sixth order in derivatives of the metric.

At lowest order in $\De$, we set $N=0$ and obtain the coefficients $h_6$ and $f_6$ as functions of $h_0,\ldots,h_5$ and $f_0,\ldots,f_5$. However, as mentioned before, this still does not mean that the remaining equation $\mathcal{E}^{r}{}_{r} = 0$ is solved at order 0. In fact, 
$\left( \mathcal{E}^{r}{}_{r} \right)_{0} = 0 $
acts as a constraint between the coefficients $f_0$, $f_1$, $f_2$, $f_3$, $f_4$, $f_5$, $h_0$, $h_1$, $h_2$, $h_3$, $h_4$, $h_5$, fixing one of them as a function of the others (and of the couplings of the model). The situation is similar to the condition~\eqref{Constfh10} found in the solutions $\lbrace 1,0 \rbrace$, but more complicated since it involves more coefficients. The constraint has the form
\bea
\label{semc}
120    \left[ 2  f_1  h_0 (8  \ga_1+3  \ga_2)+ f_0  h_1 (4  \ga_1+ \ga_2) \right] \frac{h_0^2 }{ f_0} f_5  
+
60  \left[ 2  f_1  h_0 (4  \ga_1+ \ga_2) + f_0  h_1 (2  \ga_1+ \ga_2) \right]  h_0 \, h_5 - \frac{\Phi_{rr,4}}{ f_0^5} = 0 ,
\eea
where $\Phi_{rr,4}$ depends on $f_{0,\ldots,4}$, $h_{0,\ldots,4}$, $\al$, $\beta_{1,2}$ and $\gamma_{1,\ldots,8}$; for the explicit expression, see Eq.~\eqref{ograndec}. In most cases, this constraint can be solved for the highest-order coefficient $f_5$ or $h_5$. Nevertheless, if a solution has $f_1 = h_1 = 0$, then~\eqref{semc} reduces to $\Phi_{rr,4}=0$, which must be solved for a lower-order coefficient, such as $f_4$ or $h_4$.

A solution in the class $\lbrace 0,0 \rbrace$, therefore, is characterized by twelve free parameters: $r_0$ and eleven among $h_{0,\ldots,5}$ and $f_{0,\ldots,5}$. Taking into account the residual gauge symmetry~\eqref{ResGauge}, this number can be reduced to ten free physical parameters. All the solutions in this class have regular curvature at $r=r_0$; for example, for the Kretschmann scalar we have
\beq
\label{Kret00}
R_{\mu\nu\al\be} R^{\mu\nu\al\be} \, \underset{r \to r_0}{=}  \, \frac{4}{f_0^4} \left[ 1 + f_0^4 h_2^2 + 8 f_2^2 f_0^2 h_0^2 + 4 f_1 f_2 f_0^2 h_0 h_1 + f_1^4 h_0^2 + f_1^2 \left(f_0^2 h_1^2 - 2 h_0 \right) \right]  + O(\De) .
\eeq

This class of solutions includes the particular sub-classes which we shall denote by $\lbrace 0,0 \rbrace_{f_1=0, f_2\neq 0}$, $\lbrace 0,0 \rbrace_{f_{1,2}=0,f_3\neq 0}$, $\lbrace 0,0 \rbrace_{f_{1,2,3}=0,f_4\neq 0}$, $\lbrace 0,0 \rbrace_{f_{1,2,3,4}=0,f_5\neq 0}$ and $\lbrace 0,0 \rbrace_{f_{1,2,3,4,5}=0,f_6\neq 0}$, depending on whether some  coefficients $f_{1,\ldots,5}$ are zero. As we show in what follows (see also the discussion in Sec.~\ref{Sec.Corresp}), such solutions are generally mapped to non-Frobenius solutions in standard Schwarzschild coordinates.  Nonetheless, from Eq.~\eqref{Kret00} we conclude that such solutions also have regular curvature invariants at $r=r_0$. 
Other interesting particular cases are the sub-classes $\lbrace 0,0 \rbrace_{f_{1,3,5}=h_{1,3,5}=0,f_2\neq 0}$, $\lbrace 0,0 \rbrace_{f_{1,2,3,5}=h_{1,3,5}=0,f_4\neq 0}$ and $\lbrace 0,0 \rbrace_{f_{1,2,3,4,5}=h_{1,3,5}=0,f_6\neq 0}$, for which the absence of odd-power terms among the free parameters makes the solution even in $\Delta$.

\subsubsection*{Description in standard spherically symmetric coordinates and interpretation}

From~\eqref{map1} we deduce that the solutions in the class $\lbrace 0,0 \rbrace$ correspond to expansions around a certain generic point $\bar{r}=\bar{r}_0\neq 0$ in the standard spherically symmetric coordinates~\eqref{metric-Standard}. The most general solution with the maximum number of free parameters is mapped to the class $(0,0)_{\bar{r}_0}$, following Eq.~\eqref{st2}. In this way, we conclude that a Frobenius series solution around a generic point $\bar{r}_0\neq 0$ is characterised by ten free physical parameters.
This parameter count was previously suggested in~\cite{Praha1}, based on the direct solution of the field equations to a few orders, but a rigorous proof was still pending. Here, the use of modified coordinates~\eqref{metric}, which yield autonomous field equations and a clear identification of the free parameters, was instrumental in proving the result.\footnote{The same result can also be achieved by using a metric in conformal-to-Kundt form, as reported in~\cite{Praha1}.}

In what follows, we discuss the other particular sub-classes that fall into the context of Eqs.~\eqref{Non-FrobeniusAB} and~\eqref{st3}:
\begin{itemize}
\item Generic solutions in the sub-class $\lbrace 0,0 \rbrace_{f_1=0, f_2\neq 0}$ correspond to non-Frobenius solutions of the class $(1,0)_{\bar{r}_0, 1/2}$ in standard spherically symmetric coordinates, with half-integer steps,
\beq
\label{Sol-Sch-10}
\begin{aligned}
C(\bar{r}) & = c_0 \bar{\De} + c_{1} \bar{\De}^{3/2} + c_2 \bar{\De}^2 + c_{3} \bar{\De}^{5/2} + c_4 \bar{\De}^3 + O(\bar{\De}^{7/2}) ,
\\
\frac{B(\bar{r})}{b_0} & = 1 + b_{1} \bar{\De}^{1/2} + b_2 \bar{\De} + b_{3} \bar{\De}^{3/2} + b_4 \bar{\De}^2 + b_{5} \bar{\De}^{5/2} + b_6 \bar{\De}^3 + O(\bar{\De}^{7/2}) .
\end{aligned}
\eeq
From the analysis above, we conclude that such solutions are characterised by nine physical free parameters. Indeed, by solving the field equations order by order for the functions $C(\bar{r})$ and $B(\bar{r})$, one can verify that the coefficients $c_{4,5,\ldots}$ and $b_{6,7,\ldots}$ are expressed in terms of the lower-order coefficients and the parameters of the model, and there is an additional constraint between the parameters $\bar{r}_0$, $c_0$, $c_{1}$, $c_{2}$, $c_{3}$, $b_{1}$, $b_2$, $b_{3}$, $b_4$, $b_{5}$ (furthermore, $b_0$ is not physical, as it is linked to the time re-scaling freedom of the metric).

However, if the solution is even in $\Delta$ --- which corresponds to the sub-class $\lbrace 0,0 \rbrace_{f_{1,3,5}=h_{1,3,5}=0,f_2\neq 0}$ --- then the terms with $n$ odd are going to be missing from the expansion~\eqref{Sol-Sch-10}, resulting in a class $(1,0)_{\bar{r}_0}$ of Frobenius series solutions,
\beq
\begin{aligned}
\label{Sol-Sch-10-Frob}
C(\bar{r}) & = c_0 \bar{\De} + c_2 \bar{\De}^2 + c_4 \bar{\De}^3 + O(\bar{\De}^4) ,
\\
\frac{B(\bar{r})}{b_0} & = 1 + b_2 \bar{\De} + b_4 \bar{\De}^2 + b_6 \bar{\De}^3 + O(\bar{\De}^4) .
\end{aligned}
\eeq
Such solutions are characterised by four physical free parameters (among $\bar{r}_0$, $c_0$, $c_2$, $b_2$ and $b_4$, according to the constraint given by Eq.~\eqref{Const-Sol-Sch-10-Frob} in Appendix~\ref{app2}) and describe a wormhole, since $C(\bar{r}_0) = 0$ but $B(\bar{r}_0) \neq 0$~\cite{Stelle15PRD}. This can also be viewed from the metric written in modified coordinates~\eqref{metric} by fixing $r_0=0$, as it directly gives a solution that is well behaved at $r=0$, even in $r$ and with $F(r)=f_0 + f_2 r^2 + \ldots$, allowing $r$ to be extended to negative values.

Both types of solution series~\eqref{Sol-Sch-10} and~\eqref{Sol-Sch-10-Frob} also occur in the fourth-order gravity, but with different number of free parameters. The case with half-integer steps is sometimes referred to as a ``half-integer'' or ``non-symmetric'' wormhole. Indeed, by setting $r_0=0$ we still have $F(r)=f_0 + f_2 r^2 + \ldots$, but now $F(r)$ and $H(r)$ are not symmetric under the transformation $r \to - r$. Further discussion about their interpretation in the context of quadratic gravity can be found in~\cite{Stelle15PRD,Bonanno:2022ibv}.

\item Solutions in the sub-class $\lbrace 0,0 \rbrace_{f_{1,2}=0,f_3\neq 0}$ correspond to non-Frobenius solutions of the class $\big( \tfrac{4}{3},0 \big)_{\bar{r}_0, 1/3}$ in standard spherically symmetric coordinates, with steps $\bar{\Delta}^{1/3}$ and eight free physical parameters. Quadratic gravity also admits solutions with such indicial structure, though with fewer free parameters, and their interpretation as unusual wormholes is still open~\cite{Podolsky:2019gro}.

\item Generic solutions in the sub-class $\lbrace 0,0 \rbrace_{f_{1,2,3}=0,f_4\neq 0}$ correspond to non-Frobenius solutions of the class $\big( \tfrac{3}{2},0 \big) _{\bar{r}_0, 1/4}$ in standard spherically symmetric coordinates, with  steps $\bar{\Delta}^{1/4}$ and seven free physical parameters. Moreover, in the sub-class $\lbrace 0,0 \rbrace_{f_{1,2,3,5}=h_{1,3,5}=0,f_4\neq 0}$ the solutions are even, causing $b_n=c_n=0$ for $n$ odd  in the expansion~\eqref{Non-FrobeniusAB}. This results in a solution class $\big( \tfrac{3}{2},0 \big) _{\bar{r}_0, 1/2}$ in standard coordinates with three free physical parameters, as there appears a constraint between the original parameters. 
Neither of these types of solutions were identified in quadratic gravity.\footnote{However, it is likely that a class of solutions $\big( \tfrac{3}{2},0 \big) _{\bar{r}_0, 1/4}$ with four free parameters exists for generic quadratic gravity models (but not for the Einstein--Weyl gravity).}

\item Solutions in the sub-class $\lbrace 0,0 \rbrace_{f_{1,2,3,4}=0,f_5\neq 0}$ correspond to non-Frobenius solutions of the class $\big( \tfrac{8}{5},0 \big) _{\bar{r}_0, 1/5}$ in standard spherically symmetric coordinates, with steps $\bar{\Delta}^{1/5}$ and six free physical parameters. These types of solutions are also not known to occur in quadratic gravity.

\item Generic solutions in the sub-class $\lbrace 0,0 \rbrace_{f_{1,2,3,4,5}=0,f_6\neq 0}$ correspond to non-Frobenius solutions of the class $\big( \tfrac{5}{3},0 \big) _{\bar{r}_0, 1/6}$ in standard spherically symmetric coordinates, with  steps $\bar{\Delta}^{1/6}$ and five free physical parameters. In addition, the sub-class of even solutions in $\Delta$ yields a solution class $\big( \tfrac{5}{3},0 \big) _{\bar{r}_0, 1/3}$ with two physical free parameters in standard coordinates, owing to the absence of the coefficients $b_n$ and $c_n$ with $n$ odd in the expansion~\eqref{Non-FrobeniusAB}. Like the previous cases, solutions of this type were not identified in quadratic gravity.
\end{itemize}

\subsection{Solutions in the class $\lbrace 0,1 \rbrace$}
\label{Sec-01}

The substitution of a metric of the class $\lbrace 0,1 \rbrace$ into the field equations results in an expansion starting at zeroth order, like~\eqref{ExpEoM0000}. In addition, the expansion of the generalised Bianchi identity~\eqref{bia} order by order starts at order $\De^{-1}$,
\beq
\label{bia01}
\frac{\big(\mathcal{E}^{r}{}_{r} \big)_{0} - \big(\mathcal{E}^{t}{}_{t} \big)_{0} }{2 \De} 
+
\frac{1}{2} \left[
 3 \big(\mathcal{E}^{r}{}_{r} \big)_{1} - \big(\mathcal{E}^{t}{}_{t} \big)_{1} + 
\frac{4 f_1}{f_0 } \left[ \big( \mathcal{E}^{r}{}_{r} \big)_{0} - \big( \mathcal{E}^{\th}{}_{\th} \big)_{0} \right]  + \frac{ h_1}{h_0 } \left[ \big( \mathcal{E}^{r}{}_{r} \big)_{0} - \big(\mathcal{E}^{t}{}_{t} \big)_{0} \right]    \right] + O(\De) = 0 .
\eeq
Notice that if $\big(\mathcal{E}^{t}{}_{t} \big)_{0} = 0$, then~\eqref{bia01} yields $\big(\mathcal{E}^{r}{}_{r} \big)_{0}=0$. Using the induction hypothesis that the field equations are solved up to order $\De^N$, \textit{i.e.}, $\big(\mathcal{E}^{\mu}{}_{\nu} \big)_{n}=0$ for all $n\in \lbrace 0,1,\ldots, N\rbrace$, it is straightforward to verify that~\eqref{bia01} gives
\beq
\frac{1}{2} \left[
 (2N+3) \big( \mathcal{E}^{r}{}_{r} \big)_{N+1} - \big(\mathcal{E}^{t}{}_{t} \big)_{N+1} \right] \De^{N} + O \left( \De^{N+1} \right)  = 0 \qquad \Longrightarrow \qquad \big(\mathcal{E}^{r}{}_{r} \big)_{N+1} =  \frac{1}{2N+3} \big(\mathcal{E}^{t}{}_{t} \big)_{N+1}.
\eeq
Therefore, for metrics of the class $\lbrace 0,1 \rbrace$ it suffices to solve the field equations $\mathcal{E}^{\th}{}_{\th} = 0$ and $\mathcal{E}^{t}{}_{t}=0$ (or equivalently, $\mathcal{E}^{r}{}_{r}=0$) order by order, without the need to check the remaining component of the field equations.

At a given order $\De^{N}$ (with $N = 0,1,\ldots$) of the expansion of the field equations for a metric of the class $\lbrace 0,1 \rbrace$, the coefficient $\left( \mathcal{E}^{\mu}{}_{\nu} \right)_{N}$ depends on the coefficients $f_n$ and $h_n$ with $n = 0,\ldots,N+3$.
In order to single out the highest coefficients ($f_{N+3}$ and $h_{N+3}$) in $\left( \mathcal{E}^{\mu}{}_{\nu} \right)_{N}$, 
it is useful to write the field equations in the form
\beq
\begin{aligned}
\mathcal{E}^{t}{}_{t} & =  
\tfrac{4 \ga_1 + \ga_2}{F} \Big( 
2 F^{(6)} H^3 + 17 F^{(5)} H^2 H' + 30  F^{(4)} H H'^2 + 6 F^{(3)} H'^3
\Big) 
+ (2 \ga_1 + \ga_2) \Big( H^{(6)} H^2 + \tfrac{7}{2} H^{(5)} H H' + H^{(4)} H'^2 \Big) + \cdots ,
\\
\mathcal{E}^{\th}{}_{\th} & =  
\tfrac{8\ga_1 + 3\ga_2}{F } \Big( 
F^{(6)}  H ^3 + 9 F^{(5)}  H ^2 H'  + 18 F^{(4)}  H  H'^2 + 6 F^{(3)}  H'^3 
\Big) 
+ (4\ga_1 + \ga_2) \Big( 
\tfrac{1}{2}  H^{(6)}  H ^2 + 2 H^{(5)}  H  H'  + H^{(4)}  H'^2  \Big) + \cdots .
\end{aligned}
\eeq
The terms explicitly written are those that can originate the coefficients $f_{N+3}$ and $h_{N+3}$ at the order $\De^{N}$, while the ellipsis denotes the terms that only contribute the coefficients $f_n$ and $h_n$ with $n=0,\ldots,N+2$.

Expanding the field equations in powers of $\De$, at order $\De^{N}$ we have
\beq
\begin{aligned}
\label{seeeg01}
\left( \mathcal{E}^{t}{}_{t} \right)_{N}  & =  
\tfrac{h_0^2 (N+1) (N+2)^2 (N+3) (2 N+1)}{2 f_0} \left[ 
 2 h_0 ( 4 \ga_1 + \ga_2 ) (N+3) f_{N+3} + f_0 (2 \ga_1 + \ga_2) (N+4) h_{N+3} + \Phi_{tt,N+2} \right]    = 0,
\\
\left( \mathcal{E}^{\th}{}_{\th} \right)_{N}  & =  
\tfrac{h_0^2 (N+1)^2 (N+2)^2 (N+3)}{2 f_0} \left[ 
 2 h_0 (8 \ga_1 + 3 \ga_2 ) (N+3) f_{N+3} + f_0 (4 \ga_1 + \ga_2) (N+4) h_{N+3}  + \Phi_{\th\th,N+2}  \right]   = 0,
\end{aligned}
\eeq
where $\Phi_{tt,N+2}$ and $\Phi_{\th\th,N+2}$ are functions of the coefficients $h_i$ and $f_i$ with $i = 0,\ldots, N+2$ [again, the functions that appear here are different from the ones that occur in Eqs.~\eqref{Sys10-N} and~\eqref{aquela00}].
This gives
\beq
\begin{aligned}
f_{N+3} & =  \frac{ ( 4  \ga_1 + \ga_2 )  \Phi_{tt,N+2} - ( 2  \ga_1   + \ga_2 )  \Phi_{\th\th,N+2} }{4  \ga_2 (3  \ga_1 + \ga_2 ) (N+3) h_0 },
\\
h_{N+3} & = - \frac{ (8  \ga_1  + 3  \ga_2 )  \Phi_{tt,N+2} -  ( 4  \ga_1   + \ga_2 )  \Phi_{\th\th,N+2}  }{2  \ga_2  (3  \ga_1 + \ga_2 ) (N+4) f_0 }.
\end{aligned}
\eeq
Hence, the field equations can be solved order by order starting at order $\De^0$, which fixes the coefficients $f_3$ and $h_3$, and so on.
We conclude that a solution in the family $\lbrace 0,1 \rbrace$ is characterised by a total of seven parameters: $f_0, f_1, f_2, h_0, h_1, h_2$ and $r_0$ (which can be reduced to five physical parameters).
The solutions in this class are regular at $r = r_0$, for instance,
\beq
\label{Kret01}
R_{\mu\nu\al\be} R^{\mu\nu\al\be} \underset{r \to r_0}{=}  4 \left(\frac{f_1^2 h_0^2}{f_0^2}+\frac{1}{f_0^4}+h_1^2\right) + O(\De) .
\eeq

Like in the case of the solutions $\lbrace 0,0 \rbrace$, we shall also single out sub-classes of solutions for which some of the free coefficients $f_i$ are switched off. They correspond to the sub-classes $\lbrace 0,1 \rbrace_{f_1=0, f_2\neq 0}$ and $\lbrace 0,1 \rbrace_{f_{1,2}=0, f_3\neq 0}$, depending on the triviality of the coefficients $f_1$ and $f_2$. As we show in Sec.~\ref{Sec.Corresp}, these solutions are mapped to non-Frobenius solutions in standard Schwarzschild coordinates; notwithstanding, from Eq.~\eqref{Kret01} we conclude that such solutions also have regular curvature invariants at $r=r_0$.

\subsubsection*{Description in standard spherically symmetric coordinates and interpretation}

According to~\eqref{map1} and~\eqref{st2}, a generic solution in the class $\lbrace 0,1 \rbrace$ with a maximum number of non-zero free parameters corresponds to an element of the class $(1,1)_{\bar{r}_0}$ in standard spherically symmetric coordinates~\eqref{metric-Standard}. Since $C(\bar{r}),B(\bar{r})\sim \bar{\Delta}$, such solutions represent expansions around a horizon located at $\bar{r}=\bar{r}_0$,
\beq \label{Sol11r0}
\begin{aligned}
C(\bar{r}) & =   
  c_0 \bar{\Delta}
+ c_1 \bar{\Delta}^2 
+ c_2  \bar{\Delta}^3
+ c_3  \bar{\Delta}^4
+  O(\bar{\Delta}^5),
\\
\frac{B(\bar{r})}{b_0} & =  
 \bar{\Delta}
+ b_1 \bar{\Delta}^2 
+ b_2  \bar{\Delta}^3
+ b_3  \bar{\Delta}^4
+  O(\bar{\Delta}^5).
\end{aligned}
\eeq
As shown above, they are characterised by five free physical parameters, verifying the result of the analysis carried out directly in standard Schwarzschild coordinates~\cite{Praha1}. Indeed, there are five independent parameters among $\bar{r}_0$, $c_0$, $c_1$, $c_2$, $b_1$ and $b_2$; $b_0$ represents the residual gauge freedom, and the coefficients $c_{3,\ldots}$ and $b_{3,\ldots}$ are fixed by the previous ones.

Unusual horizons of non-Frobenius form also exist for the sub-classes in the context of Eqs.~\eqref{Non-FrobeniusAB} and~\eqref{st3}:
\begin{itemize}
\item Solutions in the sub-class $\lbrace 0,1 \rbrace_{f_1=0, f_2\neq 0}$ correspond to expansions with structure  $\big( \tfrac{3}{2},\tfrac{1}{2} \big)_{\bar{r}_0,1/2}$ in standard coordinates, characterised by four physical parameters. Similar solutions also occur in quadratic gravity, but with two fewer parameters~\cite{Stelle15PRD}.

\item Solutions in the sub-class $\lbrace 0,1 \rbrace_{f_{1,2}=0, f_3\neq 0}$ correspond to the class $\big( \tfrac{5}{3},\tfrac{1}{3} \big)_{\bar{r}_0,1/3}$ in standard coordinates, characterised by three physical parameters. Similar solutions have not been identified in quadratic gravity.
\end{itemize}

\subsection{Solutions in the class $\lbrace 0,2 \rbrace$}
\label{Sec-02}

For a generic metric with indicial structure $\lbrace 0,2 \rbrace$, the expansion of the field equations starts at the zeroth order, like~\eqref{ExpEoM0000}. The generalised Bianchi identity~\eqref{bia} gives
\beq
\label{bia02}
\frac{\big(\mathcal{E}^{r}{}_{r} \big)_{0} - \big(\mathcal{E}^{t}{}_{t} \big)_{0} }{\De} 
+
 2 \big(\mathcal{E}^{r}{}_{r} \big)_{1} - \big(\mathcal{E}^{t}{}_{t} \big)_{1} + 
\frac{2 f_1}{f_0 } \left[ \big( \mathcal{E}^{r}{}_{r} \big)_{0} - \big( \mathcal{E}^{\th}{}_{\th} \big)_{0} \right]  + \frac{ h_1}{2h_0 } \left[ \big( \mathcal{E}^{r}{}_{r} \big)_{0} - \big(\mathcal{E}^{t}{}_{t} \big)_{0} \right]     + O(\De) = 0 .
\eeq
Hence, if one solves $\big(\mathcal{E}^{t}{}_{t} \big)_{0} = 0$, then it is true that $\big(\mathcal{E}^{r}{}_{r} \big)_{0}=0$. At higher orders, if the field equations are solved up to order $\Delta^N$, \textit{i.e.}, $\big(\mathcal{E}^{\mu}{}_{\nu} \big)_{n}=0$ for all $n\in \lbrace 0,1,\ldots, N\rbrace$, then~\eqref{bia02} gives
\beq
\left[
 (N+2) \big( \mathcal{E}^{r}{}_{r} \big)_{N+1} - \big(\mathcal{E}^{t}{}_{t} \big)_{N+1} \right] \De^{N} + O \left( \De^{N+1} \right)  = 0 \qquad \Longrightarrow \qquad \big(\mathcal{E}^{r}{}_{r} \big)_{N+1} =  \frac{1}{N+2} \big(\mathcal{E}^{t}{}_{t} \big)_{N+1}.
\eeq
Thus, for solutions in the class $\lbrace 0,2 \rbrace$ it is enough to consider the field equations $\mathcal{E}^{\th}{}_{\th} = 0$ and $\mathcal{E}^{t}{}_{t}=0$ (or equivalently, $\mathcal{E}^{r}{}_{r}=0$) order by order.

The solutions in the class $\lbrace 0,2 \rbrace$ are slightly different from those discussed before, inasmuch as the first term of the series of $F(r)$ and $H(r)$ is not a free parameter.
Indeed, here the lowest-order equations $\big(\mathcal{E}^{t}{}_{t} \big)_{0}=0$ and $\big(\mathcal{E}^{\th}{}_{\th} \big)_{0}=0$ must be solved for the parameters $f_0$ and $h_0$, which become fixed by the couplings of the model [see Eq.~\eqref{SolCaseVII} of Appendix~\ref{App}]. 
Already at this level, the class might split into four sub-classes depending on how $f_0$ and $h_0$ are related; the possibilities are:
\begin{subequations}\label{02Subclasses}
\begin{eqnarray} 
h_0 & = & - \frac{1}{f_0^2} ,\quad f_0^2 = \sqrt{\eta_1},
\label{02Subclasses-I}
\\
h_0 & = & \frac{1}{f_0^2} ,\quad f_0^2 = \sqrt{\eta_2},
\label{02Subclasses-II}
\\
h_0 & = & \frac{1}{f_0^2} + \frac{2}{\eta_3 + \eta_4} ,\quad f_0^2 = - \frac{1}{2 \eta_4} \left[ \eta_2 \pm \sqrt{  \eta_2 \left(\eta_2   -2 \eta_4^2 - 2 \eta_3 \eta_4 \right) } \right],
\\
h_0 & = & \frac{1}{f_0^2} + \frac{2}{\eta_3 - \eta_4} ,\quad f_0^2 =  \frac{1}{2 \eta_4} \left[ \eta_2 \pm \sqrt{  \eta_2 \left(\eta_2   -2 \eta_4^2 + 2 \eta_3 \eta_4 \right) } \right],
\end{eqnarray}
\end{subequations}
where the quantities $\eta_1$, $\eta_2$, $\eta_3$ and $\eta_4$ are related to the couplings of the model; see Eq.~\eqref{osces123}.

For any of these possibilities, once the field equations are solved at lowest order, it automatically follows 
\beq
\big(\mathcal{E}^{t}{}_{t} \big)_{1}=0,
\eeq
whereas $\big(\mathcal{E}^{\theta}{}_{\theta} \big)_{1}=0$ yields a constraint between $f_1$ and $h_1$, in the form
\begin{small}
\bea
&&2 f_1 \Big\lbrace 6 \big[4  \gamma_3 +2  \gamma_4 + \gamma_5 + \gamma_6 + 4 \gamma_7 + 4 \gamma_8 )\big]
+2 f_0^2 (2  \beta_1+ \beta_2)+f_0^2 h_0 \big[8  \gamma_1 +6  \gamma_2 +60  \gamma_3 +26  \gamma_4 +12  \gamma_5 +9  \gamma_6 +28  \gamma_7 +12  \gamma_8 +4 f_0^2 (3  \beta_1+ \beta_2)
\nonumber 
\\
&&
- \alpha  f_0^4 \big] 
+2 f_0^4 h_0^2 \big[16  \gamma_1 +6  \gamma_2 +12  \gamma_3 +6  \gamma_4 +3  \gamma_5 +2  \gamma_6 +f_0^2 (10  \beta_1+3  \beta_2)\big]
+ f_0^6 h_0^3 (32  \gamma_1 +10  \gamma_2 -108  \gamma_3 -30  \gamma_4 -9  \gamma_5 -8  \gamma_6 -12  \gamma_7 )  \Big\rbrace
\nonumber 
\\
&&
=
- 3 f_0^3 h_1 \Big\lbrace 12  \gamma_3 +2  \gamma_4 +4  \gamma_7 
+2 f_0^2 h_0 \big[4  \gamma_1 +24  \gamma_3 +4  \gamma_4 + \gamma_6 +2 f_0^2 (3  \beta_1+ \beta_2)\big]
-\alpha  f_0^4
+f_0^4 h_0^2 (16  \gamma_1 +4  \gamma_2 -60  \gamma_3 -22  \gamma_4 -9  \gamma_5 -7  \gamma_6 
\nonumber 
\\
&& \,\,\,\,\,\, -28  \gamma_7 -12  \gamma_8 ) \Big\rbrace .
\label{cons02}
\eea
\end{small}
In the general case, this constraint can be solved for $f_1$ or $h_1$, leaving one parameter free.\footnote{In particular cases, if the couplings of the model are related in a specific way, it can happen that~\eqref{cons02} is satisfied regardless of $f_1$ and $h_1$. We shall not contemplate this possibility here, as in this work we avoid branching into too specific models.} It also admits the trivial solution, $f_1= h_1 =0$. We shall refer to these two sub-classes as $\lbrace 0,2 \rbrace_{f_1,h_1\neq 0}$ and $\lbrace 0,2 \rbrace_{f_1,h_1 = 0}$.

Starting at order $\Delta^2$, at each order $\Delta^N$ the two field equations form a system that might be solved for the two parameters $f_N$ and $h_N$. Here, however, the recursive relations are more involved than in the other classes of solutions. In fact, now the coefficients $\mathfrak{c}_{1,2}(N)$ and $\mathfrak{c}_{1,2}^\prime(N)$ of the parameters $f_N$ and $h_N$ at a given order $\Delta^N$ ($N=2,3,\ldots$) could depend on \textit{all} the couplings of the model [see Eq.~\eqref{c1a4} for the explicit expressions],
\beq
\begin{aligned}
\left( \mathcal{E}^{t}{}_{t} \right)_{N}  & =  
\frac{1}{f_0^7} \left[ 
\mathfrak{c}_{1}(N,\alpha,\beta_{1,2},\gamma_{1,\ldots,8}) \, f_{N} + \mathfrak{c}_{2}(N,\alpha,\beta_{1,2},\gamma_{1,\ldots,8}) \, h_{N} + \Phi_{tt,N-1}  \right]    = 0,
\\
\left( \mathcal{E}^{\th}{}_{\th} \right)_{N}  & =  
\frac{1}{2 f_0^7} \left[ 
\mathfrak{c}_{1}^\prime(N,\alpha,\beta_{1,2},\gamma_{1,\ldots,8}) \, f_{N} + \mathfrak{c}_{2}^\prime(N,\alpha,\beta_{1,2},\gamma_{1,\ldots,8}) \, h_{N}  + \Phi_{\th\th,N-1}  \right]   = 0,
\end{aligned}
\eeq
where $\Phi_{tt,N-1}$ and $\Phi_{\th\th,N-1}$ are functions of the coefficients $h_i$ and $f_i$ with $i = 0,\ldots, N-1$.
This is a very different situation from the other classes of solutions, where we just had a simple dependence on $N$ and on the sixth-derivative couplings $\gamma_{1,2}$ [\textit{cf.} Eqs. \eqref{Sys10-N}, \eqref{aquela00} and~\eqref{seeeg01}]. 
Therefore, provided that $\mathfrak{c}_1^\prime \mathfrak{c}_2 - \mathfrak{c}_1 \mathfrak{c}_2^\prime \neq 0$ for all $N$,\footnote{For specific models, it might happen that there exists $N>1$ such that $\mathfrak{c}_1^\prime \mathfrak{c}_2 - \mathfrak{c}_1 \mathfrak{c}_2^\prime = 0$ but, again, this depends on particular combinations of the couplings of the model.} we can guarantee that the solution exists and is characterised by two free parameters: $r_0$ and one among $f_1$ and $h_1$. Due to the residual gauge freedom, it turns out that the solution does not have a free physical parameter.
The solutions in the class $\lbrace 0,2 \rbrace$ are also regular at $r = r_0$, for instance,
\bea
\label{Kret02}
R & \underset{r \to r_0}{=} & 2 \left(  \frac{1}{f_0^2}- h_0 \right)     - \left[ \frac{4 f_1 \big(2 f_0^2 h_0 + 1\big)}{f_0^3} + 6 h_1 \right] \Delta + O(\De^2) ,
\\
R_{\mu\nu\al\be} R^{\mu\nu\al\be} & \underset{r \to r_0}{=}   &
4 \left(\frac{1}{f_0^4}+h_0^2\right) + \left(24 h_0 h_1-\frac{16 f_1}{f_0^5}\right) \Delta + O(\De^2) .
\eea

It is important to note that the branch of solutions that follows from the trivial solution of~\eqref{cons02} is always present. Indeed, $f_1=h_1 =0$ implies that $\Phi_{tt,1} = \Phi_{\th\th,1} = 0$, whence $f_N=h_N=0$ for all $N=1,2,\ldots$. The solutions in this sub-class $\lbrace 0,2 \rbrace_{f_1,h_1 = 0}$ can be written in closed form, and the geometry corresponds to a direct product of two 2-dimensional metrics with constant curvature,
\beq
\label{formafechada02}
\rd s^2 = - h_0 (r-r_0)^2 \rd t^2 + \frac{\rd r^2}{h_0 (r-r_0)^2} + f_0^2 \left( \rd \th^2 + \sin^2 \th \rd \phi^2 \right) 
.
\eeq
The solution such that $f_0^2 h_0 = -1$ [see Eq.~\eqref{02Subclasses-I}] is the Nariai spacetime, while the one with
$f_0^2 h_0 = 1$ [see Eq.~\eqref{02Subclasses-II}] is the Bertotti--Robinson spacetime. The other cases are also homogeneous spacetimes, as the curvature invariants are constant (see, \textit{e.g.}, \cite{griffiths_podolsky_2009} and references therein).

The metric~\eqref{formafechada02} belongs to the class of near-horizon extreme geometries \cite{Kunduri:2013gce,Kunduri:2008rs,Kunduri:2008tk,Kunduri:2007vf}. It can also be obtained by taking the near-horizon extreme limit of the solutions with non-zero $f_1$ and $h_1$, which effectively corresponds to keeping just the leading terms in $\De$. In this vein, the solutions with $h_0 > 0$ can be considered as the near-horizon limit of a stationary extreme black hole. On the other hand, the ones with $h_0 < 0$ would correspond to the near-horizon limit of geometries with non-stationary regions above and below the extreme horizon; in particular, the metrics in the class $\lbrace 0,2 \rbrace$ with $h_0 < 0$ are not expected to describe stationary black holes.

Finally, we point out that the solutions in the class $\lbrace 0,2 \rbrace$ are dominated by the curvature-cubic terms in the action (proportional to $\ga_{3,\ldots,8}$), while the solutions described in the previous classes are more sensitive to the curvature-quadratic sixth-derivative terms (proportional to $\ga_{1,2}$). Indeed, whereas the solutions obtained in the other classes exist even if the couplings $\ga_{3,\ldots,8}$ are set to zero, the solutions $\lbrace 0,2 \rbrace$ only exist if at least one of $\ga_{3,\ldots,8}$ is non-zero --- regardless of whether $\ga_{1,2}$ are zero or not.

\subsubsection*{Description in standard spherically symmetric coordinates and interpretation}

Using~\eqref{map1} and~\eqref{st2} we conclude that the sub-class $\lbrace 0,2 \rbrace_{f_1,h_1\neq 0}$ corresponds to solutions of type $(2,2)_{\bar{r}_0}$ in standard spherically symmetric coordinates~\eqref{metric-Standard}. Therefore, $C(\bar{r}),B(\bar{r})\sim \bar{\Delta}^2$ and $\bar{r}=\bar{r}_0$ represent an extreme (double-degenerate) horizon. Although in quadratic gravity such solutions occur only in the presence of a cosmological constant~\cite{Pravda:2020zno} or matter~\cite{Pravda:2024uyv}, they exist for a large family of sixth-order pure gravity models. Their extreme nature is also revealed in the absence of free physical parameters, as they are completely determined by the couplings of the model.

On the other hand, the solutions in the sub-class $\lbrace 0,2 \rbrace_{f_1,h_1 = 0}$, given by~\eqref{formafechada02}, represent the extreme near-horizon geometry and are such that $F(r)$ is constant. Thus, as discussed in Sec.~\ref{Sec.Corresp}, they cannot be cast in the form~\eqref{metric-Standard}.

\section{Solutions expanded in powers of $r^{-1}$}
\label{Sec.SolExpInvr}

\subsection{Solutions in the class $\lbrace 1,0 \rbrace^\infty$}
\label{Sec-10Assimpt}

The relation between the coefficients $f_0$ and $h_0$ is fixed in the form $f_0^2 h_0 = 1$ for solutions in the class $\lbrace 1,0 \rbrace^\infty$. Taking this into account, one can verify that the expansion of the field equations in inverse powers of $r$ starts at order $r^{-4}$,
\beq
\label{ExpEoM0010Assimpt}
\mathcal{E}^{\mu}{}_{\nu} = \sum_{i=4}^\infty \left( \mathcal{E}^{\mu}{}_{\nu} \right)_{-i} r^{-i}  = 0 .
\eeq
Therefore, the generalised Bianchi identity~\eqref{bia} at the lowest order yields
\beq
\label{bia10Assimpt}
- 2  \left[ \big(\mathcal{E}^{r}{}_{r} \big)_{-4} + \big({\mathcal{E}^{\theta}}_{\theta} \big)_{-4}
\right]  r^{-5}
+ O \left( r^{-6} \right)  = 0 \quad \Longrightarrow \quad \big({\mathcal{E}^{\theta}}_{\theta} \big)_{-4} = - \big(\mathcal{E}^{r}{}_{r} \big)_{-4}.
\eeq
Assuming that the field equations are solved up to an order $r^{-N}$, \textit{i.e.}, $\big(\mathcal{E}^{\mu}{}_{\nu} \big)_{-n} = 0$ for all $n \in \lbrace 4, 5, \ldots , N \rbrace$, then~\eqref{bia10Assimpt} gives
\beq
\label{bia10nAssimpt}
- \left[  (N-1) \big(\mathcal{E}^{r}{}_{r} \big)_{-(N+1)} + 2 \big({\mathcal{E}^{\theta}}_{\theta} \big)_{-(N+1)} \right]   r^{-(N+2)} + O \left( r^{-(N+3)} \right)  = 0
\quad \Longrightarrow \quad  \big({\mathcal{E}^{\theta}}_{\theta} \big)_{-(N+1)} = -\frac{N-1}{2} \big(\mathcal{E}^{r}{}_{r} \big)_{-(N+1)} .
\eeq
Hence, for solutions in the class $\lbrace 1,0 \rbrace^\infty$ it suffices to solve the $tt$ and $rr$ components of the field equations, order by order.

At a given order $r^{-N}$ ($N=4,5,\ldots$), the field equations yield
\beq
\begin{aligned}
\left( \mathcal{E}^{t}{}_{t} \right)_{-N}  & =  
\frac{(N-3)\alpha}{f_0^3} \left[ 
- 2 (N-3) f_{N-2} + f_0^3 h_{N-2} + \Phi_{tt,N-3} \right]    = 0,
\\
\left( \mathcal{E}^{r}{}_{r} \right)_{-N}  & =  
\frac{(N-3)\alpha}{f_0^3} \left[ 
 2 f_{N-2} + f_0^3 h_{N-2}  + \Phi_{rr,N-3}  \right]   = 0,
\end{aligned}
\eeq
where $\Phi_{tt,N-3}$ and $\Phi_{\th\th,N-3}$ are functions of the coefficients $h_i$ and $f_i$ with $i = 0,\ldots, N-3$. It is immediate to obtain
\beq
\label{sys01inf}
\begin{aligned}
f_{N-2} & = \dfrac{ \Phi_{tt,N-3} - \Phi_{rr,N-3}}{2(N-2)},
\\
h_{N-2} & =  - \dfrac{ \Phi_{tt,N-3} + (N-3) \Phi_{rr,N-3}}{(N-2) f_0^3},
\end{aligned}
\eeq
which proves that the solution can be obtained recursively by solving for $f_{N-2}$ and $h_{N-2}$ order by order starting with $N=4$. Therefore, such a solution is characterised by three free parameters, $f_0$, $f_1$, $h_1$; among these, only one is physical.

The first few terms of the solutions read
\begin{subequations}
\bea
F(r) & = & f_0 r + f_1 - \frac{3 f_0 h_1^2 (14 \gamma_7 + 9 \gamma_8 )}{5 \alpha  r^5} + \frac{3 f_1 h_1^2 (14 \gamma_7 +9 \gamma_8)}{\alpha  r^6} 
\nonumber
\\
&& - \frac{9 h_1^2 \big[540 \beta_1 (2 \gamma_7 + \gamma_8)+30 \beta_2 (12 \gamma_7 +5 \gamma_8 )+7 \alpha  f_1^2 (14 \gamma_7 +9 \gamma_8 )\big]}{7 \alpha ^2 f_0 r^7} + O(r^{-8}) ,
\\
H(r) & = & \frac{1}{ f_0^2} +\frac{ h_1}{r} -\frac{ f_1  h_1}{ f_0 r^2} +\frac{ f_1^2  h_1}{ f_0^2 r^3} 
-\frac{ f_1^3  h_1}{ f_0^3 r^4} + \frac{ f_1^4  h_1}{ f_0^4 r^5}
-\frac{ h_1 \big(\alpha   f_1^5+12  \gamma_7   f_0^3  h_1\big)}{\alpha   f_0^5 r^6}
+ \left[ \frac{ f_1^6  h_1}{ f_0^6} + \frac{72  \gamma_7   f_1  h_1^2}{\alpha   f_0^3} + \frac{2  h_1^3 ( \gamma_8 -24  \gamma_7 )}{5 \alpha } \right] \frac{1}{r^7}
\nonumber
\\
&&
- \left[ \frac{f_1^7 h_1}{f_0^7} + \frac{36 h_1^2 \big[30 \beta_1 (2 \gamma_7 + \gamma_8)+5 \beta_2 (4 \gamma_7 +3 \gamma_8)+7 \alpha  \gamma_7 f_1^2\big]}{\alpha ^2 f_0^4} - \frac{14 f_1 h_1^3 (24 \gamma_7 - \gamma_8)}{5 \alpha  f_0} \right]  \frac{1}{r^8} + O(r^{-9}).
\eea
\end{subequations}
Notwithstanding $\gamma_7$ and $\gamma_8$ are the only six-derivative couplings written in these equations, the dependence on $\gamma_{1,\ldots,6}$ appears at higher orders.
Also, notice that via a redefinition of the series coefficients
\beq
\begin{aligned}
f_0 & =  f_0^\prime , \qquad f_i = \frac{f_i^\prime}{\big( f_0^\prime \big) ^{i-1}} \quad (i=1,2,\ldots),
\\
h_0 & = \frac{1}{f_0^{\prime 2}}  , \qquad h_i   =  \frac{h_i^\prime}{\big( f_0^\prime\big)^{i+2}}  \quad (i=1,2,\ldots),
\end{aligned}
\eeq
one can perform a simultaneous re-scaling of the coordinates $t$ and $r$ [see Eq.~\eqref{ResGauge}, with $\lambda= 1/f_0^\prime$], which is equivalent to setting $f_0=1$. Afterwards, a shift of the coordinate $r$ yields $f_1=0$, leaving $h_1$ as the only physical free parameter. 
As we show in what follows, this solution can be interpreted as asymptotic corrections to the Schwarzschild geometry.

\subsubsection*{Description in standard spherically symmetric coordinates and interpretation}

Combining~\eqref{map2} and~\eqref{st1} it follows that the solution class $\lbrace 1,0 \rbrace^\infty$ corresponds to the class $(0,0)^\infty$ representing an asymptotic expansion in standard spherically symmetric coordinates~\eqref{metric-Standard},
\begin{subequations}\label{Sol00InfStand}
\bea
C(\bar{r}) & = & 1 + \dfrac{b_1}{\bar{r}} + \dfrac{18 b_1^2 (4 \gamma_7 + 3 \gamma_8)}{\alpha \bar{r}^6} + \dfrac{b_1^3 (66 \gamma_7 + 49 \gamma_8)}{\alpha \bar{r}^7}
+ \dfrac{720 b_1^2 [8 \gamma_7 (3 \beta_1 + \beta_2 )+3 \gamma_8 (4 \beta_1 + \beta_2) ]}{\alpha ^2 \bar{r}^8}
+ O(\bar{r}^{-9}) ,
\\
\dfrac{B(\bar{r})}{b_0} & = & 1 + \dfrac{b_1}{\bar{r}} - \dfrac{12 b_1^2 \gamma_7 }{\alpha \bar{r}^6} - \dfrac{b_1^3 (18 \gamma_7 + 5 \gamma_8 )}{\alpha \bar{r}^7}
- \dfrac{180 b_1^2 [4 \gamma_7 (3 \beta_1 + \beta_2 ) + 3 \gamma_8 (2 \beta_1 + \beta_2)]}{\alpha ^2 \bar{r}^8}
+ O(\bar{r}^{-9}).
\eea
\end{subequations}
These solutions can be regarded as higher-derivative corrections to the Schwarzschild geometry in the regime of large $\bar{r}$, and the single free physical parameter $b_1$ can be interpreted as mass.
In fact, the Schwarzschild spacetime is an exact solution of the model~\eqref{mostgeneralaction} if $\gamma_7 = \gamma_8 = 0$.  
Other terms in the action~\eqref{mostgeneralaction} can affect the solution, but their contributions are sourced by the terms associated with $\gamma_7$ and $\gamma_8$. For example, the couplings $\beta_{1,2}$ start to contribute at order $\bar{r}^{-8}$, whereas $\gamma_{1,2}$ at order $\bar{r}^{-10}$, $\gamma_6$ at order $\bar{r}^{-11}$, and $\gamma_{3,4,5}$ at order $\bar{r}^{-16}$; they are always multiplied by $\gamma_7$ or $\gamma_8$.

Corrections to the Schwarzschild metric originated by higher-derivative terms have been obtained in the literature within the effective and fundamental frameworks (see, \textit{e.g.},~\cite{Anselmi:2013wha,deRham:2020ejn,Daas:2023axu,Daas:2024pxs}). Our solution~\eqref{Sol00InfStand} coincides with the results from these works after making the necessary adjustments of notation and choices of higher-derivative terms considered in the gravitational action. However, to our knowledge, the identification of the order in which the couplings $\gamma_{1,\ldots,6}$ start to contribute has not been presented before.

Finally, we point out that, as an asymptotically flat solution,~\eqref{Sol00InfStand} can be compared to the solution of the linearised field equations presented in Appendix~\ref{App-Lin}. Nonetheless, the Yukawa potentials in~\eqref{PhiPsi} and~\eqref{PhiPsi-delta} cannot be reproduced by a Frobenius series in inverse powers of $\bar{r}$ and, thus, escape the ansatz that we explored for the solutions of the full non-linear theory. Conversely, the linearised field equations around the Minkowski background are not affected by cubic curvature terms, such as those related to $\gamma_7$ and $\gamma_8$ that proved to be decisive in the higher-order terms in~\eqref{Sol00InfStand}. In conclusion, the common ground for both families of solutions comprises the $\gamma_{7,8}$-independent terms and the terms that are analytic in $\bar{r}^{-1}$, namely, the constant zeroth-order term and the leading term $\bar{r}^{-1}$. The complete field equations may also admit solutions that behave like Yukawa potentials in the regime of large $\bar{r}$, but this might only be assessed by other techniques, \textit{e.g.}, numerical calculations~\cite{Pawlowski:2023dda,Daas:2025lzr} (see also~\cite{Stelle15PRD,Daas:2022iid,Bonanno:2019rsq,Bonanno:2022ibv,Silveravalle:2022wij} for related discussions in the context of four-derivative gravity).

\subsection{Solutions in the class $\lbrace 1,2 \rbrace^\infty$}
\label{Sec-12Assimpt}

Solutions in the class $\lbrace 1,2 \rbrace^\infty$ exist provided that the couplings of the model satisfy the relation
\beq
\frac{144 \gamma_{3} +36  \gamma_{4} +9  \gamma_{5} +9  \gamma_{6} +24  \gamma_{7} +4  \gamma_{8} }{\alpha} > 0,
\eeq
in which case the indicial equations (field equations at order $r^0$) result in
\beq
\label{h0-02inf}
h_0 = \pm \sqrt{\frac{\alpha}{144 \gamma_{3} +36  \gamma_{4} +9  \gamma_{5} +9  \gamma_{6} +24  \gamma_{7} +4  \gamma_{8} }} \, ,
\eeq
see Appendix~\ref{App} for the details.
For the sake of economy of notation and since the coefficient $h_0$ is completely fixed by the parameters of the model, in what follows we shall implement condition~\eqref{h0-02inf} by rewriting $\alpha$ as a function of $h_0^2$ and $\gamma_{3,\ldots,8}$. In other words, we are trading the parameter $\alpha$ for $h_0$, without loss of generality.

Expanding the field equations in inverse powers of $r$,
\beq
\label{ExpEoM0012Assimpt}
\mathcal{E}^{\mu}{}_{\nu} = \sum_{i=0}^\infty \left( \mathcal{E}^{\mu}{}_{\nu} \right)_{-i} r^{-i}  = 0 ,
\eeq
one can verify that, if the field equations are solved up to an order $r^{-N}$, \textit{i.e.}, $\big(\mathcal{E}^{\mu}{}_{\nu} \big)_{-n} = 0$ for all $n \in \lbrace 0, 1, \ldots , N \rbrace$, then the generalised Bianchi identity~\eqref{bia} implies
\beq
\label{bia12nAssimpt}
\left[ \big(\mathcal{E}^{t}{}_{t} \big)_{-(N+1)} + (N-2) \big(\mathcal{E}^{r}{}_{r} \big)_{-(N+1)} + 2 \big({\mathcal{E}^{\theta}}_{\theta} \big)_{-(N+1)} \right]   r^{-(N+2)} + O \left( r^{-(N+3)} \right)  = 0.
\eeq
Thus, one can choose any two components of the field equations to solve order by order and make sure that $\big(\mathcal{E}^{r}{}_{r} \big)_{-3} = 0$ also holds. To avoid this extra step, we shall work directly with the equation $\mathcal{E}^{r}{}_{r} = 0$; as for the other component, we choose the $tt$ one.

At a given order $r^{-N}$ ($N=1,2,\ldots$), the field equations yield
\beq
\begin{aligned}
\label{otalsis}
\left( \mathcal{E}^{t}{}_{t} \right)_{-N}  & =  
\frac{(N-3)h_0}{f_0} \left[ 
2 N \mathfrak{d}_{1}(N) \, h_0 f_{N} + \mathfrak{d}_{2}(N) \, f_0 h_{N} \right] + \Phi_{tt,N-1}   = 0,
\\
\left( \mathcal{E}^{r}{}_{r} \right)_{-N}  & =  
\frac{\mathfrak{d}_{3}(N) \, h_0}{f_0} \left[ 
4 N h_0 f_N + (N-3) f_0 h_N   \right] + \Phi_{rr,N-1}   = 0,
\end{aligned}
\eeq
where the coefficients $\mathfrak{d}_{1,2,3}(N)$ actually depend on the couplings of the model [see Eq.~\eqref{d1a4} for the explicit expressions] and $\Phi_{tt,N-1}$ and $\Phi_{rr,N-1}$ are functions of the coefficients $h_i$ and $f_i$ with $i = 0,\ldots, N-1$. This system has the unique solution 
\beq
\label{sys12inf}
\begin{aligned}
f_{N} & = - \dfrac{ \mathfrak{d}_{3} \Phi_{tt,N-1} - \mathfrak{d}_{2} \Phi_{rr,N-1}  }{2N \mathfrak{d}_{3} \mathfrak{d}_{4} h_0^2}f_0,
\\
h_{N} & =  \dfrac{ 2 \mathfrak{d}_{3} \Phi_{tt,N-1} - (N-3) \mathfrak{d}_{1} \Phi_{rr,N-1}}{(N-3) \mathfrak{d}_{3} \mathfrak{d}_{4} h_0} ,
\end{aligned}
\eeq
with 
\beq
\mathfrak{d}_{4}(N) \equiv (N-3) \mathfrak{d}_{1}(N) - 2 \mathfrak{d}_{2}(N)  ,
\eeq
provided that  $(N-3) \mathfrak{d}_{3}(N) \mathfrak{d}_{4}(N) \neq 0$. Since $\mathfrak{d}_{4}(N)$ is proportional to $(N-1)$ [see Eq.~\eqref{d1a4.4}], it is clear that the cases of $N=1$ and $N=3$ have to be considered separately. For the other values of $N$, the system~\eqref{otalsis} has a unique solution unless the couplings of the model satisfy specific relations (such a possibility will not be considered in the present work).

For $N=1$, it turns out that $\mathfrak{d}_{2}(1) = - \mathfrak{d}_{1}(1)$ and $\Phi_{tt,0}=\Phi_{rr,0}=0$; thence, the system~\eqref{otalsis} is underdetermined and its solution imposes a relation between $f_1$ and $h_1$:
\beq
h_1 = \frac{2 f_1 h_0}{f_0} .
\eeq
For $N=2$, the system~\eqref{otalsis} has a single solution,
\beq
f_2 = 0 , \qquad h_2 = \frac{1+ f_1^2 h_0}{f_0^2} .
\eeq
For $N=3$, it also happens that $\Phi_{tt,2}=\Phi_{rr,2}=0$ so that the $tt$ component of~\eqref{otalsis} is automatically satisfied, whilst the $rr$ component gives
\beq
f_3 = 0,
\eeq
leaving $h_3$ as a free parameter.

Therefore, solutions in the class $\lbrace 1,2 \rbrace^\infty$ are characterised by three free parameters: $f_0$, $h_1$ and $h_3$. The first few terms of the solution read
\begin{subequations}
\bea
F(r) & = & f_0 r + f_1 + \frac{f_6(f_0,h_0,f_1,h_3,\beta_{1,2},\gamma_{1,\ldots,8})}{r^5}  + O(r^{-6})
\\
H(r) & = & h_0 r^2 + \frac{2 f_1 h_0}{f_0} r + \frac{1+ f_1^2 h_0}{f_0^2} + \frac{h_3}{r} - \frac{f_1 h_3}{f_0 r^2}
+ \frac{f_1^2 h_3}{f_0^2 r^3} + \frac{h_6(f_0,h_0,f_1,h_3,\beta_{1,2},\gamma_{1,\ldots,8})}{r^4} + O(r^{-5}).
\eea
\end{subequations}
Among the free parameters, only one is physical. 
For instance, by redefining the series coefficients 
\beq
\begin{aligned}
f_0 & =  f_0^\prime , \qquad f_i = \frac{f_i^\prime}{\big( f_0^\prime \big) ^{i-1}} \quad (i=1,2,\ldots),
\\
h_i  & =  \frac{h_i^\prime}{\big( f_0^\prime\big)^{i}}  \quad (i=0,1,\ldots),
\end{aligned}
\eeq
it is straightforward to verify that the simultaneous rescaling of the coordinates $t$ and $r$, following~\eqref{ResGauge} with $\lambda= 1/f_0^\prime$, is equivalent to setting $f_0=1$. Similarly, a shift of the coordinate $r$ [see~\eqref{ResGauge}] can be applied to produce a metric with $f_1=0$. Thus, the only physical free parameter is $h_3$, which allows us to divide the class into two sub-classes, $\lbrace 1,2 \rbrace^\infty_{h_3\neq 0}$ and $\lbrace 1,2 \rbrace^\infty_{h_3=0}$, depending on whether $h_3$ vanishes or not.

The solution with $h_3 = 0$ turns out to be either the de~Sitter or anti-de~Sitter spacetime, depending on the sign of $h_0$,
\beq
\label{formafechada12inf}
\rd s^2 = - ( 1 +  h_0 r^2 ) \rd t^2 + \frac{\rd r^2}{ 1 +  h_0 r^2 } + r^2 \left( \rd \th^2 + \sin^2 \th \rd \phi^2 \right) 
,
\eeq
with $h_0$ given by~\eqref{h0-02inf} playing the role of an effective cosmological constant.
Note that this solution is also contained in the sub-class $\lbrace 1,0 \rbrace_{f_{1,2,3,4,5}=h_{3,4,5}=0}$.

\subsubsection*{Description in standard spherically symmetric coordinates and interpretation}

The solutions in the sub-class $\lbrace 1,2 \rbrace^\infty_{h_3=0}$, which are equivalent to de~Sitter or anti-de~Sitter geometries~\eqref{formafechada12inf}, are already in standard spherically symmetric coordinates. Hence, although the discussion involving~\eqref{map2} and~\eqref{st1} leads to a mapping of the class $\lbrace 1,2 \rbrace^\infty$ to an asymptotic expansion of type $(2,2)^\infty$, for this sub-class the correspondence actually holds for any value of $\bar{r}\geqslant 0$. In fact, as mentioned above, the metric~\eqref{formafechada12inf} is also contained in the class $\lbrace 1,0 \rbrace_{f_{1,2,3,4,5}=h_{3,4,5}=0}$, which is mapped to $(0,0)_0$.

For the sub-class $\lbrace 1,2 \rbrace^\infty_{h_3\neq 0}$, however, we obtain solutions that do manifest their asymptotic nature through negative powers of~$\bar{r}$,
\begin{subequations}
\bea
C(\bar{r}) & = & c_0 \bar{r}^2 + 1 + \frac{c_3}{\bar{r}} + \frac{c_6}{\bar{r}^4}  +  O(\bar{r}^{-6})
\\
\frac{B(\bar{r})}{b_0} & = & \bar{r}^2 + \frac{1}{c_0} + \frac{c_3}{c_0 \bar{r}} + \frac{b_6}{\bar{r}^4}   + O(\bar{r}^{-6}).
\eea
\end{subequations}
The sole free physical parameter of the solution is $c_3$, while $c_0=h_0$ is fixed by the couplings of the model according to Eq.~\eqref{h0-02inf} and $b_0$ is not physical.
The higher-order coefficients $c_{6,\ldots}$ and $b_{6,\ldots}$ can be expressed in terms of $c_3$ and the parameters of the model; more specifically, they are proportional to $c_3$ and to the couplings $\gamma_7$ and $\gamma_8$. 
These solutions can be viewed as asymptotic corrections to the Schwarzschild--(anti-)de~Sitter geometry in the same way that the solution~\eqref{Sol00InfStand} represents corrections to the Schwarzschild spacetime. Indeed, one can verify that the Schwarzschild--(anti-)de~Sitter spacetime
\beq
\label{Sch(A)dS}
\rd s^2 = - \left( 1 + h_0 \bar{r}^2 + \frac{c_3}{\bar{r}} \right)  \rd t^2 + \frac{\rd \bar{r}^2}{ 1 + h_0 \bar{r}^2 + \frac{c_3}{\bar{r}} } + \bar{r}^2 \left( \rd \th^2 + \sin^2 \th \rd \phi^2 \right) 
,
\eeq
with $h_0$ playing the role of an effective cosmological constant, 
is an exact solution of the model~\eqref{mostgeneralaction} if $\gamma_7=\gamma_8=0$.

\subsection{Solutions in the class $\lbrace 0,2 \rbrace^\infty$}
\label{Sec-02Assimpt}

From the physical point of view, the solution class $\lbrace 0,2 \rbrace^\infty$ does not introduce any new solutions to the ones already mentioned in the class $\lbrace 0,2 \rbrace$. Its only solutions are the direct product spacetimes of Eq.~\eqref{formafechada02}. Nevertheless, for the sake of completeness, in what follows, we prove this statement.

Solving the field equations at the lowest order $r^0$ fixes the coefficients $f_0$ and $h_0$ as functions of the parameters of the model [see Eq.~\eqref{SolCaseVII} of Appendix~\ref{App}]. At higher orders, assuming that the field equations are solved up to order $r^{-N}$, \textit{i.e.}, $\big(\mathcal{E}^{\mu}{}_{\nu} \big)_{-n}=0$ for all $n\in \lbrace 0,1,\ldots, N\rbrace$, then the generalised Bianchi identity~\eqref{bia} yields
\beq
\left[
\big(\mathcal{E}^{t}{}_{t} \big)_{-(N+1)} + N \big( \mathcal{E}^{r}{}_{r} \big)_{-(N+1)} \right]r^{-(N+2)} + O \left( r^{-(N+3)} \right)  = 0 \qquad \Longrightarrow \qquad \big(\mathcal{E}^{t}{}_{t} \big)_{-(N+1)} = - N \big(\mathcal{E}^{r}{}_{r} \big)_{-(N+1)}.
\eeq
Thus, in this discussion we shall only consider the components $rr$ and $\theta\theta$ of the field equations.

At a given order $r^{-N}$ ($N=1,2,\ldots$), the field equations have the form
\beq
\begin{aligned}
\label{otalsis02inf}
\left( \mathcal{E}^{r}{}_{r} \right)_{-N}  & =  
\frac{1}{f_0^7} \left[ 
2 \mathfrak{e}_{1}(N) \, f_{N} - (N-2)(N-1) \mathfrak{e}_{2}(N) \, h_{N} \, +  \, \Phi_{rr,N-1} \right]   = 0,
\\
\left( \mathcal{E}^{\theta}{}_{\theta} \right)_{-N}  & =  
\frac{1}{2f_0^7} \left[ 
2 \mathfrak{e}_{3}(N) \, f_{N}  + (N-2)(N-1) \mathfrak{e}_{4}(N) \, h_{N} \, +  \, \Phi_{\theta\theta,N-1} \right]   = 0,
\end{aligned}
\eeq
where the coefficients $\mathfrak{e}_{1,2,3,4}(N)$ actually depend on the couplings of the model [see Eq.~\eqref{e1a4} for the explicit expressions] and $\Phi_{rr,N-1}$ and $\Phi_{\theta\theta,N-1}$ are functions of the coefficients $h_i$ and $f_i$ with $i = 0,\ldots, N-1$. For $N=3,4,\ldots$, this system has the unique solution 
\beq
\begin{aligned}
f_{N} & = - \dfrac{ \mathfrak{e}_{4} \Phi_{rr,N-1} + \mathfrak{e}_{2} \Phi_{\theta\theta,N-1}  }{2(\mathfrak{e}_{2} \mathfrak{e}_{3} + \mathfrak{e}_{1} \mathfrak{e}_{4})},
\\
h_{N} & =  \dfrac{ \mathfrak{e}_{3} \Phi_{rr,N-1} - \mathfrak{e}_{1} \Phi_{\theta\theta,N-1}}{(N-2)(N-1) (\mathfrak{e}_{2} \mathfrak{e}_{3} + \mathfrak{e}_{1} \mathfrak{e}_{4})} ,
\end{aligned}
\eeq
provided that  
\beq
\label{dete}
\mathfrak{e}_{2} \mathfrak{e}_{3} + \mathfrak{e}_{1} \mathfrak{e}_{4} \neq 0  .
\eeq
Since in the present work we do not consider models whose couplings satisfy particular relations, we shall assume that~\eqref{dete} holds true for all $N$. 
In the same spirit, for $N=1$ and $N=2$ we have $\Phi_{rr,0} = \Phi_{\theta\theta,0} = \Phi_{rr,1} = \Phi_{\theta\theta,1} = 0$, and the solution of the system~\eqref{otalsis02inf} is
\beq
f_1 = f_2 = 0 , \qquad h_1, h_2 \in \mathbb{R}.
\eeq
Therefore, a solution in the class $\lbrace 0,2 \rbrace^\infty$ is characterised by two free parameters, $h_1$ and $h_2$, none of which is physical.

One can solve the field equations to higher orders to obtain $f_3=h_3=0$, $f_4=h_4=0$ and so on. Indeed, by direct substitution into the field equations, one can verify that the metric
\beq
\label{formafechada02inf}
\rd s^2 = - ( h_0 r^2  +  h_1 r + h_2 ) \rd t^2 + \frac{\rd r^2}{h_0 r^2  +  h_1 r + h_2} + f_0^2 \left( \rd \th^2 + \sin^2 \th \rd \phi^2 \right) 
.
\eeq
with $f_0$ and $h_0$ given by any of the possibilities of Eq.~\eqref{SolCaseVII} is an exact solution. In particular, there is no loss of generality in taking $h_1 = h_2 = 0$, since the same effect can be achieved by redefining the coordinates according to the transformations~\eqref{ResGauge}. This completes the proof that, without assuming special relations between the couplings of the model, the geometries represented by the solutions in the class $\lbrace 0,2 \rbrace^\infty$ are contained in the class $\lbrace 0,2 \rbrace$ --- more precisely, in the sub-class $\lbrace 0,2 \rbrace_{f_1,h_1 = 0}$. In addition, they cannot be expressed using standard spherically symmetric coordinates~\eqref{metric-Standard}.

\section{Summary and conclusions}
\label{Sec.SumCon}

In the present work, we studied exact static spherically symmetric vacuum solutions in the generic six-derivative gravity, \textit{i.e.}, without assuming any special relation between the couplings. We systematically analysed the possible solutions admitting Frobenius expansions around ${r=r_0}$ and ${r=\infty}$ (that is, $1/r$ expansion) in the modified Schwarzschild coordinates with coupling-independent series exponents. The classes of solutions found are summarised in Tables~\ref{Tab1} and~\ref{Tab2}. Due to the field equations being autonomous in these coordinates,\footnote{Although the field equations are also autonomous in conformal-to-Kundt coordinates, as shown in the contexts of Einstein--Weyl~\cite{Podolsky:2018pfe,Svarc:2018coe,Podolsky:2019gro} and Weyl-cubic~\cite{Daas:2023axu} gravity models, these coordinates are not particularly useful in general six-derivative gravity, where many terms in the field equations do not transform well under conformal transformation.} we were able to prove the existence of the solutions from \cite{Praha1}, instead of relying on the observation that no more extra parameters seem to appear after solving the field equations up to a certain order. Returning to standard Schwarzschild coordinates, this confirmed that the only solutions of this type are regular at the origin $\bar{r}=0$. The main result of our work, however, is the discovery of novel classes of solutions, including some that cannot be covered by Schwarzschild coordinates. 

\begin{table}[b]
    \centering
    \begin{tabular}{|c|c|c|c|c|}
            \hline
             \multicolumn{2}{|c|}{Solution family}  & \multirow{2}{*}{Parameters} & Number of free & \multirow{2}{*}{Interpretation} \\ 
             $\,(s,t)_0\,$ or $\,(-s,t)_{\bar{r}_0}\,$  & $\lbrace \si,\ta \rbrace$   & 	       & parameters &  \\ 
            \hline
            \hline
            $(0,0)_0$ & $\lbrace 1,0 \rbrace$ & $f_0$, $f_2$, ($f_3$, $h_3$), $f_4$, $h_2$, $h_4$, $r_0$ & $7 \to 5$ & regular core \\ 
            \hline
            \hline
            $(0,0)_{\bar{r}_0}$ & $\lbrace 0,0 \rbrace$ & ($f_0$, $f_1$, $f_2$, $f_3$, $f_4$, $f_5$, $h_0$, $h_1$, $h_2$, $h_3$, $h_4$, $h_5$), $r_0$ & $12 \to 10$ & generic solution \\ 
            \hline  
            $(1,0)_{\bar{r}_0, 1/2}$ & $\lbrace 0,0 \rbrace_{f_1=0, \,f_2\neq 0}$ & ($f_0$, $f_2$, $f_3$, $f_4$, $f_5$, $h_0$, $h_1$, $h_2$, $h_3$, $h_4$, $h_5$), $r_0$ & $11 \to 9$ & non-symmetric wormhole \\ 
            \hline  
            $\left( 1,0 \right) _{\bar{r}_0}$ & $\lbrace 0,0 \rbrace_{f_{1,3,5}=h_{1,3,5}=0, \,f_6\neq 0}$ & ($f_0$, $f_2$, $f_4$, $h_0$, $h_2$, $h_4$), $r_0$ & $6 \to 4$ & symmetric wormhole \\ 
            \hline
            $\left( \tfrac{4}{3},0 \right) _{\bar{r}_0, 1/3}$ & $\lbrace 0,0 \rbrace_{f_{1,2}=0, \, f_3\neq 0}$ & ($f_0$, $f_3$, $f_4$, $f_5$, $h_0$, $h_1$, $h_2$, $h_3$, $h_4$, $h_5$), $r_0$ & $10 \to 8$ & \multirow{6}{*}[-10pt]{unusual wormhole} \\ 
            \cline{1-4}%
            $\left( \tfrac{3}{2},0 \right) _{\bar{r}_0, 1/4}$ & $\lbrace 0,0 \rbrace_{f_{1,2,3}=0, \, f_4\neq 0}$ & ($f_0$, $f_4$, $f_5$, $h_0$, $h_1$, $h_2$, $h_3$, $h_4$, $h_5$), $r_0$ & $9 \to 7$ &  \\ 
            \cline{1-4}%
            $\left( \tfrac{3}{2},0 \right) _{\bar{r}_0, 1/2}$ & $\lbrace 0,0 \rbrace_{f_{1,2,3,5}=h_{1,3,5}=0, \, f_6\neq 0}$ & ($f_0$, $f_4$, $h_0$, $h_2$, $h_4$), $r_0$ & $5 \to 3$ & \\
            \cline{1-4}%
            $\left( \tfrac{8}{5},0 \right) _{\bar{r}_0, 1/5}$ & $\lbrace 0,0 \rbrace_{f_{1,2,3,4}=0, \, f_5\neq 0}$ & ($f_0$, $f_5$, $h_0$, $h_1$, $h_2$, $h_3$, $h_4$, $h_5$), $r_0$ & $8 \to 6$ &  \\ 
            \cline{1-4}%
            $\left( \tfrac{5}{3},0 \right) _{\bar{r}_0, 1/6}$ & $\lbrace 0,0 \rbrace_{f_{1,2,3,4,5}=0, \, f_6\neq 0}$ & ($f_0$, $h_0$, $h_1$, $h_2$, $h_3$, $h_4$, $h_5$), $r_0$ & $7 \to 5$ & \\ 
            \cline{1-4}%
            $\left( \tfrac{5}{3},0 \right) _{\bar{r}_0, 1/3}$ & $\lbrace 0,0 \rbrace_{f_{1,2,3,4,5}=h_{1,3,5}=0, \, f_6 \neq 0}$ & ($f_0$, $h_0$, $h_2$, $h_4$), $r_0$ & $4 \to 2$ & \\ 
            \hline
            \hline
            $(1,1)_{\bar{r}_0}$ & $\lbrace 0,1 \rbrace$ & $f_0$, $f_1$, $f_2$, $h_0$, $h_1$, $h_2$, $r_0$ & $7 \to 5$ & black hole horizon \\ 
            \hline 
            $\left( \tfrac{3}{2},\tfrac{1}{2} \right)_{\bar{r}_0,1/2}$ & $\lbrace 0,1 \rbrace_{f_1=0, \, f_2 \neq 0}$ & $f_0$, $f_2$, $h_0$, $h_1$, $h_2$, $r_0$ & $6 \to 4$ &  \multirow{2}{*}[-3pt]{unusual horizon}\\ 
            \cline{1-4}
            $\left( \tfrac{5}{3},\tfrac{1}{3} \right)_{\bar{r}_0,1/3}$ & $\lbrace 0,1 \rbrace_{f_{1,2}=0, \, f_3\neq 0}$ & $f_0$, $h_0$, $h_1$, $h_2$, $r_0$ & $5 \to 3$ & \\ 
            \hline
            \hline
            $(2,2)_{\bar{r}_0}$ & $\lbrace 0,2 \rbrace$ & ($f_1$, $h_1$), $r_0$  & $2 \to 0$ & extreme (double) horizon \\ 
            \hline
            $\nexists$ & $\lbrace 0,2 \rbrace_{f_1=h_1=0}$ & $r_0$  & $1 \to 0$ & extreme near-horizon geometry  \\ 
            \hline
        \end{tabular}    
        \caption{\small Summary of solutions in Frobenius series around $r=r_0$. The first column indicates the indicial structure in standard spherically symmetric coordinates~\eqref{metric-Standard}. The second column indicates the class or sub-class in modified coordinates~\eqref{metric}. The third column lists the free parameters of the solution in the form~\eqref{FrobeniusHF}; parentheses indicate that there exists a constraint between the coefficients, so the total number of free parameters is one less. The count of free parameters that characterise the solution is listed in the fourth column, with the arrow indicating the reduction of parameters to the physical ones after taking into account the residual gauge freedom for the metric~\eqref{metric}. The interpretation of the solution is provided in the last column.}
\label{Tab1}
\end{table}

\begin{table}[t]
    \centering
    \begin{tabular}{|c|c|c|c|c|}
            \hline
             \multicolumn{2}{|c|}{Solution family}  & \multirow{2}{*}{Parameters} & Number of free & \multirow{2}{*}{Interpretation} \\ 
             $\,(-s,t)^\infty\,$  & $\lbrace \si,\ta \rbrace^\infty$   & 	       & parameters  & \\ 
            \hline
            \hline
            $(0,0)^\infty$ & $\lbrace 1,0 \rbrace^\infty$ & $f_0$, $f_1$, $h_1$ & $3 \to 1$ & corrections to Schwarzschild \\ 
            \hline
            $(0,0)^\infty$ & $\lbrace 1,0 \rbrace^\infty_{h_1=0}$ & $f_0$, $f_1$ & $2 \to 0$ & Minkowski \\ 
            \hline
            \hline
            $(2,2)^\infty$ & $\lbrace 1,2 \rbrace^\infty$ & $f_0$, $h_1$, $h_3$ & $3 \to 1$ & corrections to Schwarzschild--(A)dS \\ 
            \hline
            $(2,2)^\infty$ & $\lbrace 1,2 \rbrace^\infty_{h_3=0}$ & $f_0$, $h_1$ & $2 \to 0$ & de~Sitter or anti-de~Sitter \\ 
            \hline  
            \hline  
            $\nexists$ & $\lbrace 0,2 \rbrace^\infty$ & $h_1$, $h_2$ & $2 \to 0$ & extreme near-horizon geometry \\
            \hline  
        \end{tabular}    
        \caption{\small Summary of solutions in Frobenius series in powers of $r^{-1}$. The first column indicates the indicial structure in standard spherically symmetric coordinates~\eqref{metric-Standard}. The second column indicates the class or sub-class in modified coordinates~\eqref{metric}. The third column lists the free parameters of the solution in the form~\eqref{FrobeniusHF-Infinity}, while the fourth column indicates their number and the reduction to the physical parameters after taking into account the residual gauge freedom. The last column provides the interpretation of the solution.
         }
\label{Tab2}
\end{table}

Especially interesting new solutions are those within families that are not present in four-derivative gravity (therefore, not in general relativity either). Specifically, we found static solutions admitting extreme (double degenerate) horizons in the class $\lbrace 0,2 \rbrace$ with ${h_0>0}$; such solutions in general relativity and quadratic gravity always require some matter content such as electromagnetic field (\textit{e.g.}, the Reissner--Nordstr\"om solution or solutions in~\cite{Pravda:2024uyv}). We believe that the presence of extreme horizons may actually hint at the possible existence of regular black holes as it is well known that every asymptotic flat regular black hole must have an even number of horizons (see, \textit{e.g.},~\cite{Maeda:2021jdc}). Although beyond the local analysis capabilities of the Frobenius expansion, one could hope that there may in principle also exist two-horizon solutions that degenerate into the extreme ones we found for some choices of parameters. Confirming or disproving this hypothesis would probably require numerical treatment. Owing to the modified Schwarzschild coordinates, we could also easily identify the solutions corresponding to the near-horizon limits of the above solution, which gives rise to the direct product spacetimes of 2-spaces of constant curvature (\textit{e.g.}, the Bertotti--Robinson). It is also worth mentioning that no solutions with triple or more degenerate horizons exist in the generic six-derivative gravity; these would be needed for regular black holes bypassing the mass inflation instability \cite{Carballo-Rubio:2022kad}. Other solutions that are unique to six-derivative gravity are the classes of asymptotically (anti-)de~Sitter spacetimes, with an effective cosmological constant
\beq
\Lambda_{\text{eff}} = \pm \sqrt{\frac{\alpha}{144 \gamma_{3} +36  \gamma_{4} +9  \gamma_{5} +9  \gamma_{6} +24  \gamma_{7} +4  \gamma_{8} }} .
\eeq

Given that most papers on spherically symmetric static solutions in six-derivative gravity have focused on the weak-field regime, specifically, linearised solutions around Minkowski spacetime \cite{Accioly:2016qeb,Accioly:2016etf,Frolov:Poly,BreTib1,Nos6der}, it is important to explore how our results relate to these prior findings. Our exact solutions should match the linearised solutions in the domains in which both are sufficiently close to the flat spacetime. This is certainly the case for the asymptotic solutions; however, the well-known Yukawa-like terms associated with the massive modes in the linearised model are not accessible by any Frobenius expansion (in standard or modified Schwarzschild coordinates), as they are not analytic in $1/\bar{r}$. On the other hand, the extra terms that we found solving the complete field equations are dominated by the non-linear part; as a consequence, the comparison can only be made up to the leading term $1/\bar{r}$.

Even more interesting is the comparison between exact and linearised regular solutions around ${\bar{r}=0}$. Exact solutions have (anti-)de~Sitter-like cores, if the functions $A$ and $B$ in the metric~\eqref{metric-Standard} behave as ${1-\mu\bar{r}^2 + O(\bar{r}^3)}$, or Minkowski cores, if $\mu =0$. In the first situation, the solution can be compared with the known linearised solutions around Minkowski, but only for ${\bar{r}\ll \vert\mu\vert^{-1/2}}$. It indeed happens that the global vacuum linearised solution is able to reproduce the exact solution in this regime. One could extend the comparison to higher orders based on linearised vacuum solutions around de~Sitter, were they available. 
Another option is to compare with the linearised solutions with sources; for instance, a delta source can generate a cubic term in $\bar{r}$ in the expansions of $A$ and $B$~\cite{Nos6der}. Finally, in the situation of exact solutions with Minkowski cores, there is the possibility of having the leading term at order $\bar{r}^3$ or $\bar{r}^4$. While the latter case is compatible with the linearised global vacuum solution, the former case cannot be reproduced by such a solution. This might suggest the breakdown of the linear approximation at $O(\bar{r}^3)$ or the existence of a source at $\bar{r}=0$ (recall that a cubic term implies that the solution is not smooth at $\bar{r}=0$, although being regular).

Last but not least, our analysis relies on several key assumptions each of which could be relaxed in the future works. We performed the Frobenius expansion in the modified Schwarzschild coordinates while restricting ourselves to the general six-derivative gravity and the solutions with coupling-independent series exponents. As we have seen, by doing the Frobenius analysis in different coordinates, one might discover non-Frobenius solutions in other coordinates. Apart from using different coordinates, there are also many interesting non-Frobenius series one might consider (such as expansions capable of capturing the Yukawa-like terms present in the linearised weak-field regime or some more exotic solutions). Furthermore, we assumed the independence of exponents on the coupling constants in order to classify the solution classes that are common to generic six-derivative gravities. This is probably a strong constraint in the space of solutions, given that a six-derivative gravity model is defined by eleven parameters. Solutions with coupling-dependent indices are known to exist in the simpler quadratic gravity~\cite{Stelle15PRD} and must be even more numerous here (including singular solutions), but their analysis is significantly more difficult.  Although one might not be able to visualise all interesting solutions working in the whole parameter space, there exist classes of theories that are worth analysing within, as well as beyond, the six-order gravity.

\begin{acknowledgements}
We thank Robert \v{S}varc for fruitful discussions (including the suggestion of using modified Schwarzschild coordinates) and Ji\v{r}í Podolský and Pavel Krtou\v{s} for various comments on our work.
Both authors acknowledge financial support by Primus grant PRIMUS/23/SCI/005 from Charles University and the support from the Charles University Research Center Grant No. UNCE24/SCI/016.
\end{acknowledgements}


\appendix

\section{Linearised six-derivative gravity}
\label{App-Lin}

In this appendix, we briefly review some results concerning the linearised limit of the six-derivative gravity~\eqref{mostgeneralaction}. For details, we refer the interested reader to~\cite{Accioly:2016qeb,BreTibLiv}; further applications and extensions can be found, \textit{e.g.}, in~\cite{QuandtSchmidt,Newton-MNS,Newton-BLG,Frolov:Poly,BreTib1,Nos6der,BreTib2,Accioly:2016etf,BreTibLiv}.
The linearisation of the field equations~\eqref{EoM} around the Minkowski background can be done by writing
\beq
\label{linear}
g_{\mu\nu} = \eta_{\mu\nu} + \kappa h_{\mu\nu},
\eeq
where $\kappa$ is a bookkeeping parameter, and dropping the terms that are $O(\kappa^2)$.
At the level of action, the same result is obtained by substituting~\eqref{linear} into~\eqref{mostgeneralaction} and only keeping the terms quadratic in the metric perturbation. Since all the terms with coefficients $\gamma_{3,\ldots,8}$ are cubic in curvature, their expansion around a flat background is $O(\kappa^3)$; as a consequence, they are not relevant in the linear limit.

Besides the massless graviton, a general six-derivative gravity also propagates two pairs of massive particles of spin 0 and spin 2. Their masses $m_{0\pm}$ and $m_{2\pm}$, respectively, are related to the roots $z = - m_{i\pm}^2$ of the equations
\beq
\label{p0-p2}
\begin{aligned}
&  \alpha - 2 ( 3 \beta_1 + \beta_2 ) z - 2 \left( 3 \gamma_1 + \gamma_2 \right) z^2 = 0
,
\\
& \alpha + \beta_2 z + \gamma_2 z^2  = 0
,
\end{aligned}
\eeq
namely,
\beq
m_{0\pm}^2 = \frac{3 \beta_1 + \beta_2 \pm \sqrt{(3 \beta_1 + \beta_2)^2 + 2 \alpha (3 \gamma_1 + \gamma_2)}}{2 (3 \gamma_1 +\gamma_2)} ,
\qquad 
m_{2\pm}^2 = \frac{ \beta_2 \pm \sqrt{\beta_2^2 - 4 \alpha \gamma_2}}{2 \gamma_2} .
\eeq
There are several scenarios for these ``masses'' depending on the relation between the couplings $\beta_{1,2}$ and $\gamma_{1,2}$. For example, if $\alpha,\beta_2,\gamma_2>0$ and $\beta_2^2 - 4 \alpha \gamma_2>0$, then $m_{2+}^2$ and $m_{2-}^2$ are distinct positive quantities; if $\beta_2^2 - 4 \alpha \gamma_2 < 0$ then $m_{2+}$ and $m_{2-}$ form a complex conjugate pair. In the extreme case $\beta_2^2 - 4 \alpha \gamma_2 = 0$, the two masses degenerate into a double non-tachyonic mode if $\beta_2 / \gamma_2 > 0$ (or into a tachyonic mode if $\beta_2 / \gamma_2 < 0$). The same reasoning can be applied to the scalar sector with similar results~\cite{Accioly:2016qeb}. Although the case of real masses has a more direct physical interpretation, in recent years the possibility of complex-conjugate pairs has attracted some interest as it can lead to a way of combining renormalizability and unitarity in the framework of perturbative quantum gravity~\cite{ModestoShapiro16} (see also, \textit{e.g.},~\cite{Modesto16,AnselmiPiva2,Asorey:2024mkb,Rachwal:2021bgb} for further developments).

It is also important to note that the polynomials in~\eqref{p0-p2} become linear if $3\gamma_1+\gamma_2 \to 0$ and $\gamma_2 \to 0$. In this case, $m_{i+}^2\to \infty$ and we recover linearised four-derivative gravity. If only one of these conditions holds, there is an imbalance of the number of derivatives in the spin-0 and spin-2 sectors of the model. For this reason, throughout this work, we always assume that $\gamma_2(3\gamma_1+\gamma_2) \neq 0$ [see the discussion related to Eq.~\eqref{ga1ga2}].

Using the results of~\citep{Accioly:2016qeb} regarding the decomposition of the metric perturbation into its massless and massive spin-2 and spin-0 modes, it can be shown that the static spherically symmetric general solution of the vacuum field equations of linearised six-derivative gravity in Schwarzschild coordinates is given by
\beq
\label{metric-Lin}
\rd s^2 = - \left[ 1 + 2 \kappa \Phi(\bar{r}) \right]  \rd t^2 + \left[ 1 + 2 \kappa \bar{r} \Psi^\prime(\bar{r}) \right]  \rd \bar{r}^2 + \bar{r}^2 \left( \rd \th^2 + \sin^2 \th \rd \phi^2 \right) ,
\eeq
where the functions $\Phi$ and $\Psi$ are combinations of the Newton and Yukawa-like potentials,\footnote{The case of complex quantities $m_{i\pm}$ are also covered by Eq.~\eqref{PhiPsi} via the complexification of the constants $k_{i\pm}$ and $\tilde{k}_{i\pm}$ such that $k_{i+} =  k_{i-}^*$ and $\tilde{k}_{i+} =  \tilde{k}_{i-}^*$ (the star denotes complex conjugation), to guarantee that the final expression is real~\cite{Newton-BLG}. The number of free parameters is thus preserved. From a phenomenological point of view, complex modes generally cause the functions $\Phi$ and $\Psi$ to oscillate~\cite{Accioly:2016qeb}. 
}
\beq
\label{PhiPsi}
\begin{aligned}
\Phi(\bar{r}) & =  k_0 - \frac{1}{\bar{r}} \left[ k_{1}  +  k_{0+} e^{-m_{0+}\bar{r}} + k_{0-} e^{-m_{0-}\bar{r}} + \tilde{k}_{0+} e^{m_{0+}\bar{r}} + \tilde{k}_{0-} e^{m_{0-}\bar{r}}  - 2 \left( k_{2+} e^{-m_{2+}\bar{r}} + k_{2-} e^{-m_{2-}\bar{r}} + \tilde{k}_{2+} e^{m_{2+}\bar{r}} + \tilde{k}_{2-} e^{m_{2-}\bar{r}} \right) \right] 
\\
\Psi(\bar{r}) & = k_0 - \frac{1}{\bar{r}} \left[ k_{1}  -  \left( k_{0+} e^{-m_{0+}\bar{r}} + k_{0-} e^{-m_{0-}\bar{r}} + \tilde{k}_{0+} e^{m_{0+}\bar{r}} + \tilde{k}_{0-} e^{m_{0-}\bar{r}} \right) -  \left( k_{2+} e^{-m_{2+}\bar{r}} + k_{2-} e^{-m_{2-}\bar{r}} + \tilde{k}_{2+} e^{m_{2+}\bar{r}} + \tilde{k}_{2-} e^{m_{2-}\bar{r}} \right) \right] .
\end{aligned}
\eeq
This solution is characterised by ten free parameters $k_{0}$, $k_{1}$, $k_{0\pm}$, $\tilde{k}_{0\pm}$, $k_{2\pm}$ and $\tilde{k}_{2\pm}$, with $k_0$ related to the freedom to re-scale the time coordinate. Comparing the two functions in~\eqref{PhiPsi}, the graviton (massless) contribution is the same in both of them, the contributions of the scalars have opposite signs, and the massive spin-2 contributions are related by a factor 2. 
Note that the degenerate case $m_{i-}^2 = m_{i+}^2$ causes a reduction in the total number of free parameters. A similar situation occurs if $\beta_2=-2\beta_1$ and $\gamma_2 = -2 \gamma_1$, as this gives $m_{0\pm}^2=m_{2\pm}^2$. Nevertheless, such a reduction of free parameters is a feature of the linearised model and might not manifest in the solutions of the full non-linear theory; indeed, none of the exact solutions we found in this paper is changed by these particular choices for the couplings $\beta_{1,2}$ and $\gamma_{1,2}$. For this reason, in what follows we shall assume that $m_{i-}^2 \neq m_{i+}^2$ and $m_{0\pm}^2 \neq m_{2\pm}^2$.

\subsection{Field generated by a point-like mass in rest}

The requirement of asymptotic flatness of the solution~\eqref{metric-Lin} fixes $\tilde{k}_{0\pm}=\tilde{k}_{2\pm}=0$. If, in addition, a point-like mass $M$ in the form of a Dirac delta sitting at $\bar{r}=0$ is introduced as a source, more parameters are fixed~\cite{Newton-MNS,Accioly:2016qeb}:
\beq
k_1 = \frac{M}{16 \pi \gamma} , \qquad
k_{0\pm} = \frac{k_1}{3} \frac{m_{0\mp}^2}{m_{0\mp}^2 - m_{0\pm}^2} , \qquad 
k_{2\pm} = \frac{2 k_1 }{3} \frac{m_{2\mp}^2}{m_{2\mp}^2 - m_{2\pm}^2} .
\eeq
Hence, the solution
\beq
\begin{aligned}
\label{PhiPsi-delta}
\Phi_\delta(\bar{r}) & =   k_0 - \frac{M}{16 \pi \gamma \bar{r}}  \left( 1  + \frac{1}{3}   \frac{m_{0+}^2e^{-m_{0-}\bar{r}} -m_{0-}^2e^{-m_{0+}\bar{r}}}{m_{0+}^2 - m_{0-}^2}    - \frac{4}{3} \frac{m_{2+}^2e^{-m_{2-}\bar{r}} -m_{2-}^2e^{-m_{2+}\bar{r}} }{m_{2+}^2 - m_{2-}^2}  \right) ,
\\
\Psi_\delta(\bar{r}) & =  k_0 - \frac{M}{16 \pi \gamma \bar{r}}  \left( 1  - \frac{1}{3}   \frac{m_{0+}^2e^{-m_{0-}\bar{r}} -m_{0-}^2e^{-m_{0+}\bar{r}}}{m_{0+}^2 - m_{0-}^2}    - \frac{2}{3} \frac{m_{2+}^2e^{-m_{2-}\bar{r}} -m_{2-}^2e^{-m_{2+}
\bar{r}} }{m_{2+}^2 - m_{2-}^2}  \right)  
\end{aligned}
\eeq
becomes physically characterised by the sole parameter $M$ (recall that one can always set $k_0=0$). This solution corresponds to the weak field generated by a point particle in rest and has a natural interpretation as the higher-derivative corrections to the Schwarzschild geometry in the regime of large $\bar{r}$. In particular, $\Phi_\delta$ is the modified Newton potential in six-derivative gravity.

Extrapolating~\eqref{PhiPsi-delta} to the regime of small $\bar{r}$, it has been observed that the solution is bounded at $\bar{r}=0$~\cite{Newton-MNS,Newton-BLG} (like the analogous solution in four-derivative gravity~\cite{Stelle77}). Moreover --- this time unlike the four-derivative gravity analogue ---~\eqref{PhiPsi-delta} does not yield curvature singularities in the sense that scalars formed by contractions of any number of Riemann and metric tensors (evaluated at leading order $\kappa^2$) are bounded~\cite{BreTib1,Nos6der}. This happens because the expansion of the metric~\eqref{metric-Lin} has a de~Sitter core,
\beq
\begin{aligned}
A_\delta(\bar{r}) & \equiv 1 + 2 \kappa \bar{r} \Psi_\delta^\prime(\bar{r}) =  1 + \bar{a}_2 \bar{r}^2 + \bar{a}_3 \bar{r}^3 + O(\bar{r}^4) ,
\\
B_\delta(\bar{r}) & \equiv  1 + 2 \kappa \Phi_\delta(\bar{r}) = \bar{b}_0 + \bar{b}_2 \bar{r}^2 + \bar{b}_3 \bar{r}^3 + O(\bar{r}^4) ,
\end{aligned}
\eeq
where the coefficients $\bar{a}_{2,3,\ldots}$ and $\bar{b}_{0,2,\ldots}$ depend on $M$ and the parameters of the model. Also, $\bar{a}_{3},\bar{b}_{3} \neq 0$, which implies a mild singularity (at $\bar{r}= 0$), which shows up in scalar invariants with derivatives of curvature, such as $R\Box R$~\cite{Nos6der}.

\subsection{Global vacuum solution}

Although~\eqref{PhiPsi} is the general vacuum solution of the linearised field equations, 
it may not correspond to a global vacuum, as 
it can actually be sourced by Dirac deltas (and its derivatives) sitting at $\bar{r}=0$. 
Extending the reasoning of~\cite{Stelle15PRD,Perkins:2016imn} to six-derivative gravity, one concludes that the coefficients must satisfy
\beq
\label{globvacsol}
k_1 = 0 , \qquad k_{0\pm} = - \tilde{k}_{0\pm}, \qquad k_{2\pm} = - \tilde{k}_{2\pm} 
\eeq
in order to guarantee that such sources are not present.
These conditions also make the functions in~\eqref{PhiPsi} to be regular and even in $\bar{r}$, and the metric to be smooth at $\bar{r}=0$. We shall refer to the solution with~\eqref{globvacsol}  as the global vacuum solution, namely,
\beq
\label{A0B0linSC}
\begin{aligned}
A_0(\bar{r}) & \equiv 1 + 2 \kappa \bar{r} \Psi_0^\prime(\bar{r}) = 1 + 4 \kappa \sum_{i=0,2} \left[ \tilde{k}_{i+} \left(  m_{i+} \cosh (m_{i+} \bar{r}) -\frac{\sinh (m_{i+} \bar{r})}{\bar{r}} \right)  + \tilde{k}_{i-} \left(  m_{i-} \cosh (m_{i-} \bar{r}) -\frac{\sinh (m_{i-} \bar{r})}{\bar{r}} \right) \right] ,
\\
B_0(\bar{r}) & \equiv  1 + 2 \kappa \Phi_0(\bar{r}) = 1 + 2 \kappa \left[ k_0 - 2  \tilde{k}_{0+} \frac{ \sinh (m_{0+} \bar{r}) }{\bar{r}} - 2  \tilde{k}_{0-} \frac{ \sinh (m_{0-} \bar{r}) }{\bar{r}}  + 4 \tilde{k}_{2+} \frac{\sinh (m_{2+} \bar{r}) }{\bar{r}} + 4 \tilde{k}_{2-} \frac{\sinh (m_{2-} \bar{r}) }{\bar{r}}   \right] ,
\end{aligned}
\eeq
which has a total of five free parameters. Expanding around $\bar{r}=0$ it follows
\beq
\label{A0B0lin}
\begin{aligned}
A_0(\bar{r}) & = 1 + \bar{a}_2 \bar{r}^2 + \bar{a}_4 \bar{r}^4 + O(\bar{r}^6) ,
\\
B_0(\bar{r}) & = \bar{b}_0 + \bar{b}_2 \bar{r}^2 + \bar{b}_4 \bar{r}^4 + O(\bar{r}^6)  ,
\end{aligned}
\eeq
where
\beq
\label{systemlin}
\begin{aligned}
\bar{a}_2 & = \frac{4\kappa }{3}  \, \big( \tilde{k}_{0+} m_{0+}^3 + \tilde{k}_{0-} m_{0-}^3 + \tilde{k}_{2+} m_{2+}^3 + \tilde{k}_{2-} m_{2-}^3 \big) ,
\\
\bar{a}_4 & = \frac{2\kappa }{15} \, \big( \tilde{k}_{0+} m_{0+}^5 + \tilde{k}_{0-} m_{0-}^5 + \tilde{k}_{2+} m_{2+}^5 + \tilde{k}_{2-} m_{2-}^5 \big) ,
\\
\bar{b}_0 & = 1 + 2 \kappa \, \big[ k_0 - 2 \big( \tilde{k}_{0+} m_{0+} + \tilde{k}_{0-} m_{0-} \big)  + 4 \big(  \tilde{k}_{2+} m_{2+} + \tilde{k}_{2-} m_{2-} \big)\big]   ,
\\
\bar{b}_2 & =  - \frac{3\kappa}{2}  \, \big[ \tilde{k}_{0+} m_{0+}^3 + \tilde{k}_{0-} m_{0-}^3 - 2 \big(\tilde{k}_{2+} m_{2+}^3 + \tilde{k}_{2-} m_{2-}^3\big) \big]   ,
\\
\bar{b}_4 & =  - \frac{\kappa}{30} \,  \big[ \tilde{k}_{0+} m_{0+}^5 + \tilde{k}_{0-} m_{0-}^5 - 2 \big(\tilde{k}_{2+} m_{2+}^5 + \tilde{k}_{2-} m_{2-}^5\big) \big]  .
\end{aligned}
\eeq
The linear system~\eqref{systemlin} can be solved for $k_0$ and $\tilde{k}_{i\pm}$, allowing us to treat $\bar{a}_{2,4}$ and $\bar{b}_{0,2,4}$ as free parameters of the solution.

\section{System of indicial equations and its solutions}
\label{App}

Here we consider the explicit form for the system of indicial equations and its solutions. We shall divide the analysis into the nine cases described in Table~\ref{Tab0}. 
Whenever $\sigma \neq 0$, we shall only discuss the components $tt$ and $rr$ of the field equations. There is no loss in generality because, if the leading term of the field equations is of order $p(\sigma,\tau)$ [defined in Eq.~\eqref{pfeld}], then the Bianchi identities~\eqref{bia} yield
\beq \label{bia-app}
4\sigma \, \mathcal{E}^\theta{}_\theta \, \overset{\text{LO}}{=} \, - \tau \, \mathcal{E}^t{}_t  +  ( 2 p(\sigma,\tau) + 4\sigma + \tau ) \, \mathcal{E}^r{}_r    ,
\eeq
where the equality is valid \emph{at the leading order} (LO), \textit{i.e.}, ignoring the terms of order higher than $\De^{p(\sigma,\tau)}$ (or, in the case of asymptotic solutions, lower than $r^{p(\sigma,\tau)}$).
Hence, if the components $tt$ and $rr$ of the field equations are solved at the leading order, the remaining component will be automatically solved as well.
Only in the cases of $\sigma = 0$ shall we discuss the component $\theta\theta$.

In the following analysis, we use the freedom to shift the coordinate $r$ to fix $r_0=0$ in the cases of the expansions~\eqref{FrobeniusHF}. This is solely motivated by the simplification of notation, as in this way the leading term of the field equations is written in terms of the coordinate $r$ for both the expansions around $r=r_0=0$ and the asymptotic ones. Nonetheless, to recover the expressions for expansions around $r=r_0 \neq 0$, it suffices to replace $r \mapsto \De=r-r_0$ in the expansion of the field equations.

\subsection{Cases I, II and III}

In these first three cases, the leading term in the field equations comes from the Einstein--Hilbert term in the action.
In Case I, we have
\beq 
\label{IndCaseI}
\mathcal{E}^t{}_t = \frac{\alpha }{f_0^2} r^{-2 \sigma }  + \ldots  ,
\qquad 
\mathcal{E}^r{}_r = \frac{\alpha }{f_0^2} r^{-2 \sigma }  + \ldots ,
\eeq
from which it is clear that the field equations cannot be solved at this order for any $\alpha \neq 0$.

In Case III we have 
\beq 
\label{IndCaseIII}
\mathcal{E}^t{}_t = - \alpha  h_0 \sigma   (3 \sigma +\tau -2) r^{\tau -2}  + \ldots  ,
\qquad 
\mathcal{E}^r{}_r = - \alpha  h_0 \sigma  (\sigma +\tau ) r^{\tau -2}  + \ldots ,
\eeq
therefore, to solve the field equations at this order we must have either $\sigma =0$ or $\sigma = - \tau = 1$. These values, however, are outside the validity domain of Case III, which is $\sigma < 0$ and $2<\tau <2-2 \sigma$  or $\sigma > 0$ and $2-2 \sigma <\tau <2$ (see Table~\ref{Tab0}).

The leading term of the expansion of the field equations in the Case II is obtained by combining Eqs.~\eqref{IndCaseI} and~\eqref{IndCaseIII}, for both types of terms have the same order when $\tau = 2 - 2 \sigma$. The result is
\beq 
\label{IndCaseII}
\mathcal{E}^t{}_t = \frac{\alpha}{f_0^2} (1 -   \sigma^2 z ) r^{-2 \sigma } + \ldots  ,
\qquad 
\mathcal{E}^r{}_r = \frac{\alpha}{f_0^2} [1 + (\sigma -2) \sigma  z ] r^{-2 \sigma }  + \ldots ,
\eeq
where we defined $z \equiv f_0^2 h_0$. 
The only solution for the field equations at this order is 
\beq
\sigma = z = 1, \quad \text{ which implies } \quad \tau = 0. 
\eeq
This solution is within the domain of validity of Case II \emph{for asymptotic expansions}, being therefore a legit solution and an indication of the existence of the solution class $\lbrace 1,0 \rbrace^\infty$ (see discussion in Sec.~\ref{Sec-10Assimpt}).

\subsection{Case IV}

In this case, we have $\sigma \neq 0$ and $\tau = 2$, and the field equations yield 
\beq 
\label{IndCaseIV}
\mathcal{E}^t{}_t = \left[ a_1 + b_1 h_0 + c_1 h_0^2  \right] h_0  + \ldots  = 0,
\qquad 
\mathcal{E}^r{}_r = \left[ a_2 + b_2 h_0 + c_2 h_0^2  \right] h_0  + \ldots = 0 ,
\eeq
where the quantities $a_{1,2}$, $b_{1,2}$ and $c_{1,2}$ depend on $\sigma$ and 
receive contributions from the two-, four- and six-derivative structures in the action, in this order,
{\allowdisplaybreaks 
\begin{subequations}\label{abc12-CaseIV}
\begin{eqnarray}
a_1 & = & -3 \alpha  \sigma ^2  ,
\\
a_2 & = & -\alpha  \sigma  (\sigma +2)  ,
\\
b_1 & = &  (\sigma -1) (3 \sigma +1) \big[2 \sigma ^2 (3 \beta_1 + \beta_2 )+4 \beta_1 \sigma +2 \beta_1 + \beta_2 \big] ,
\\
b_2 & = & -(\sigma -1)^2 \big[(2 \sigma ^2 (3  \beta_1 + \beta_2 )+4  \beta_1  \sigma +2  \beta_1 + \beta_2 \big]  ,
\\
c_1 & = &  
2 [4  \ga_3 +2  \ga_4 + \ga_5 + \ga_6 +4 ( \ga_7 + \ga_8 )]
+2 \sigma  (24  \ga_3 +8  \ga_4 +3  \ga_5 +2  \ga_6 +8  \ga_7 )
\nonumber
\\
&&
+\sigma ^2 [14  \ga_2 +132  \ga_3 +30  \ga_4 +9  \ga_5 -2 ( \ga_6 +2  \ga_7 +24  \ga_8 )]
+2 \sigma ^3 [6  \ga_2 +104  \ga_3 +22  \ga_4 +5 ( \ga_5 + \ga_6 )+8  \ga_8 ]
\nonumber
\\
&&
-\sigma ^4 (66  \ga_2 -144  \ga_3 -26  \ga_4 +12  \ga_5 -29  \ga_6 +8  \ga_7 -48  \ga_8 )
+4 \sigma ^5 (10  \ga_2 +6  \ga_4 +6  \ga_5 - \ga_6 +24  \ga_7 )
\nonumber
\\
&&
-12 \sigma ^6 (9  \ga_3 +3  \ga_4 + \ga_5 + \ga_6 +3  \ga_7 + \ga_8 ) ,
\\
c_2 & = &  
2 [4  \ga_3 +2  \ga_4 + \ga_5 + \ga_6 +4 ( \ga_7 + \ga_8 )]
+2 \sigma  [12  \ga_3 +2  \ga_4 - \ga_6 -4 ( \ga_7 +3  \ga_8 )]
\nonumber
\\
&&
+\sigma ^2 (34  \ga_2 +60  \ga_3 +18  \ga_4 +15  \ga_5 -6  \ga_6 +20  \ga_7 )
+2 \sigma ^3 (-30  \ga_2 +32  \ga_3 +4  \ga_4 -7  \ga_5 +11  \ga_6 -8  \ga_7 +20  \ga_8 )
\nonumber
\\
&&
+\sigma ^4 (18  \ga_2 +96  \ga_3 +38  \ga_4 +12  \ga_5 +11  \ga_6 +72  \ga_7 )
+4 \sigma ^5 (2  \ga_2 +18  \ga_3 -3  \ga_6 -10  \ga_7 -6  \ga_8 )
\nonumber
\\
&&
+12 \sigma ^6 (9  \ga_3 +3  \ga_4 + \ga_5 + \ga_6 +3  \ga_7 + \ga_8 ) .
\end{eqnarray}
\end{subequations}}

At leading order, each of the equations in~\eqref{IndCaseIV} can be formally solved for $h_0$. The requirement that there exists a common solution for both quadratic equations translates into the relation
\beq
\label{2sides}
(a_2 b_1 - a_1 b_2) (b_2 c_1 - b_1 c_2) = (a_1 c_2 - a_2 c_1)^2 .
\eeq
While the left-hand side of this equation is proportional to $\alpha$, the right-hand side is proportional to $\alpha^2$.  Therefore, for the field equations to be solved at leading order for a coupling-independent $\sigma$ and arbitrary values of $\alpha$, $\beta_{1,2}$ and $\gamma_{1,\ldots,8}$, each side of~\eqref{2sides} must vanish independently:
\bea
\left\{ 
\begin{array}{l l}
2 \alpha  \sigma  (3 \sigma ^2+2 \sigma +1) (\sigma -1)^2  \big[2 \sigma ^2 (3  \beta_1 + \beta_2 )+4  \beta_1  \sigma +2  \beta_1 + \beta_2 \big]^2 [(\sigma - 1) c_1 +  (3\sigma + 1) c_2 ]  =  0 , 
\\
\alpha ^2 \sigma ^2 [3 \sigma c_2 - (\sigma + 2) c_1]^2 =  0 . 
\\
\end{array} \right .
\eea
It can be verified that the only solutions of the system, independent of the values of $\gamma_{1,\ldots,8}$, are $\sigma = 0$ (which lies outside the domain of this Case IV) and $\sigma = 1$, which yields [see Eq.~\eqref{abc12-CaseIV}]
\beq
a_1=a_2=-3\alpha , \qquad b_1 = b_2 = 0 , \qquad c_1 = c_2 = 3 (144 \gamma_{3} +36  \gamma_{4} +9  \gamma_{5} +9  \gamma_{6} +24  \gamma_{7} +4  \gamma_{8}).
\eeq

We conclude that Case IV only has a solution if the parameters of the model satisfy the constraint
\beq
\frac{144 \gamma_{3} +36  \gamma_{4} +9  \gamma_{5} +9  \gamma_{6} +24  \gamma_{7} +4  \gamma_{8} }{\alpha} > 0,
\eeq
in which case we have
\beq
\label{CaseIVsol}
\sigma = 1 , \qquad \tau = 2 , \qquad h_0 = \pm \sqrt{\frac{\alpha}{144 \gamma_{3} +36  \gamma_{4} +9  \gamma_{5} +9  \gamma_{6} +24  \gamma_{7} +4  \gamma_{8} }} .
\eeq
This indicial structure indicates the existence of a solution class $\lbrace 1,2 \rbrace^\infty$ (see discussion in Sec.~\ref{Sec-12Assimpt}).

\subsection{Case V}

In this case, the expansion of the field equations at lowest order only depends on the parameters of the six-derivative terms in the action, and it has the form
\begin{subequations}
\begin{eqnarray}
\mathcal{E}^t{}_t & = & \frac{h_0^3 }{4 \, r^{6 - 3\tau}} \sum_{i=1}^8 \ga_i \, g_i^{(t)}(\si,\ta)   + \ldots  ,
\\
\mathcal{E}^r{}_r & = & \frac{h_0^3}{4 \, r^{6 - 3\tau}} \sum_{i=1}^8 \ga_i \, g_i^{(r)}(\si,\ta)   + \ldots ,
\\
\mathcal{E}^\theta{}_\theta & = & \frac{h_0^3}{4 \, r^{6 - 3\tau}} \sum_{i=1}^8 \ga_i \, g_i^{(\theta)}(\si,\ta)   + \ldots ,
\end{eqnarray}
\end{subequations}
where the coefficients $g_i^{(t)}(\si,\ta)$, $g_i^{(r)}(\si,\ta)$ and $g_i^{(\theta)}(\si,\ta)$ depend only on the quantities $\si$ and $\tau$. Moreover, these three sets of coefficients are related by the Bianchi identities [see Eq.~ \eqref{bia-app}], namely,
\beq
\label{ththC1}
4 \sigma \, g_i^{(\theta)}(\si,\ta) = - \tau \, g_i^{(t)}(\si,\ta) + (4 \sigma +7 \tau -12) \, g_i^{(r)}(\si,\ta) .
\eeq
As a consequence, if $\sigma \neq 0$, the coefficient $g_i^{(\theta)}(\si,\ta)$ can be expressed as a linear combination of $g_i^{(t)}(\si,\ta)$ and $g_i^{(r)}(\si,\ta)$. Only if $\sigma = 0$, the coefficients $g_i^{(\theta)}(\si,\ta)$ are independent of the others. For this particular case, we have
{\allowdisplaybreaks 
\begin{subequations}\label{Os_G_thetha_ffgghh><}
\begin{eqnarray}
g_1^{(\theta)}(0,\ta) & = & 2 \tau (\tau -2)^2 (\tau -1) \big( 47 \tau ^2-151 \tau +120 \big ) ,
\\
g_2^{(\theta)}(0,\ta) & = & \tau (\tau -2)^2 (\tau -1) \big(  23 \tau ^2-75 \tau +60 \big ),
\\
g_3^{(\theta)}(0,\ta) & = & - 2 \tau ^2 (\tau -1)^2 \big( 37 \tau ^2 - 133 \tau + 120  \big )  ,
\label{Ag3theta}
\\
g_4^{(\theta)}(0,\ta) & = & - \tau ^2 (\tau -1)^2 \big(  25 \tau ^2-89 \tau +80 \big ) ,
\\
g_5^{(\theta)}(0,\ta) & = &  -\frac{\tau ^2}{2}  (\tau -1)^2  \big( 19 \tau ^2-67 \tau +60  \big ) ,
\\
g_6^{(\theta)}(0,\ta) & = &  -\frac{\tau ^2}{2} (\tau -1)^2   \big( 13 \tau ^2-45 \tau +40  \big ) ,
\\
g_7^{(\theta)}(0,\ta) & = &  -2 \tau ^2 (\tau -1)^2   \big(  13 \tau ^2-45 \tau +40 \big ) ,
\\
g_8^{(\theta)}(0,\ta) & = &  -2 \tau ^3 (\tau -1)^3 ,
\end{eqnarray}
\end{subequations}
}
whereas the general expressions for the coefficients $g_i^{(t)}(\si,\ta)$ and $g_i^{(r)}(\si,\ta)$ read
{\allowdisplaybreaks 
\begin{small}
\begin{subequations}\label{Os_G_ffgghh><}
\begin{eqnarray}
g_1^{(t)}(\si,\ta) & = & 2 (\tau -2) \big[ 6 \sigma^2 + (4\si + \ta) (\tau -1)  \big] \big[ \tau  (18 \sigma ^2+56 \sigma  \tau +35 \tau ^2) -52 \sigma ^2-244 \sigma  \tau -207 \tau ^2 +8  (31 \sigma +49 \tau - 30) \big]   ,
\\
g_1^{(r)}(\si,\ta) & = & -2 (\tau -2) \big[ 6 \sigma^2 + (4\si + \ta) (\tau -1)  \big] \big[ 16 \sigma ^3-14 \sigma ^2 \tau -32 \sigma  \tau ^2-5 \tau ^3 +3 \big(12 \sigma ^2+36 \sigma  \tau +7 \tau ^2\big)-20 (4 \sigma +\tau ) \big]  ,
\\
g_2^{(t)}(\si,\ta) & = & 
\tau  \big(32 \sigma ^5+4 \sigma ^4 \tau +40 \sigma ^3 \tau ^2+206 \sigma ^2 \tau ^3+152 \sigma  \tau ^4+35 \tau ^5\big)
+4 \big(24 \sigma ^5-16 \sigma ^4 \tau -174 \sigma ^3 \tau ^2-379 \sigma ^2 \tau ^3 -299 \sigma  \tau ^4 
\nonumber
\\
&&
-78 \tau ^5\big)
-152 \sigma ^4+1800 \sigma ^3 \tau +4222 \sigma ^2 \tau ^2+3620 \sigma  \tau ^3+1083 \tau ^4 
-2 \big(544 \sigma ^3+2544 \sigma ^2 \tau +2624 \sigma  \tau ^2+915 \tau ^3\big)
\nonumber
\\
&&
+16 \big(136 \sigma ^2+227 \sigma  \tau +94 \tau ^2\big)
-480 (2 \sigma +\tau )
,
\\
g_2^{(r)}(\si,\ta) & = & 
-\tau  \big(64 \sigma ^5-28 \sigma ^4 \tau -104 \sigma ^3 \tau ^2-90 \sigma ^2 \tau ^3-32 \sigma  \tau ^4-5 \tau ^5\big)
+4 (2 \sigma +3 \tau ) \big(20 \sigma ^4-44 \sigma ^3 \tau -37 \sigma ^2 \tau ^2-15 \sigma  \tau ^3-3 \tau ^4\big)
\nonumber
\\
&&
+184 \sigma ^4+1560 \sigma ^3 \tau +1370 \sigma ^2 \tau ^2+468 \sigma  \tau ^3+93 \tau ^4
-2 \big(448 \sigma ^3+688 \sigma ^2 \tau +228 \sigma  \tau ^2+51 \tau ^3\big)
+40 \big(12 \sigma ^2+4 \sigma  \tau +\tau ^2\big)  ,
\\
g_3^{(t)}(\si,\ta) & = & - 4 \big[ 6 \sigma ^2+ (4 \sigma + \tau)  (\tau -1) \big]^2 \big[ 3 \sigma ^2+11 \sigma  \tau +14 \tau ^2 -26 \sigma -59 \tau +60 \big] ,
\label{Ag3t}
\\
g_3^{(r)}(\si,\ta) & = & 4 \big[ 6 \sigma ^2+ (4 \sigma + \tau)  (\tau -1) \big]^2 \big[ 3 \sigma ^2-11 \sigma  \tau -2 \tau ^2 +5 (4 \sigma +\tau ) \big]  ,
\label{Ag3r}
\\
g_4^{(t)}(\si,\ta) & = & - 2 \big[
72 \sigma ^6+264 \sigma ^5 \tau +528 \sigma ^4 \tau ^2+520 \sigma ^3 \tau ^3+291 \sigma ^2 \tau ^4+91 \sigma  \tau ^5+14 \tau ^6
-576 \sigma ^5-1976 \sigma ^4 \tau -2428 \sigma ^3 \tau ^2 
\nonumber
\\
&&
-1542 \sigma ^2 \tau ^3
-532 \sigma  \tau ^4-87 \tau ^5
+1788 \sigma ^4+3616 \sigma ^3 \tau +2819 \sigma ^2 \tau ^2+1111 \sigma  \tau ^3+192 \tau ^4
-1768 \sigma ^3-2108 \sigma ^2 \tau 
\nonumber
\\
&&
-990 \sigma  \tau ^2-179 \tau ^3
+20 \big(28 \sigma ^2+16 \sigma  \tau +3 \tau ^2\big)
\big] ,
\\
g_4^{(r)}(\si,\ta) & = & 2 \big[
72 \sigma ^6-200 \sigma ^5 \tau -304 \sigma ^4 \tau ^2-200 \sigma ^3 \tau ^3-77 \sigma ^2 \tau ^4-19 \sigma  \tau ^5-2 \tau ^6
+(2 \sigma +3 \tau ) \big(200 \sigma ^4+184 \sigma ^3 \tau +86 \sigma ^2 \tau ^2
\nonumber
\\
&&
+24 \sigma  \tau ^3
+3 \tau ^4\big)
-644 \sigma ^4-760 \sigma ^3 \tau -365 \sigma ^2 \tau ^2-99 \sigma  \tau ^3-12 \tau ^4
+5 (4 \sigma +\tau ) \big(12 \sigma ^2+4 \sigma  \tau +\tau ^2\big)
\big]  ,
\\
g_5^{(t)}(\si,\ta) & = &
-48 \sigma ^6-144 \sigma ^5 \tau -276 \sigma ^4 \tau ^2-296 \sigma ^3 \tau ^3-195 \sigma ^2 \tau ^4-75 \sigma  \tau ^5-14 \tau ^6
+3 \big(128 \sigma ^5+384 \sigma ^4 \tau +460 \sigma ^3 \tau ^2+334 \sigma ^2 \tau ^3
\nonumber
\\
&&
+142 \sigma  \tau ^4+29 \tau ^5\big)
-3 \big(416 \sigma ^4+744 \sigma ^3 \tau +601 \sigma ^2 \tau ^2+289 \sigma  \tau ^3+64 \tau ^4\big)
+1352 \sigma ^3+1416 \sigma ^2 \tau +756 \sigma  \tau ^2+179 \tau ^3
\nonumber
\\
&&
-60 \big(8 \sigma ^2+4 \sigma  \tau +\tau ^2\big)
,
\\
g_5^{(r)}(\si,\ta) & = &  
48 \sigma ^6-144 \sigma ^5 \tau -204 \sigma ^4 \tau ^2-128 \sigma ^3 \tau ^3-45 \sigma ^2 \tau ^4-15 \sigma  \tau ^5-2 \tau ^6
+3 (2 \sigma +3 \tau ) \big(48 \sigma ^4+40 \sigma ^3 \tau +18 \sigma ^2 \tau ^2
\nonumber
\\
&&
+6 \sigma  \tau ^3+\tau ^4\big)
-3 \big(160 \sigma ^4+184 \sigma ^3 \tau +87 \sigma ^2 \tau ^2+25 \sigma  \tau ^3+4 \tau ^4\big)
+5 \big(40 \sigma ^3+24 \sigma ^2 \tau +6 \sigma  \tau ^2+\tau ^3\big)
,
\\
g_6^{(t)}(\si,\ta) & = & 
-48 \sigma ^6-128 \sigma ^5 \tau -228 \sigma ^4 \tau ^2-224 \sigma ^3 \tau ^3-142 \sigma ^2 \tau ^4-59 \sigma  \tau ^5-14 \tau ^6
+240 \sigma ^5+824 \sigma ^4 \tau +1064 \sigma ^3 \tau ^2+716 \sigma ^2 \tau ^3
\nonumber
\\
&&
+320 \sigma  \tau ^4+87 \tau ^5
-620 \sigma ^4-1492 \sigma ^3 \tau -1258 \sigma ^2 \tau ^2-623 \sigma  \tau ^3-192 \tau ^4
+560 \sigma ^3+864 \sigma ^2 \tau +522 \sigma  \tau ^2+179 \tau ^3
\nonumber
\\
&&
-20 \big(8 \sigma ^2+8 \sigma  \tau +3 \tau ^2\big) ,
\\
g_6^{(r)}(\si,\ta) & = &  
48 \sigma ^6-160 \sigma ^5 \tau -172 \sigma ^4 \tau ^2-80 \sigma ^3 \tau ^3-38 \sigma ^2 \tau ^4-11 \sigma  \tau ^5-2 \tau ^6
+(2 \sigma +3 \tau ) \big(136 \sigma ^4+88 \sigma ^3 \tau +36 \sigma ^2 \tau ^2
\nonumber
\\
&&
+12 \sigma  \tau ^3+3 \tau ^4\big)
-436 \sigma ^4-388 \sigma ^3 \tau -138 \sigma ^2 \tau ^2-51 \sigma  \tau ^3-12 \tau ^4
+5 (4 \sigma +\tau ) \big(8 \sigma ^2+\tau ^2\big)
 ,
\\
g_7^{(t)}(\si,\ta) & = & -4 \big[
36 \sigma ^6+36 \sigma ^5 \tau +56 \sigma ^4 \tau ^2+144 \sigma ^3 \tau ^3+145 \sigma ^2 \tau ^4+59 \sigma  \tau ^5+14 \tau ^6
-168 \sigma ^5-320 \sigma ^4 \tau -572 \sigma ^3 \tau ^2-686 \sigma ^2 \tau ^3
\nonumber
\\
&&
-320 \sigma  \tau ^4-87 \tau ^5
+424 \sigma ^4+768 \sigma ^3 \tau +1105 \sigma ^2 \tau ^2+623 \sigma  \tau ^3+192 \tau ^4
-400 \sigma ^3-704 \sigma ^2 \tau -522 \sigma  \tau ^2-179 \tau ^3
\nonumber
\\
&&
+20 \big(8 \sigma ^2+8 \sigma  \tau +3 \tau ^2\big)
\big] ,
\label{Ag7t}
\\
g_7^{(r)}(\si,\ta) & = & 4\big[
36 \sigma ^6-132 \sigma ^5 \tau -136 \sigma ^4 \tau ^2-72 \sigma ^3 \tau ^3-23 \sigma ^2 \tau ^4-11 \sigma  \tau ^5-2 \tau ^6
+(2 \sigma +3 \tau ) \big(112 \sigma ^4+80 \sigma ^3 \tau +22 \sigma ^2 \tau ^2
\nonumber
\\
&&
+12 \sigma  \tau ^3+3 \tau ^4\big)
-376 \sigma ^4-368 \sigma ^3 \tau -103 \sigma ^2 \tau ^2-51 \sigma  \tau ^3-12 \tau ^4
+5 (4 \sigma +\tau ) \big(8 \sigma ^2+\tau ^2\big)
\big]  ,
\\
g_8^{(t)}(\si,\ta) & = & -4 \big[
12 \sigma ^6+12 \sigma ^5 \tau -6 \sigma ^4 \tau ^2-16 \sigma ^3 \tau ^3+12 \sigma ^2 \tau ^4+27 \sigma  \tau ^5+14 \tau ^6
-3 \big(8 \sigma ^5+8 \sigma ^4 \tau -8 \sigma ^3 \tau ^2+8 \sigma ^2 \tau ^3+36 \sigma  \tau ^4
\nonumber
\\
&&
+29 \tau ^5\big)
+3 \big(8 \sigma ^4+4 \sigma ^3 \tau +4 \sigma ^2 \tau ^2+45 \sigma  \tau ^3+64 \tau ^4\big)
-8 \sigma ^3-54 \sigma  \tau ^2-179 \tau ^3 +60 \tau ^2
\big] ,
\\
g_8^{(r)}(\si,\ta) & = & 4\big[
12 \sigma ^6-60 \sigma ^5 \tau -54 \sigma ^4 \tau ^2-8 \sigma ^3 \tau ^3-3 \sigma  \tau ^5-2 \tau ^6
+3 (2 \sigma +3 \tau ) \big(16 \sigma ^4+8 \sigma ^3 \tau +\tau ^4\big)
-3 \big(56 \sigma ^4+44 \sigma ^3 \tau 
\nonumber
\\
&&
+\sigma  \tau ^3+4 \tau ^4\big)
+5 \big(16 \sigma ^3+\tau ^3\big)
\big]  .
\end{eqnarray}
\end{subequations}
\end{small}
}
Hence, for the field equations to be solved at lowest order in $r$ for $\sigma$ and $\tau$ independent of $\ga_{1,\ldots,8}$, we must have\footnote{Recall that, for $\sigma \neq 0$, if $g_i^{(t)}(\si,\ta) = g_i^{(r)}(\si,\ta) =  0$ is solved for a given $i$, then the remaining equation $g_i^{(\theta)}(\si,\ta) = 0$ is automatically solved, owing to~\eqref{ththC1}. In the remaining case of $\sigma = 0$, one must solve the three equations in~\eqref{ffgghh><}.}
\beq
\label{ffgghh><}
g_i^{(t)}(\si,\ta) = g_i^{(r)}(\si,\ta) = g_i^{(\theta)}(\si,\ta) =  0, \quad  \forall \, i=1,\ldots,8 ,
\eeq
which constitute the system of indicial equations for Case V. By direct substitution in Eqs.~\eqref{Os_G_thetha_ffgghh><} and~\eqref{Os_G_ffgghh><}, one can verify that~\eqref{ffgghh><} admits the two solutions
\beq
\label{Solffgghh><}
\sigma = \tau = 0 \qquad \text{and} \qquad \sigma = 0, \, \tau = 1 .
\eeq

In order to prove that the system~\eqref{ffgghh><} does not have other solutions, let us first consider the subsystem [see Eqs.~\eqref{Ag3t} and~\eqref{Ag3r}]
\beq
g_3^{(t)}(\si,\ta) = g_3^{(r)}(\si,\ta) =  0 .
\eeq
It is straightforward to verify that its only real solutions 
are either the points $(\sigma,\tau) \in \left\lbrace \left( \tfrac{10}{9},\tfrac{20}{9}\right) , \left( 0,\tfrac{5}{2}\right)\right\rbrace $
or the ellipse
\beq
\label{elli}
6 \sigma^2 + 4 \sigma  (\tau -1)+\tau  (\tau -1) = 0 ,
\eeq 
in which case they have the form
\beq
\label{belli}
\tau = \frac{1}{2} \left( 1 -4 \sigma \pm \sqrt{1 +8 \sigma -8 \sigma ^2} \right) , \qquad \frac{1}{4} \big(2 - \sqrt{6} \big) \leqslant \sigma \leqslant \frac{1}{4} \big(2 + \sqrt{6} \big) .
\eeq
As for the solution $\sigma = \tfrac{10}{9}$ and $\tau = \tfrac{20}{9}$, by direct substitution one can verify that it is not a zero of any of the other functions in~\eqref{Os_G_ffgghh><}; moreover, $\sigma = 0$ and $\tau = \tfrac{5}{2}$ is not a zero of~\eqref{Ag3theta}.
Now, it suffices to show that the zeros of at least one of the functions in~\eqref{Os_G_ffgghh><} only intersect the ellipse~\eqref{elli} at the points in Eq.~\eqref{Solffgghh><}. Although this can be verified numerically, here we offer an analytic proof. Let us focus on the coefficient in Eq.~\eqref{Ag7t}; \textit{i.e.}, consider the system
\beq
\left\{ 
\begin{array}{l l}
6 \sigma^2 + 4 \sigma  (\tau -1)+\tau  (\tau -1)  =  0 , 
\\
g_7^{(t)}(\si,\ta) =  0 . 
\\
\end{array} \right .
\eeq
Applying the first equation to substitute all the occurrences of $\sigma^2$ in the second one, we obtain the equivalent system
\beq
\label{82sys}
\left\{ 
\begin{array}{l l}
6 \sigma^2 + 4 \sigma  (\tau -1)+\tau  (\tau -1)  =  0 , \\
(\tau - 1) (3\tau - 8) \big[ 2 \sigma \big(6 \tau ^3-41 \tau ^2+70 \tau -44\big) -  \tau  \big(33 \tau ^3-95 \tau ^2+58 \tau +22\big) \big] = 0 . \\
\end{array} \right .
\eeq
The last equation admits solutions in the following form:
\beq
\tau = 1 , \qquad \tau = \frac{8}{3} , \qquad \text{or} \qquad \sigma = \frac{\tau  \big(33 \tau ^3-95 \tau ^2+58 \tau +22\big)}{2 \big(6 \tau ^3-41 \tau ^2+70 \tau -44\big)},
\eeq
where it is assumed that $6 \tau ^3-41 \tau ^2+70 \tau -44 \neq 0$ (there is no relevant solution otherwise).
For the first two solutions, the other equation in~\eqref{82sys} yields
\begin{subequations}
\begin{eqnarray}
\tau = 1 \qquad  \Longrightarrow  \qquad \sigma = 0 ,
\\
\tau = \frac{8}{3} \qquad  \Longrightarrow  \qquad \sigma \notin \mathbb{R} ,
\end{eqnarray}
\end{subequations}
while for the third family of solutions, we find
\beq
\label{81tau}
\tau ^2 \big( 17 \tau^2 - 60 \tau + 55 \big) \big(3 \tau ^4-10 \tau ^3+12 \tau ^2-8 \tau +4 \big) = 0 .
\eeq
It is easy to check that the only real solution of~\eqref{81tau} is $\tau = 0$, for which $\sigma = 0$. Therefore, the curve defined by $g_7^{(t)}(\si,\ta) =  0$ and the ellipse~\eqref{elli} do not have intersections other than the points in Eq.~\eqref{Solffgghh><}, which turn out to be the only solutions of the indicial system~\eqref{ffgghh><}.

Taking into account the regime of validity of the Case V (see Table~\ref{Tab0}), we conclude that the indicial structures of~\eqref{Solffgghh><} correspond to potential families of solutions of the types $\lbrace 0,0 \rbrace$ and $\lbrace 0,1 \rbrace$ (see discussion in Secs.~\ref{Sec-00} and~\ref{Sec-01}, respectively).

\subsection{Cases VI and VIII}

Owing to their similarity, let us analyse cases VI and VIII together; afterwards, we shall return to Case VII. 
The expansion of the field equations at the leading order yields
\beq
\label{ffgghh>>1}
\mathcal{E}^t{}_t  =  \frac{g + x_0}{f_0^6 \, r^{6\si}} + \ldots ,
\qquad
\mathcal{E}^r{}_r  =  \frac{g + x_0}{f_0^6 \, r^{6\si}} + \ldots ,
\qquad
\mathcal{E}^\theta{}_\theta  =  - \frac{2g + y_0}{f_0^6 \, r^{6\si}} + \ldots ,
\eeq
where we defined
\beq \label{oG}
g = 4  \ga_3 +2  \ga_4 + \ga_5 + \ga_6 +4 ( \ga_7 + \ga_8 )
\eeq
and
\bea
x_0 & = &
\left\{ 
\begin{array}{l l}
(2 \beta_1 + \beta_2)f_0^2  + \alpha  f_0^4, & \text{if } \, \sigma = 0 \, \text{ (Case VI)}, \\
0, & \text{if } \, \sigma >0 \, \text{ (Case VIII)},
\end{array} \right .
\\
y_0 & = &
\left\{ 
\begin{array}{l l}
(2 \beta_1 + \beta_2)f_0^2 , & \text{if } \, \sigma = 0 \, \text{ (Case VI)}, \\
0, & \text{if } \, \sigma >0 \, \text{ (Case VIII)}.
\end{array} \right .
\eea
It is immediate to see that~\eqref{ffgghh>>1} cannot be solved if the parameters $\ga_{1,\cdots,8}$, $\beta_{1,2}$ and $\alpha$ are completely arbitrary, so no solution can arise from Cases VI and VIII.

\subsection{Case VII}

In this special case, both the parameters $\sigma = 0$ and $\tau = 2$ are fixed, and the leading-order term of the field equations constitutes a system to be solved for the coefficients $f_0$ and $h_0$,
\begin{subequations}
\begin{eqnarray}
\mathcal{E}^t{}_t & \overset{\text{LO}}{=} & \mathcal{E}^r{}_r \, = \, \frac{1}{f_0^6} \left[ 
g 
+ (2 \beta_1 + \beta_2)f_0^2
+ \alpha  f_0^4 
- (12 \ga_3 +2 \ga_4 +4 \ga_7 ) f_0^4 h_0^2 - (2 \beta_1 + \beta_2) f_0^6 h_0^2 
+2 g f_0^6 h_0^3 
\right]    + \ldots  = 0,
\\
\mathcal{E}^\theta{}_\theta & = & - \frac{1}{f_0^6} \left[ 
2g
+ (2 \beta_1 + \beta_2)f_0^2
- 2 (6 \ga_3 + \ga_4 +2 \ga_7 ) f_0^2 h_0 
+ \alpha  f_0^6 h_0
- (2 \beta_1 + \beta_2)f_0^6 h_0^2
+ g f_0^6 h_0^3 
\right] + \ldots  = 0,
\label{CaseVIIb}
\end{eqnarray}
\end{subequations}
where $g$ is given by Eq.~\eqref{oG}, and the equality between $\mathcal{E}^t{}_t$ and $\mathcal{E}^r{}_r$ is valid only at the leading order. If we multiply the second equation by $f_0^2 h_0$ and add it to the first equation, we obtain
\beq 
\left( f_0^4 h_0^2-1\right) \left[ g + \big(2 \beta_1 + \beta_2 + \alpha  f_0^2 \big) f_0^2  - \big[ 2g + (2 \beta_1 + \beta_2 ) f_0^2 \big] f_0^2 h_0 + g f_0^4 h_0^2 \right] = 0 ,
\eeq
which can be formally solved for $h_0$,
\beq
h_0 = \pm \frac{1}{f_0^2}  \qquad \text{ or } \qquad
h_0 = \frac{1}{f_0^2} + \frac{2 \beta_1 + \beta_2 \pm \sqrt{(2 \beta_1 + \beta_2)^2 - 4 \alpha  g}}{2 g} .
\eeq
By substituting each of these four solutions into~\eqref{CaseVIIb}, we obtain  quadratic equations for $f_0^2>0$, whose formal solutions read
\begin{subequations}\label{SolCaseVII}
\begin{eqnarray} 
h_0 = - \frac{1}{f_0^2} \qquad & \Longrightarrow & \qquad f_0^2 = \sqrt{\eta_1},
\\
h_0 = \frac{1}{f_0^2} \qquad & \Longrightarrow & \qquad f_0^2 = \sqrt{\eta_2},
\\
h_0 = \frac{1}{f_0^2} + \frac{2 }{\eta_3 + \eta_4} \qquad & \Longrightarrow & \qquad f_0^2 = - \frac{1}{2 \eta_4} \left[ \eta_2 \pm \sqrt{  \eta_2 \left(\eta_2   -2 \eta_4^2 - 2 \eta_3 \eta_4 \right) } \right],
\\
h_0 = \frac{1}{f_0^2} + \frac{2 }{\eta_3 - \eta_4} \qquad & \Longrightarrow & \qquad f_0^2 =  \frac{1}{2 \eta_4} \left[ \eta_2 \pm \sqrt{  \eta_2 \left(\eta_2   -2 \eta_4^2 + 2 \eta_3 \eta_4 \right) } \right],
\end{eqnarray}
\end{subequations}
where we defined
\begin{subequations}\label{osces123}
\begin{eqnarray} 
\eta_1 & = & \frac{1}{\al} \left( 16 \ga_3 + 4 \ga_4 + \ga_5 + \ga_6 + 8 \ga_7 + 4 \ga_8 \right) 
\\
\eta_2 & = & - \frac{1}{\al} \left( 4 \ga_4 + 3 \ga_5 + 3  \ga_6 + 8 \ga_7 + 12 \ga_8   \right) 
\\
\eta_3 & = & \frac{1}{\al} \left( 2 \be_1 + \be_2   \right) 
\end{eqnarray}
and the additional combination
\beq \label{osces4}
\eta_4 \equiv \sqrt{\eta_3^2 +  \eta_2 - \eta_1} .
\eeq
\end{subequations}
Similar to the solution of Case IV [see Eq.~\eqref{CaseIVsol}], the above formal solutions exist only if the parameters of the model satisfy constraints in the form of inequalities to guarantee that $f_0^2$ is real and positive. 
In such a case, the model might admit solutions in the classes $\lbrace 0,2 \rbrace$ and $\lbrace 0,2 \rbrace^\infty$ (see discussion in Secs.~\ref{Sec-02} and~\ref{Sec-02Assimpt}, respectively).

\subsection{Case IX}

In this case, we have $\sigma \neq 0$ and $\tau = 2 - 2\si$ and the field equations yield 
\beq
\mathcal{E}^t{}_t = \frac{1}{f_0^6 \, r^{6\sigma}} \sum_{i=1}^8 \ga_i \, g_i^{(t)}(\si,z)   + \ldots ,
\qquad \qquad
\mathcal{E}^r{}_r = \frac{1}{f_0^6 \, r^{6\sigma}} \sum_{i=1}^8 \ga_i \, g_i^{(r)}(\si,z)   + \ldots ,
\eeq
where $z\equiv f_0^2 h_0 \neq 0$ (by definition) and the coefficients $g_i^{(t)}(\si,z)$ and $g_i^{(r)}(\si,z)$ read
{\allowdisplaybreaks 
\begin{subequations}\label{Os_G_ffgghh>=}
\begin{eqnarray}
g_1^{(t)}(\si,z) & = & 8 z \sigma   \big[z\big(\sigma ^2-\sigma +1\big) -1\big] \big[ 
z\big(23 \sigma ^3-9 \sigma ^2-4 \sigma +1\big) + \sigma  \big] ,
\\
g_1^{(r)}(\si,z) & = & - 8 z \sigma \big[z\big(\sigma ^2-\sigma +1\big) -1\big] \big[ z \big(11 \sigma ^3-3 \sigma ^2-1\big) + \sigma  \big] ,
\\
g_2^{(t)}(\si,z) & = & 2 z \sigma   \big[ 
 z^2 \big(38 \sigma ^5-112 \sigma ^4+122 \sigma ^3-34 \sigma ^2-7 \sigma +2\big)
-2 z \sigma  \big(10 \sigma ^2-8 \sigma +1\big)  
-3 \sigma 
\big]
,
\label{Bg2t}
\\
g_2^{(r)}(\si,z) & = &  2 z \sigma   \big[ 
z^2 \big(18 \sigma ^5-116 \sigma ^4+154 \sigma ^3-78 \sigma ^2+11 \sigma +2\big)
+2 z \sigma  \big(14 \sigma ^2-12 \sigma +1\big) 
+3 \sigma 
\big] ,
\label{Bg2r}
\\
g_3^{(t)}(\si,z) & = & -4 \big[z\big(\sigma ^2-\sigma +1\big) -1\big]^2 \big[z\big(37 \sigma ^2+2 \sigma -2\big) -1\big] ,
\\
g_3^{(r)}(\si,z) & = & 4 \big[z\big(\sigma ^2-\sigma +1\big) -1\big]^2 \big[ z \big(17 \sigma ^2+4 \sigma +2\big) +1 \big] ,
\\
g_4^{(t)}(\si,z) & = & 
-2 \big[ 
z^3 \big(34 \sigma ^6-96 \sigma ^5+124 \sigma ^4-56 \sigma ^3+5 \sigma ^2+10 \sigma -2\big) 
- z^2 \big(18 \sigma ^4+20 \sigma ^2+2 \sigma -1\big) +21 z \sigma ^2-1 
\big],
\\
g_4^{(r)}(\si,z) & = & 
2 \big[
z^3 \big(26 \sigma ^6-60 \sigma ^5+76 \sigma ^4-40 \sigma ^3+17 \sigma ^2-4 \sigma +2\big) 
- z^2\big(22 \sigma ^4-8 \sigma ^3+16 \sigma ^2+2 \sigma +1\big)
+ z\sigma  (13 \sigma +2)
+1
\big],
\nonumber
\\
\\
g_5^{(t)}(\si,z) & = & 
-z^3 \big(28 \sigma ^6-72 \sigma ^5+81 \sigma ^4-28 \sigma ^3-9 \sigma ^2+12 \sigma -2\big)
+3 z^2 \sigma ^2 (8 \sigma -1)
-12 z\sigma ^2 +1 ,
\\
g_5^{(r)}(\si,z) & = &  
 z^3 \big(44 \sigma ^6-144 \sigma ^5+189 \sigma ^4-98 \sigma ^3+27 \sigma ^2-6 \sigma +2\big)
-3 z^2 \sigma ^2 (8 \sigma +1)
+12 z \sigma ^2+1,
\\
g_6^{(t)}(\si,z) & = & 
- z^3 \big(48 \sigma ^6-168 \sigma ^5+181 \sigma ^4-48 \sigma ^3-18 \sigma ^2+14 \sigma -2\big)
- z^2 \sigma  \big(16 \sigma ^3-24 \sigma ^2-9 \sigma +2\big)
-9 z \sigma ^2+1 ,
\\
g_6^{(r)}(\si,z) & = & 
-z^3 \big(16 \sigma ^6-88 \sigma ^5+103 \sigma ^4-38 \sigma ^3-10 \sigma ^2+8 \sigma -2\big) 
- z^2 \sigma  \big(32 \sigma ^3-28 \sigma ^2+15 \sigma +2\big)
+9 z \sigma ^2+1  ,
\\
g_7^{(t)}(\si,z) & = & -4 \big[
 z^3 \big(91 \sigma ^6-200 \sigma ^5+169 \sigma ^4-40 \sigma ^3-19 \sigma ^2+14 \sigma -2\big)
-z^2 \big(33 \sigma ^4-8 \sigma ^3-3 \sigma ^2+6 \sigma -1\big) 
+15 z \sigma ^2 -1
\big] ,
\nonumber
\\
\\
g_7^{(r)}(\si,z) & = &  4 \big[
z^3 \big(47 \sigma ^6-98 \sigma ^5+93 \sigma ^4-46 \sigma ^3+21 \sigma ^2-8 \sigma +2\big)
- z^2\big(19 \sigma ^4-4 \sigma ^3+7 \sigma ^2-2 \sigma +1\big)
+ z \sigma  (7 \sigma +2)+1
\big] ,
\nonumber
\\
\\
g_8^{(t)}(\si,z) & = & -4 \big[
z^3 \big(79 \sigma ^6-192 \sigma ^5+132 \sigma ^4+14 \sigma ^3-48 \sigma ^2+18 \sigma -2\big)
-3 z^2 \sigma ^4 +3 z \sigma ^2 -1
\big] ,
\\
g_8^{(r)}(\si,z) & = &  - 4 \big[
z^3 \big(13 \sigma ^6-54 \sigma ^5+48 \sigma ^4+2 \sigma ^3-24 \sigma ^2+12 \sigma -2\big) 
+9 z^2 \sigma ^4-3 z \sigma ^2-1
\big]   .
\end{eqnarray}
\end{subequations}
}
Hence, for the field equations to be solved at lowest order in $r$ for $\sigma$ independent of $\ga_{1,\ldots,8}$, we must have
\beq
\label{ffgghh>=}
g_i^{(t)}(\si,z) = g_i^{(r)}(\si,z) =  0, \quad  \forall \, i=1,\ldots,8 ,
\eeq
which constitute the system of indicial equations for the Case IX.

To prove that the only solution for~\eqref{ffgghh>=} is $\sigma = 1$ and $z=1$, notice that the subsystem
\beq
g_3^{(t)}(\si,z) = g_3^{(r)}(\si,z) =  0 , \qquad \sigma \neq 0, \qquad z \neq 0  
\eeq
only admits the real solutions $(\sigma , z ) = \left(  - \tfrac{1}{9} , - \tfrac{81}{143} \right)$
or in the form 
\beq
\label{CaseBz}
z=\frac{1}{\sigma ^2-\sigma +1} , \qquad \sigma \neq 0.
\eeq
Now, substituting these solutions into the subsystem
\beq
g_2^{(t)}(\si,z) = g_2^{(r)}(\si,z) =  0 , \qquad \sigma \neq 0, \qquad z \neq 0  ,
\eeq
we realise that $(\sigma , z ) = \left(  - \tfrac{1}{9} , - \tfrac{81}{143} \right)$ is not a solution, while for~\eqref{CaseBz} we obtain [see Eqs.~\eqref{Bg2t} and~\eqref{Bg2r}]
\beq
\left\{ 
\begin{array}{l l}
\sigma ( \sigma - 1 ) \big( 15 \sigma ^4-55 \sigma ^3+20 \sigma ^2+10 \sigma -2 \big) &= 0 , \\
\sigma ( \sigma - 1 ) \big( 49 \sigma ^4-125 \sigma ^3+92 \sigma ^2-18 \sigma -2 \big) &= 0 ,\\
\sigma \neq 0 ,
\end{array} \right .
\eeq
whose only solution is $\sigma = 1$, which through~\eqref{CaseBz} implies $z=1$. It is then straightforward to verify that $\sigma=z=1$ is also a zero of all the other functions in~\eqref{Os_G_ffgghh>=}, being, therefore, the only solution of~\eqref{ffgghh>=}. 
To conclude, the Case IX suggests a possible solution family of class $\lbrace 1,0 \rbrace$ (see discussion in Sec.~\ref{Sec-10}).

\section{Some explicit formulas}
\label{app2}

Explicit formulas for the quantities $\mathfrak{a}_{1,2,3}$ and $\mathfrak{a}_{1,2,3}^\prime$ used in Sec.~\ref{Sec-10}:
\beq
\label{osas123elinha}
\begin{split}
\mathfrak{a}_1 & =  4 (64  \ga_1 +27  \ga_2 + 384  \ga_3 + 92  \ga_4 + 27  \ga_5 + 16  \ga_6 +48  \ga_7) ,
\\
\mathfrak{a}_2  & = 4 (52  \ga_1 +21  \ga_2 + 288  \ga_3 + 70  \ga_4 + 18  \ga_5 + 15  \ga_6 + 40  \ga_7) ,
\\
\mathfrak{a}_3  & =  42  \ga_1 + 15  \ga_2 + 216  \ga_3 + 56  \ga_4 + 15  \ga_5 + 14  \ga_6+40  \ga_7 + 6  \ga_8  ,
\\
\mathfrak{a}_1^\prime & =  12 (32  \ga_1 + 13  \ga_2 + 256  \ga_3 + 88  \ga_4 + 31  \ga_5 + 30  \ga_6 + 96  \ga_7 + 40  \ga_8 ) ,
\\
\mathfrak{a}_2^\prime  & =  4 (80  \ga_1 + 33  \ga_2 + 576  \ga_3 + 178  \ga_4 + 60  \ga_5 + 53  \ga_6 + 168  \ga_7 + 60  \ga_8 ) ,
\\
\mathfrak{a}_3^\prime & =  66  \ga_1 +27  \ga_2 +432  \ga_3 +124  \ga_4 +39  \ga_5 +34  \ga_6 +104  \ga_7 +30  \ga_8  .
\end{split}
\eeq
\\

Explicit formula for $\Phi_{rr,4}$ in Eq.~\eqref{semc}:
{\allowdisplaybreaks 
\begin{small}
\bea
\Phi_{rr,4} \, && =  
-4  \ga_3 -2  \ga_4 - \ga_5 - \ga_6 -4 ( \ga_7 + \ga_8 )
-  f_0^2 (2  \be_1 + \be_2 )
-  f_1^2  h_0 (8  \ga_1 +6  \ga_2 +84  \ga_3 +30  \ga_4 +12  \ga_5 +9  \ga_6 +36  \ga_7 +12  \ga_8 )
\nonumber
\\
&& + 2  f_0 f_1 h_1 (6  \ga_3 + \ga_4 +2  \ga_7 )
- f_0^4 \alpha  
-4  f_0^2  f_1^2  h_0 (3  \be_1 + \be_2 )
+3  f_1^4  h_0^2 (-16  \ga_1 -4  \ga_2 +60  \ga_3 +22  \ga_4 +9  \ga_5 +7  \ga_6 +28  \ga_7 +12  \ga_8 )
\nonumber
\\
&&
+4  f_2^2  f_0^2  h_0^2 (16  \ga_1 +4  \ga_2 +48  \ga_3 +14  \ga_4 +3  \ga_5 +5  \ga_6 +8  \ga_7 )-12  f_1  f_3  f_0^2  h_0^2 (16  \ga_1 +4  \ga_2 +48  \ga_3 +14  \ga_4 +3  \ga_5 +5  \ga_6 +8  \ga_7 )
\nonumber
\\
&&
-2  f_1  f_2  f_0^2  h_0  h_1 (20  \ga_1 +4  \ga_2 +96  \ga_3 +32  \ga_4 +6  \ga_5 +13  \ga_6 +32  \ga_7 ) +8  f_1^2  f_2  f_0  h_0^2 (16  \ga_1 +4  \ga_2 +48  \ga_3 +14  \ga_4 +3  \ga_5 +5  \ga_6 +8  \ga_7 )
\nonumber
\\
&&
+2  f_1^3  f_0  h_0  h_1 (24  \ga_1 +10  \ga_2 +132  \ga_3 +34  \ga_4 +9  \ga_5 +8  \ga_6 +20  \ga_7 )+ f_1^2  f_0^2  h_1^2 (24  \ga_1 +6  \ga_2 +60  \ga_3 +14  \ga_4 +3  \ga_5 +4  \ga_6 )
\nonumber
\\
&&
-2  f_1^2  f_0^2  h_0  h_2 (36  \ga_1 +12  \ga_2 +48  \ga_3 +16  \ga_4 +6  \ga_5 +5  \ga_6 )-4  f_2  f_0^3  h_1^2 (4 ( \ga_1 +6  \ga_3 + \ga_4 )+ \ga_6 )-6  f_3  f_0^3  h_0  h_1 (4 ( \ga_1 +6  \ga_3 + \ga_4 )+ \ga_6 )
\nonumber
\\
&&
+4  f_2  f_0^3  h_0  h_2 (4 ( \ga_1 +6  \ga_3 + \ga_4 )+ \ga_6 )-6  f_1  f_0^3  h_0  h_3 (4 ( \ga_1 +6  \ga_3 + \ga_4 )+ \ga_6 )+2  f_0^4  h_2^2 (6  \ga_3 + \ga_4 +2  \ga_7 )-6  f_0^4  h_1  h_3 (6  \ga_3 + \ga_4 +2  \ga_7 )
\nonumber
\\
&&
+\alpha   f_0^4  f_1 ( f_1  h_0+ f_0  h_1)
+  f_0^2 [  f_0^4  h_2^2 (2  \be_1 + \be_2 )-3  f_0^4  h_1  h_3 (2  \be_1 + \be_2 )-4  f_2  f_0^3  h_1^2 (4  \be_1 + \be_2 )-6  f_3  f_0^3  h_0  h_1 (4  \be_1 + \be_2 )
\nonumber
\\
&&
+4  f_2  f_0^3  h_0  h_2 (4  \be_1 + \be_2 )
-2  f_1  f_0^3  h_1  h_2 (2  \be_1 + \be_2 )-6  f_1  f_0^3  h_0  h_3 (4  \be_1 + \be_2 )+4  f_2^2  f_0^2  h_0^2 (8  \be_1 +3  \be_2 )-12  f_1  f_3  f_0^2  h_0^2 (8  \be_1 +3  \be_2 )
\nonumber
\\
&&
+ f_1^2  f_0^2  h_1^2 (2  \be_1 + \be_2 )-6  f_1  f_2  f_0^2  h_0  h_1 (8  \be_1 +3  \be_2 )-2  f_1^2  f_0^2  h_0  h_2 (16  \be_1 +5  \be_2 )+2  f_1^3  f_0  h_0  h_1 (4  \be_1 + \be_2 )+ f_1^4  h_0^2 (14  \be_1 +5  \be_2 ) ]
\nonumber
\\
&&
-2 (4  \ga_3 +2  \ga_4 + \ga_5 + \ga_6 +4 ( \ga_7 + \ga_8 ))  h_2^3  f_0^6-9 (2  \ga_1 + \ga_2 )  h_0  h_3^2  f_0^6+9 (4  \ga_3 +2  \ga_4 + \ga_5 + \ga_6 +4 ( \ga_7 + \ga_8 ))  h_1  h_2  h_3  f_0^6
\nonumber
\\
&&
-24 (2  \ga_1 + \ga_2 )  h_1^2  h_4  f_0^6+24 (2  \ga_1 + \ga_2 )  h_0  h_2  h_4  f_0^6-36 (4  \ga_1 + \ga_2 )  f_3  h_1^3  f_0^5-144 (4  \ga_1 + \ga_2 )  f_4  h_0  h_1^2  f_0^5 +4 (24  \ga_1 +6  \ga_2 -24  \ga_3 
\nonumber
\\
&&
-8  \ga_4 -3  \ga_5 -2  \ga_6 -8  \ga_7 )  f_2  h_0  h_2^2  f_0^5+3 (4  \ga_3 +2  \ga_4 + \ga_5 + \ga_6 +4 ( \ga_7 + \ga_8 ))  f_1  h_1  h_2^2  f_0^5+48 (4  \ga_1 + \ga_2 )  f_4  h_0^2  h_2  f_0^5+4 (-24  \ga_1 
\nonumber
\\
&&
-6  \ga_2 +24  \ga_3 +8  \ga_4 +3  \ga_5 +2  \ga_6 +8  \ga_7 )  f_2  h_1^2  h_2  f_0^5-6 (24  \ga_1 +6  \ga_2 -24  \ga_3 -8  \ga_4 -3  \ga_5 -2  \ga_6 -8  \ga_7 )  f_3  h_0  h_1  h_2  f_0^5-36 (4  \ga_1  
\nonumber
\\
&&
+ \ga_2 )  f_3  h_0^2  h_3  f_0^5+3 (-16  \ga_1 -6  \ga_2 +24  \ga_3 +8  \ga_4 +3  \ga_5 +2  \ga_6 +8  \ga_7 )  f_1  h_1^2  h_3  f_0^5+6 (-44  \ga_1 -12  \ga_2 +24  \ga_3 +8  \ga_4 +3  \ga_5 +2  \ga_6 
\nonumber
\\
&&
+8  \ga_7 )  f_2  h_0  h_1  h_3  f_0^5+6 (-4  \ga_1 +24  \ga_3 +8  \ga_4 +3  \ga_5 +2  \ga_6 +8  \ga_7 )  f_1  h_0  h_2  h_3  f_0^5+48 (4  \ga_1 + \ga_2 )  f_2  h_0^2  h_4  f_0^5-96 (3  \ga_1 + \ga_2 )  f_1  h_0  h_1  h_4  f_0^5
\nonumber
\\
&&
-36 (8  \ga_1 +3  \ga_2 )  f_3^2  h_0^3  f_0^4+96 (8  \ga_1 +3  \ga_2 )  f_2  f_4  h_0^3  f_0^4+4 (12  \ga_1 +5  \ga_2 +3 (16  \ga_3 +4  \ga_4 + \ga_5 + \ga_6 )+4  \ga_7 )  f_1  f_2  h_1^3  f_0^4+4 (-4  \ga_1 +2  \ga_2 
\nonumber
\\
&&
+96  \ga_3 +26  \ga_4 +9  \ga_5 +5  \ga_6 +12  \ga_7 )  f_2^2  h_0  h_1^2  f_0^4+6 (-60  \ga_1 -22  \ga_2 +3 (16  \ga_3 +4  \ga_4 + \ga_5 + \ga_6 )+4  \ga_7 )  f_1  f_3  h_0  h_1^2  f_0^4+(-88  \ga_1 
\nonumber
\\
&&
-30  \ga_2 
+180  \ga_3 +50  \ga_4 +15  \ga_5 +14  \ga_6 +12  \ga_7 )  f_1^2  h_0  h_2^2  f_0^4+24 (14  \ga_1 +7  \ga_2 +24  \ga_3 +7  \ga_4 +3  \ga_5 + \ga_6 +4  \ga_7 )  f_2  f_3  h_0^2  h_1  f_0^4
\nonumber
\\
&&
-264 (8  \ga_1 
+3  \ga_2 )  f_1  f_4  h_0^2  h_1  f_0^4+8 (48  \ga_1 +13  \ga_2 -2 (24  \ga_3 +7  \ga_4 +3  \ga_5 + \ga_6 +4  \ga_7 ))  f_2^2  h_0^2  h_2  f_0^4-24 (62  \ga_1 +20  \ga_2 -24  \ga_3 -7  \ga_4
\nonumber
\\
&&
-3  \ga_5 - \ga_6 
-4  \ga_7 )  f_1  f_3  h_0^2  h_2  f_0^4+(24  \ga_1 +10  \ga_2 +36  \ga_3 +10  \ga_4 +3  \ga_5 +2  \ga_6 +12  \ga_7 )  f_1^2  h_1^2  h_2  f_0^4+(-448  \ga_1 -116  \ga_2 +288  \ga_3 +88  \ga_4 
\nonumber
\\
&&
+42  \ga_5 
+8  \ga_6 
+80  \ga_7 )  f_1  f_2  h_0  h_1  h_2  f_0^4-12 (48  \ga_1 +13  \ga_2 -2 (24  \ga_3 +7  \ga_4 +3  \ga_5 + \ga_6 +4  \ga_7 ))  f_1  f_2  h_0^2  h_3  f_0^4+6 [-22  \ga_1 -5  \ga_2 +54  \ga_3 
\nonumber
\\
&&
+13  \ga_4 
+3 ( \ga_5 
+ \ga_6 +2  \ga_7 )]  f_1^2  h_0  h_1  h_3  f_0^4-24 (24  \ga_1 +7  \ga_2 )  f_1^2  h_0^2  h_4  f_0^4-16 ( \ga_2 +32  \ga_3 +12  \ga_4 +5  \ga_5 +4  \ga_6 +16  \ga_7 +8  \ga_8 )  f_2^3  h_0^3  f_0^3
\nonumber
\\
&&
+72 ( \ga_2 
+32  \ga_3 +12  \ga_4 +5  \ga_5 +4  \ga_6 +16  \ga_7 +8  \ga_8 )  f_1  f_2  f_3  h_0^3  f_0^3
-(2  \ga_2 +40  \ga_3 +12  \ga_4 +4  \ga_5 +3  \ga_6 +16  \ga_7 +4  \ga_8 )  f_1^3  h_1^3  f_0^3
\nonumber
\\
&&
+2 (104  \ga_1 +22  \ga_2 +312  \ga_3 +84  \ga_4 +24  \ga_5 +25  \ga_6 +80  \ga_7 +36  \ga_8 )  f_1^2  f_2  h_0  h_1^2  f_0^3
+8 (12  \ga_1 +6  \ga_2 +120  \ga_3 +44  \ga_4 +18  \ga_5  
\nonumber
\\
&&
+15  \ga_6 
+56  \ga_7 +30  \ga_8 )  f_1  f_2^2  h_0^2  h_1  f_0^3+6 (52  \ga_1 +14  \ga_2 +216  \ga_3 +60  \ga_4 +18  \ga_5 +17  \ga_6 +64  \ga_7 +24  \ga_8 )  f_1^2  f_3  h_0^2  h_1  f_0^3+4 (12  \ga_1  
\nonumber
\\
&&
+6  \ga_2 +168  \ga_3 
+52  \ga_4 +18  \ga_5 +15  \ga_6 +64  \ga_7 +24  \ga_8 )  f_1^2  f_2  h_0^2  h_2  f_0^3+(80  \ga_1 +28  \ga_2 +336  \ga_3 +88  \ga_4 +24  \ga_5 +22  \ga_6 +96  \ga_7 
\nonumber
\\
&&
+24  \ga_8 )  f_1^3  h_0  h_1  h_2  f_0^3+6 (4 (5  \ga_1 + \ga_2 +6  \ga_3 + \ga_4 )+ \ga_6 )  f_1^3  h_0^2  h_3  f_0^3+4 (48  \ga_1 +13  \ga_2 -48  \ga_3 -22  \ga_4 -12  \ga_5 -7  \ga_6 -40  \ga_7 
\nonumber
\\
&&
-24  \ga_8 )  f_1^2  f_2^2  h_0^3  f_0^2+12 (48  \ga_1 +16  \ga_2 +48  \ga_3 +14  \ga_4 +3  \ga_5 +5  \ga_6 +8  \ga_7 )  f_1^3  f_3  h_0^3  f_0^2-(34  \ga_1 +7  \ga_2 +156  \ga_3 +38  \ga_4 +9  \ga_5 
\nonumber
\\
&&
+10  \ga_6 +24  \ga_7 +6  \ga_8 )  f_1^4  h_0  h_1^2  f_0^2+2 (148  \ga_1 +54  \ga_2 +4  \ga_4 -6  \ga_5 +7  \ga_6 -16  \ga_7 -24  \ga_8 )  f_1^3  f_2  h_0^2  h_1  f_0^2+2 (68  \ga_1 +24  \ga_2 
\nonumber
\\
&&
+48  \ga_3 +16  \ga_4 +6  \ga_5 +5  \ga_6 )  f_1^4  h_0^2  h_2  f_0^2-8 (48  \ga_1 +16  \ga_2 +48  \ga_3 +14  \ga_4 +3  \ga_5 +5  \ga_6 +8  \ga_7 )  f_1^4  f_2  h_0^3  f_0-2 (56  \ga_1 +22  \ga_2 
\nonumber
\\
&&
+126  \ga_3 +33  \ga_4 +9  \ga_5 +8  \ga_6 +18  \ga_7 )  f_1^5  h_0^2  h_1  f_0+(56  \ga_1 +18  \ga_2 -92  \ga_3 -34  \ga_4 -14  \ga_5 -11 ( \ga_6 +4  \ga_7 )-20  \ga_8 )  f_1^6  h_0^3.
\label{ograndec}
\eea
\end{small}
}
\\

Explicit formula for the constraint between the parameters $c_0$, $c_2$, $b_2$, $b_4$ and $\bar{r}_0$ in Eq.~\eqref{Sol-Sch-10-Frob}:
{\allowdisplaybreaks 
\begin{small}
\beq
\label{Const-Sol-Sch-10-Frob}
\begin{split}
0 = & \,
128  \ga_3 +64  \ga_4 +32  \ga_5 +32  \ga_6 +128  \ga_7 +128  \ga_8 
+32  \bar{r}_0 ^2 (2  \be_1 + \be_2 )
+32  \al    \bar{r}_0 ^4 
-8  c_0 ^2  \bar{r}_0 ^4 \left[ 8  \be_1 +3  \be_2 + 2 c_2 (8  \ga_1 +3  \ga_2 ) \right] 
\\
& 
-8  b_2   c_0 ^2  \bar{r}_0 ^5 \left[ 4  \be_1 + \be_2 +2  c_2  (4  \ga_1 + \ga_2 ) \right] 
-2  b_2 ^2  c_0 ^2  \bar{r}_0 ^6 \left[ 2  \be_1 + \be_2 +2  c_2  (2  \ga_1 + \ga_2 ) \right] 
-4  b_2 ^2  c_0 ^2  \bar{r}_0 ^4 (6  \ga_3 + \ga_4 +2  \ga_7 )
\\
& 
-8  c_0 ^2  \bar{r}_0 ^2 (16  \ga_1 +4  \ga_2 +48  \ga_3 +14  \ga_4 +3  \ga_5 +5  \ga_6 +8  \ga_7 )
+8  c_0 ^3  \bar{r}_0 ^3 ( \ga_2 +32  \ga_3 +12  \ga_4 +5  \ga_5 +4  \ga_6 +16  \ga_7 +8  \ga_8 )
\\
& 
+ b_2 ^3  c_0 ^3  \bar{r}_0 ^6 (8  \ga_1 +4  \ga_2 +4  \ga_3 +2  \ga_4 + \ga_5 + \ga_6 +4  \ga_7 +4  \ga_8 )
+2  b_2 ^2  c_0 ^3  \bar{r}_0 ^5 (8  \ga_1 +2  \ga_2 +24  \ga_3 +8  \ga_4 +3  \ga_5 +2  \ga_6 +8  \ga_7 )
\\
& 
-4  b_2   c_0 ^3  \bar{r}_0 ^4 \left[ 16  \ga_1 + \ga_2 -2 (24  \ga_3 +7  \ga_4 +3  \ga_5 + \ga_6 +4  \ga_7 ) \right] 
-8  b_2   c_0 ^2  \bar{r}_0 ^3 (4  \ga_1 +24  \ga_3 +4  \ga_4 + \ga_6 )
\\
&
-12  b_2   b_4   c_0 ^3  \bar{r}_0 ^6 (2  \ga_1 + \ga_2 )
-24  b_4   c_0 ^3  \bar{r}_0 ^5 (4  \ga_1 + \ga_2 ) .
\end{split}
\eeq
\end{small}
}
\\

Explicit formulas for the quantities $\mathfrak{c}_{1,2}(N)$ and $\mathfrak{c}_{1,2}^\prime(N)$ used in Sec.~\ref{Sec-02}:\footnote{In the these equations, the coefficients $f_0$ and $h_0$ are fixed by the parameters of the model according to the possibilities listed in Eq.~\eqref{SolCaseVII}. Although the scenarios with $f_0^2 h_0 = \pm 1$ yield considerable simplification of the quantities $\mathfrak{c}_{1,2}$, $\mathfrak{c}_{1,2}^\prime$ and the combination $\mathfrak{c}_1^\prime \mathfrak{c}_2 - \mathfrak{c}_1 \mathfrak{c}_2^\prime$, they can only vanish for $N>1$ if particular relations between the parameters $\alpha$, $\beta_{1,2}$ and $\gamma_{1,\cdots,8}$ are assumed.}
\begin{small}
{\allowdisplaybreaks 
\begin{subequations}\label{c1a4}
\begin{eqnarray}
\mathfrak{c}_1 & = &  
2 \alpha f_0^4  \Big(f_0^2 h_0 N^2+ 1\Big) + 8 \beta _1 f_0^2 \Big[ 1 - f_0^4 h_0^2 N \big(N^3+N^2-1\big)\Big] + 2 \beta _2 f_0^2 \Big[2 - f_0^4 h_0^2 N \big(N^3+N^2+N-1\big)\Big]
\nonumber
\\
&& 
- 8 \gamma _1 f_0^4 h_0^2 (N-1) N (N+1)^2 \Big[f_0^2 h_0 N (N+1) + 1\Big]
-2 \gamma _2 f_0^6 h_0^3 (N-1) N^2 (N+1)^3
\nonumber
\\
&&
+24 \gamma _3 \big(f_0^2 h_0-1\big) \Big\lbrace f_0^2 h_0 \big[f_0^2 h_0 (2 N^3+2 N^2-N-2) N+N^2-1\big]-1\Big]
\nonumber
\\
&&
+4\gamma_4 \Big\lbrace 
f_0^6 h_0^3 N \big(4 N^3+4 N^2-N-4\big)-f_0^4 h_0^2 \big[(2 (N-1) N (N+1)^2+1\big] -f_0^2 h_0 N^2 +3
\Big\rbrace
\nonumber
\\
&&
+6\gamma _5 \Big[ f_0^6 h_0^3 N \big(N^3+N^2-1\big) + 1 \Big]
+2\gamma _6 \Big\lbrace f_0^4 h_0^2 N \big[f_0^2 h_0 \big(2 (N+1) N^2+N-2\big)-(N-1) (N+1)^2\big]+3\Big\rbrace
\nonumber
\\
&&
+8\gamma _7 \Big\lbrace f_0^6 h_0^3 N \big[2 (N+1) N^2+N-2\big] -f_0^4 h_0^2 -f_0^2 h_0 N^2 + 3 \Big\rbrace
+24 \gamma _8 \Big(f_0^6 h_0^3 N^2+1\Big) ,
\\
\mathfrak{c}_2 & = & 
- f_0^5 h_0 (N+1)^2 \big(N^2+N-2\big) \Big\lbrace
(2 \beta _1+\beta _2) f_0^2
+ (2 \gamma _1 +\gamma _2 ) f_0^2 h_0 N (N+1)
+12 \gamma _3 \big(1- f_0^2 h_0\big)
+2\gamma _4 \big(1-3 f_0^2 h_0\big)
\nonumber
\\
&&
-3 ( \gamma _5 + \gamma _6 ) f_0^2 h_0
+\gamma _7 \big(4-12 f_0^2 h_0\big)
-12 \gamma _8 f_0^2 h_0
\Big\rbrace ,
\\
\mathfrak{c}_1^\prime & = & 
-2 \alpha  f_0^6 h_0 N (N+1) 
+8 \beta _1 f_0^2 \Big\lbrace f_0^2 h_0 N (N+1) \big[f_0^2 h_0 (2 N (N+1)+1)+3\big]+2\Big\rbrace
+2\beta _2 f_0^2 \Big\lbrace f_0^2 h_0 N (N+1) \big[3 f_0^2 h_0 N (N+1)+4\big]+4\Big\rbrace
\nonumber
\\
&&
+16 \gamma _1 h_0 N (N+1) \Big[f_0^3 h_0 N (N+1)+f_0\Big]^2
+2 \gamma _2 f_0^2 h_0 N (N+1) \Big\lbrace 4 f_0^2 h_0 \big(N^2+N+1\big)+f_0^4 h_0^2 [3 (N-1) N (N+1) (N+2)+10]+6\Big\rbrace
\nonumber
\\
&&
-24 \gamma _3 \Big\lbrace f_0^2 h_0 \big[N (N+1) \big(f_0^2 h_0-1\big) \big(f_0^2 h_0 (2 N+1)^2+7\big)+4\big]-4 \Big\rbrace
-\gamma _4 \Big\lbrace 4 f_0^2 h_0 \big[N (N+1) \big(f_0^4 h_0^2 (7 N^2+7 N+1)
\nonumber
\\
&&
 - f_0^2 h_0 (7 N^2+7 N-8) - 15\big) + 4\big] -48 \Big\rbrace
- 6 \gamma _5 \Big\lbrace f_0^2 h_0 N \big[f_0^4 h_0^2 \big(2 (N+2) N^2+N-1\big) - f_0^2 h_0 N (N+1)^2 - 4(N+1) \big] - 4 \Big\rbrace
\nonumber
\\
&&
-2 \gamma _6 \Big\lbrace f_0^2 h_0 N (N+1) \big[ 2 f_0^4 h_0^2 \big(N^2+N+2\big) - f_0^2 h_0 (5 N^2 + 5 N - 6)-9\big]-12 \Big\rbrace
-8\gamma _7 \Big\lbrace f_0^2 h_0 \big[N (N+1) \big(
f_0^4 h_0^2 (2 N^2+2 N-1)
\nonumber
\\
&&
- 2 f_0^2 h_0 \big(N^2+N-2\big) 
-9
\big)+4\big]-12 \Big\rbrace
+24 \gamma _8 \Big[ f_0^2 h_0 N (N+1)+4 \Big] ,
\\
\mathfrak{c}_2^\prime & = & 
- f_0^3 (N+1) (N+2) \Big\lbrace
\alpha  f_0^4 - f_0^4 h_0 \big[4 \beta _1 \big(N^2+N+1\big)+\beta _2 \big(N^2+N+2\big)\big]
- 4 \gamma _1 f_0^2 h_0 N (N+1) \Big[f_0^2 h_0 N (N+1)+1\Big]
\nonumber
\\
&&
-\gamma _2 f_0^4 h_0^2 N^2 (N+1)^2
+12 \gamma _3 \Big[ f_0^4 h_0^2 (2 N (N+1)+1) - 2 f_0^2 h_0 N (N+1)-1\big]
+2\gamma _4 \Big[ f_0^4 h_0^2 (4 N (N+1)+3)-2 f_0^2 h_0 N (N+1)-1\Big]
\nonumber
\\
&&
+3 \gamma _5 f_0^4 h_0^2 \big(N^2+N+1\big)
+\gamma _6 f_0^2 h_0 \Big[ f_0^2 h_0 (2 N (N+1)+3)-N (N+1)\Big]
+4\gamma _7 \Big[f_0^4 h_0^2 (2 N (N+1)+3)-1\Big]
+12 \gamma _8 f_0^4 h_0^2
\Big\rbrace .
\end{eqnarray}
\end{subequations}}\end{small}
\\

Explicit formulas for the quantities $\mathfrak{d}_{1,2,3,4}(N)$ used in Sec.~\ref{Sec-12Assimpt}:
\begin{small}
{\allowdisplaybreaks 
\begin{subequations}\label{d1a4}
\begin{eqnarray}
\mathfrak{d}_1 & = &  
( N -2) ( N -1) (4  \beta_1  + \beta_2  ) 
+ ( N -3) N \lbrace 
4 ( N -4)  ( N +1) \gamma_1   + [( N -3)  N -6] \gamma_2  
-144  \gamma_3  - 36 \gamma_4   -9 \gamma_5    - 9  \gamma_6 
\rbrace h_0 
\nonumber
\\
&&
-16  [( N -3)  N -1 ] \gamma_7  h_0   + 8  \gamma_8   h_0   ,
\\
\mathfrak{d}_2   & = &  
( N -1)  N  [2 ( N -5) \beta_1   + ( N -4) \beta_2   ]
+ ( N -3) N [
2 ( N -4) ( N -3) ( N +1) \gamma_1    + ( N^3 - 5 N^2 + 2N +10) \gamma_2  
] h_0 
\nonumber
\\
&&
- [
72 ( N ^3-6  N ^2+5  N +4 ) \gamma_3    + 12  (2  N ^3-11  N ^2+9  N +6)  \gamma_4 
+ 9  ( N ^3-5  N ^2+4  N +2 ) \gamma_5   
\nonumber
\\
&&
 + (7  N ^3-37  N ^2+30  N +18 ) \gamma_6   
+ 16  (2  N ^3-9  N ^2+7  N +3 ) \gamma_7   
+ 4  (3  N ^3-12  N ^2+9  N +2 )  \gamma_8   
] h_0  ,
\\
\mathfrak{d}_3   & = & 
-2 ( N -3)  N  \lbrace  3  \beta_1  + \beta_2 + 3  \gamma_1   h_0  ( N -4) ( N +1) + \gamma_2   h_0  [( N -3)  N -5] \rbrace
+ \lbrace
72    [3 ( N -3)  N -4] \gamma_3 + 12    [5 ( N -3)  N -6]  \gamma_4 
\nonumber
\\
&&
+18     [( N -3)  N -1] \gamma_5  
+ 2   [8 ( N -3)  N -9]  \gamma_6
+48    [( N -3)  N -1]  \gamma_7 +4     [3 ( N -3)  N -2] \gamma_8 
\rbrace h_0 ,
\\
\mathfrak{d}_4   & = & (N-1) \big\lbrace 24  \beta_1  - [( N -3)  N - 6 ] \beta_2   
+ \lbrace
(3-N)  N  [( N -3)  N -2] \gamma_2    
-576  \gamma_3   
+12 [( N -3)  N -12]  \gamma_4   
\nonumber
\\
&&
+9  [( N -3)  N -4]  \gamma_5  
+  [5 ( N -3)  N -36] \gamma_6 
+48  [( N -3)  N -1]  \gamma_7 
+8   [3 ( N -3)  N +1]  \gamma_8 
\rbrace h_0 \big\rbrace  ,
\label{d1a4.4}
\end{eqnarray}
\end{subequations}}
\end{small}
\\

Explicit formulas for the quantities $\mathfrak{e}_{1,2,3,4}(N)$ used in Sec.~\ref{Sec-02Assimpt}:
\begin{small}
{\allowdisplaybreaks 
\begin{subequations}\label{e1a4}
\begin{eqnarray}
\mathfrak{e}_1 & = &  
\alpha  f_0^4 \big(f_0^2 h_0 N-1\big) - 2 (2 \beta _1+\beta _2) f_0^2 - f_0^6 h_0^2 N \big[\beta _2+(4 \beta _1+\beta _2) N^2\big]
-4 \gamma _1 f_0^4 h_0^2 N \Big[f_0^2 h_0 (N-1)^2 (N+1) N+N^2-1\Big]
\nonumber
\\
&&
- \gamma _2 f_0^6 h_0^3 (N-1)^2 N^2 (N+1)
+12 \gamma _3 \big(f_0^2 h_0-1\big) \Big[f_0^4 h_0^2 N \big(2 N^2-1\big)+f_0^2 h_0 (N+1)+1\Big]
\nonumber
\\
&&
+2\gamma _4 \Big[2 f_0^4 h_0^2 N^3 \big(2 f_0^2 h_0-1\big)- f_0^2 h_0 N \big(f_0^2 h_0-1\big)^2+ f_0^4 h_0^2-3\Big]
+3\gamma _5 \Big(f_0^6 h_0^3 N^3-1\Big)
\nonumber
\\
&&
+\gamma _6 \Big[f_0^6 h_0^3 \big(2 N^3+N\big)-f_0^4 h_0^2 N \big(N^2-1\big)-3\Big]
+4\gamma_7 \Big[ f_0^6 h_0^3 \big(2 N^3+N\big)+f_0^4 h_0^2-f_0^2 h_0 N-3 \Big]
+12 \gamma _8 \Big(f_0^6 h_0^3 N-1\Big) ,
\\
\mathfrak{e}_2   & = &  
f_0^5 h_0 (N+1) \Big\lbrace 2 \left(6 \gamma _3+\gamma _4+2 \gamma _7\right) 
+f_0^2 (2 \beta _1+\beta _2)
+ f_0^2 h_0 [
(N-1) N(2 \gamma _1+\gamma _2)
-3 \left(4 \gamma _3+2 \gamma _4+\gamma _5+\gamma _6+4 \gamma _7+4\gamma _8\right) 
]
\Big\rbrace ,
\\
\mathfrak{e}_3   & = & 
- \alpha  f_0^6 h_0 (N-1) N
+4 (2 \beta _1+\beta _2) f_0^2+4 (3 \beta _1+\beta _2) f_0^4 h_0 (N-1) N
+ f_0^6 h_0^2 (N-1) N \lbrace 4 \beta _1 [2 (N-1) N+1]+3 \beta _2 (N-1) N \rbrace
\nonumber
\\
&&
+8 \gamma _1 h_0 (N-1) N \Big[f_0^3 h_0 (N-1) N+f_0\Big]^2
+ \gamma_2 f_0^2 h_0 (N-1) N \Big[ f_0^4 h_0^2 \big(3 N^4-6 N^3-3 N^2+6 N+10\big) + 4 f_0^2 h_0 \big(N^2-N+1\big) +6 \Big]
\nonumber
\\
&&
-12 \gamma_3 \big(f_0^2 h_0-1\big) \Big[ f_0^4 h_0^2 (N-1) N (1-2 N)^2+7 f_0^2 h_0 (N-1) N+4 \Big]
- 2 \gamma_4 \Big\lbrace
f_0^6 h_0^3 (N-1) N [7 (N-1) N+1] 
\nonumber
\\
&&
- f_0^4 h_0^2 (N-1) N [7 (N-1) N-8] - f_0^2 h_0 [15 (N-1) N-4] - 12
\Big\rbrace
+3 \gamma_5 \Big\lbrace 
f_0^6 h_0^3 (N-1) N [1-2 (N-1) N] + f_0^4 h_0^2 (N-1)^2 N^2 
\nonumber
\\
&&
+ 4 f_0^2 h_0 (N-1) N + 4
\Big\rbrace
- \gamma_6 \Big\lbrace 
2 f_0^6 h_0^3 (N-1) N [(N-1) N+2] - f_0^4 h_0^2 (N-1) N [5 (N-1) N-6] - 9 f_0^2 h_0 (N-1) N - 12
\Big\rbrace
\nonumber
\\
&&
+ \gamma_7 \Big\lbrace 
4 f_0^6 h_0^3 (N-1) N [1-2 (N-1) N] + 8 f_0^4 h_0^2 (N-2) (N-1) N (N+1) + 4 f_0^2 h_0 [9 (N-1) N-4] + 48 \Big\rbrace
\nonumber
\\
&&
+ \gamma _8 \Big[12 f_0^2 h_0 (N-1) N+48\Big] ,
\\
\mathfrak{e}_4   & = & 
-\alpha  f_0^7+4 \beta _1 f_0^7 h_0 [(N-1) N+1]+\beta _2 f_0^7 h_0 [(N-1) N+2]
+ 4 \gamma _1 f_0^5 h_0 (N-1) N \Big(f_0^2 h_0 (N-1) N+1\Big)
+ \gamma _2 f_0^7 h_0^2 (N-1)^2 N^2
\nonumber
\\
&&
-12 \gamma _3 f_0^3 \Big\lbrace f_0^4 h_0^2 [2 (N-1) N + 1] - 2 f_0^2 h_0 (N-1) N - 1 \Big\rbrace
- \gamma _4 f_0^3 \Big\lbrace 2 f_0^4 h_0^2 [4 (N-1) N+3] - 4 f_0^2 h_0 (N-1) N - 2 \Big\rbrace
\nonumber
\\
&&
-3 \gamma _5 f_0^7 h_0^2 \big(N^2-N+1\big)
-\gamma _6 f_0^5 h_0 \Big\lbrace f_0^2 h_0 [2 (N-1) N + 3] - (N-1) N \Big\rbrace
-\gamma _7 f_0^3 \Big[  4 f_0^4 h_0^2 \big(2 N^2-2 N+3\big) - 4 \Big]
-12 \gamma _8 f_0^7 h_0^2.
\end{eqnarray}
\end{subequations}}
\end{small}

\end{document}